\pdfoutput=1
\documentclass[11pt,twoside,a4paper,cmspaper,final,collab]{cms-tdr}
\def\svnVersion{7daba56}\def\svnDate{2024/05/17}\def\cmsCernNoTag{CERN-EP-2023-201}\def\cmsCernDate{\today}\def\cmsMessage{Published in the Journal of High Energy Physics as \href{http://dx.doi.org/10.1007/JHEP05(2024)042}{\doi{10.1007/JHEP05(2024)042}.}}
\begin{document}\cmsNoteHeader{TOP-22-009}

\newcommand{\Vjets}{\ensuremath{\PV{+}\text{jets}}\xspace}
\newcommand{\Wjets}{\ensuremath{\PW{+}\text{jets}}\xspace}
\newcommand{\Zjets}{\ensuremath{\PZ{+}\text{jets}}\xspace}
\newcommand{\ttV}{\ensuremath{\ttbar\PV}\xspace}

\newcommand{\pp}{\ensuremath{\Pp\Pp}\xspace}
\newcommand{\PQj}{{\HepParticle{j}{}{}}\xspace}
\newcommand{\PQl}{{\HepParticle{l}{}{}}\xspace}
\newcommand{\PQC}{{\HepParticle{C}{}{}}\xspace}
\newcommand{\bb}{\ensuremath{\PQb\PQb}\xspace}
\newcommand{\ttlight}{\ensuremath{\ttbar{+}\textrm{light}}\xspace}
\newcommand{\ttC}{\ensuremath{\ttbar\PQC}\xspace}
\newcommand{\ttjets}{\ensuremath{\ttbar{+}\textrm{jets}}\xspace}
\newcommand{\ttb}{\ensuremath{\ttbar\PQb}\xspace}
\newcommand{\ttbb}{\ensuremath{\ttbar\bbbar}\xspace}
\newcommand{\ttbbj}{\ensuremath{\ttbar\bbbar\PQj}\xspace}
\newcommand{\ttB}{\ensuremath{\ttbar\PQB}\xspace}
\newcommand{\ttbj}{\ensuremath{\ttbar\PQb\PQj}\xspace}

\newcommand{\gtobb}{\ensuremath{\Pg\to\bbbar}\xspace}
\newcommand{\ZtoLL}{\ensuremath{\PZ\to\Pell^+\Pell^-}\xspace}

\newcommand{\ttH}{\ensuremath{\ttbar\PH}\xspace}
\newcommand{\ttZ}{\ensuremath{\ttbar\PZ}\xspace}
\newcommand{\ttW}{\ensuremath{\ttbar\PW}\xspace}
\newcommand{\ttX}{\ensuremath{\ttbar\PX}\xspace}
\newcommand{\tW}{\ensuremath{\PQt\PW}\xspace}

\newcommand{\deltaR}{\ensuremath{\DR}\xspace}
\newcommand{\dPhi}{\ensuremath{\Delta\phi}\xspace}
\newcommand{\dPhibb}{\ensuremath{\Delta\phi_{\bb}}\xspace}
\newcommand{\dEtabb}{\ensuremath{\Delta\eta_{\bb}}\xspace}
\newcommand{\dRbb}{\ensuremath{\deltaR_{\bb}}\xspace}
\newcommand{\abseta}{\ensuremath{\abs{\eta}}\xspace}
\newcommand{\absetaSC}{\ensuremath{\abs{\eta_{\mathrm{SC}}}}\xspace}
\newcommand{\mbb}{\ensuremath{m_{\bb}}\xspace}

\newcommand{\sigmatot}{\ensuremath{\sigma_{\text{fid}}}\xspace}
\newcommand{\Nj}{\ensuremath{N_{\text{jets}}}\xspace}
\newcommand{\Nb}{\ensuremath{N_{\PQb}}\xspace}

\newcommand{\dRbbavg}{\ensuremath{\deltaR_{\bb}^{\text{avg}}}\xspace}
\newcommand{\mbbmax}{\ensuremath{\mbb^{\text{max}}}\xspace}
\newcommand{\bbextra}{\smash[t]{\ensuremath{\bb^{\text{extra}}}}\xspace}
\newcommand{\dRbbextra}{\ensuremath{\deltaR(\bbextra)}\xspace}
\newcommand{\mbbextra}{\ensuremath{m(\bbextra)}\xspace}
\newcommand{\ptbbextra}{\ensuremath{\pt(\bbextra)}\xspace}
\newcommand{\absetabbextra}{\ensuremath{\abs{\eta(\bbextra)}}\xspace}

\newcommand{\bN}[1]{\ensuremath{\PQb_{#1}}\xspace}
\newcommand{\bXN}[1]{\ensuremath{\PQb^{\text{extra}}_{#1}}\xspace}
\newcommand{\ptbN}[1]{\ensuremath{\pt(\bN{#1})}\xspace}
\newcommand{\absetabN}[1]{\ensuremath{\abs{\eta(\bN{#1})}}\xspace}
\newcommand{\ptbXN}[1]{\ensuremath{\pt(\bXN{#1})}\xspace}
\newcommand{\absetabXN}[1]{\ensuremath{\abs{\eta(\bXN{#1})}}\xspace}

\newcommand{\HTl}{\smash{\ensuremath{\HT^{\text{light}}}}\xspace}
\newcommand{\HTb}{\smash{\ensuremath{\HT^{\PQb}}}\xspace}
\newcommand{\HTj}{\smash{\ensuremath{\HT^{\PQj}}}\xspace}

\newcommand{\ljXN}[1]{\ensuremath{\text{lj}^{\text{extra}}_{#1}}\xspace}
\newcommand{\ptljXN}[1]{\ensuremath{\pt(\ljXN{#1})}\xspace}
\newcommand{\dPhiljb}{\ensuremath{\abs{\dPhi(\ljXN{1},\PQb_{\text{soft}})}}\xspace}

\newcommand{\bbadd}{\ensuremath{\bb^{\text{add.}}}\xspace}
\newcommand{\dRbbadd}{\ensuremath{\deltaR(\bbadd)}\xspace}
\newcommand{\mbbadd}{\ensuremath{m(\bbadd)}\xspace}
\newcommand{\ptbbadd}{\ensuremath{\pt(\bbadd)}\xspace}
\newcommand{\absetabbadd}{\ensuremath{\abs{\eta(\bbadd)}}\xspace}
\newcommand{\bAN}[1]{\ensuremath{\PQb^{\text{add.}}_{#1}}\xspace}
\newcommand{\ptbAN}[1]{\ensuremath{\pt(\bAN{#1})}\xspace}
\newcommand{\absetabAN}[1]{\ensuremath{\abs{\eta(\bAN{#1})}}\xspace}

\newcommand{\cmsTable}[1]{\resizebox{\textwidth}{!}{#1}}
\newlength\cmsTabSkip\setlength\cmsTabSkip{1ex}

\newcommand{\muR}{\ensuremath{\mu_{\mathrm{R}}}\xspace}
\newcommand{\muF}{\ensuremath{\mu_{\mathrm{F}}}\xspace}
\newcommand{\hdamp}{\ensuremath{h_{\mathrm{damp}}}\xspace}
\newcommand{\fullcorr}{\ensuremath{\checkmark}\xspace}
\newcommand{\partcorr}{\ensuremath{\sim}\xspace}
\newcommand{\nocorr}{\ensuremath{\times}\xspace}
\newcommand{\chitwo}{\ensuremath{\chi^2}\xspace}

\newcommand{\POWHEGBOX}{{\POWHEG}\textsc{-box-res}\xspace}
\newcommand{\OPENLOOPS}{\textsc{OpenLoops}\xspace}
\newcommand{\fxfx}{\text{FxFx}\xspace}
\newcommand{\MadSpin}{\textsc{MadSpin}\xspace}

\newcommand{\sjfb}{\ensuremath{6\PQj4\PQb}\xspace}
\newcommand{\fjtb}{\ensuremath{5\PQj3\PQb}\xspace}
\newcommand{\sjtbtl}{\ensuremath{6\PQj3\PQb3\PQl}\xspace}
\newcommand{\sejfbtl}{\ensuremath{7\PQj4\PQb3\PQl}\xspace}

\newcommand{\sjfbLONG}{$\geq$6 jets: $\geq$4\PQb\xspace}
\newcommand{\fjtbLONG}{$\geq$5 jets: $\geq$3\PQb\xspace}
\newcommand{\sjtbtlLONG}{$\geq$6 jets: $\geq$3\PQb, $\geq$3 light\xspace}
\newcommand{\sejfbtlLONG}{$\geq$7 jets: $\geq$4\PQb, $\geq$3 light\xspace}

\newcommand{\fourFS}{\ensuremath{\text{4FS}}\xspace}
\newcommand{\fiveFS}{\ensuremath{\text{5FS}}\xspace}

\newcommand{\ttbbPP}{\textsc{powheg+ol+p8} \ttbb \fourFS}
\newcommand{\ttbarPP}{\textsc{powheg+p8} \ttbar \fiveFS}
\newcommand{\ttbarPH}{\textsc{powheg+h7} \ttbar \fiveFS}
\newcommand{\ttbbAMC}{\textsc{mg5\_}a\textsc{mc+p8} \ttbb \fourFS}
\newcommand{\ttbbSherpa}{\textsc{sherpa+ol} \ttbb \fourFS}
\newcommand{\ttjetsAMC}{\textsc{mg5\_}a\textsc{mc+p8} \ttjets \fxfx \fiveFS}

\newcommand{\mtop}{\ensuremath{m_{\PQt}}\xspace}
\newcommand{\mTtop}{\ensuremath{m_{\mathrm{T},\PQt}}\xspace}
\newcommand{\mTi}{\ensuremath{m_{\mathrm{T},i}}\xspace}
\newcommand{\mTidef}{\ensuremath{\mTi=\sqrt{\smash[b]{m_i^2+p_{\mathrm{T},i}^2}}}\xspace}
\newcommand{\mTisqdef}{\ensuremath{\smash{\mTi^2=m_i^2+p_{\mathrm{T},i}^2}}\xspace}
\newcommand{\sumttbbg}{\ensuremath{\sum_{i=\PQt,\PAQt,\PQb,\PAQb,\Pg}}\xspace}
\newcommand{\prodttbb}{\ensuremath{\prod_{i=\PQt,\PAQt,\PQb,\PAQb}}\xspace}

\newcommand{\checkmarkast}{\ensuremath{\checkmark^\ast}\xspace}

\newcommand{\vecmu}{\ensuremath{\vec{\mu}}\xspace}
\newcommand{\vecalpha}{\ensuremath{\vec{\alpha}}\xspace}
\newcommand{\Poi}{\ensuremath{\text{Poi}}\xspace}
\newcommand{\Nalpha}{\ensuremath{\mathcal{N}(\vecalpha)}\xspace}
\newcommand{\sigmazj}{\ensuremath{\sigma^0_j}\xspace}
\newcommand{\mufid}{\ensuremath{\mu_{\text{fid}}}\xspace}
\newcommand{\sigmazfid}{\ensuremath{\sigma^0_{\text{fid}}}\xspace}

\cmsNoteHeader{TOP-22-009}

\title{Inclusive and differential cross section measurements of \texorpdfstring{\ttbb}{ttbb} production in the lepton+jets channel at \texorpdfstring{$\sqrt{s}=13\TeV$}{sqrt(s) = 13 TeV}}
\date{\today}

\abstract{Measurements of inclusive and normalized differential cross sections of the associated production of top quark-antiquark and bottom quark-antiquark pairs, \ttbb, are presented.  The results are based on data from proton-proton collisions collected by the CMS detector at a centre-of-mass energy of 13\TeV, corresponding to an integrated luminosity of 138\fbinv.  The cross sections are measured in the lepton+jets decay channel of the top quark pair, using events containing exactly one isolated electron or muon and at least five jets.  Measurements are made in four fiducial phase space regions, targeting different aspects of the \ttbb process.  Distributions are unfolded to the particle level through maximum likelihood fits, and compared with predictions from several event generators.  The inclusive cross section measurements of this process in the fiducial phase space regions are the most precise to date.  In most cases, the measured inclusive cross sections exceed the predictions with the chosen generator settings.  The only exception is when using a particular choice of dynamic renormalization scale, $\smash{\muR=\frac12 \prodttbb \mTi^{1/4}}$, where \mTisqdef are the transverse masses of top and bottom quarks.  The differential cross sections show varying degrees of compatibility with the theoretical predictions, and none of the tested generators with the chosen settings simultaneously describe all the measured distributions.}

\hypersetup{%
pdfauthor={CMS Collaboration},%
pdftitle={Inclusive and differential cross section measurements of ttbb production in the lepton+jets channel at sqrt(s)=13 TeV},%
pdfsubject={CMS},%
pdfkeywords={CMS, ttbb, unfolding, top, differential}}

\maketitle

\section{Introduction}
\label{sec:intro}

The associated production of top and bottom quark-antiquark pairs, \ttbb, in proton-proton (\pp) collisions at the CERN LHC is notoriously challenging to model because of the nonnegligible mass of the \PQb quark, and the difference in the typical energy scales of interactions involving top and \PQb quarks~\cite{Buccioni:2019plc, Jezo:2018yaf}.
Comparing predictions for \ttbb production cross sections with inclusive and differential measurements is an important test of perturbative quantum chromodynamics (QCD) calculations.
Furthermore, \ttbb production is a leading background for searches and other measurements, such as the measurement of the associated production of top quark pairs with Higgs bosons (\ttH), where the Higgs boson decays to a pair of \PQb quarks ($\PH \to \bbbar$)~\cite{Aad:2016zqi, Aaboud:2017rss, Aaboud:2018urx, Sirunyan:2018ygk, Sirunyan:2018hoz, Sirunyan:2018mvw}, and measurements of the simultaneous production of four top quarks ($\ttbar\ttbar$)~\cite{Aaboud:2018jsj, ATLAS:2020hpj, ATLAS:2021kqb, Khachatryan:2014sca, Sirunyan:2017roi, Sirunyan:2017tep, CMS:2019jsc, CMS:2019rvj, CMS:2023zdh, ATLAS:2023ajo, CMS:TOP-22-013}.
These two processes provide direct access to the top quark Yukawa coupling, a crucial parameter of the standard model (SM)~\cite{Cao:2016wib, Cao:2019ygh}.
An improved understanding of \ttbb production will help reduce the uncertainties in these important measurements.

Fixed-order calculations of inclusive and differential cross sections have been obtained at next-to-leading order (NLO) in QCD for \ttbb production~\cite{Worek:2009jg, Bevilacqua:2009zn, Bredenstein:2009aj, Bredenstein:2010rs, Worek:2011rd} and for the associated production of \ttbb with one additional jet (\ttbbj)~\cite{Buccioni:2019plc}.
In addition, full NLO QCD corrections to off-shell \ttbb production are available~\cite{Denner:2020orv, Bevilacqua:2021cit}.
While these fixed-order calculations provide important insights into the dynamics of the \ttbb process, they cannot easily be compared with measurements because of the large extrapolations to the full parton-level phase space that this would involve.
Instead, predictions obtained by matching matrix-element (ME) generators to parton showers (PSs) can be more directly compared to the data and can be used to model the \ttbb background in other measurements.
Such predictions have been obtained using different modelling approaches~\cite{Hoeche:2014qda, Garzelli:2014aba, Hoche:2016elu, Alwall:2014hca, Frederix:2012ps, Frixione:2007nw}.
The current state of the art in simulating \ttbb production uses ME calculations at NLO in QCD with massive \PQb quarks and matching to a PS~\cite{Cascioli:2013era, Bevilacqua:2017cru, Jezo:2018yaf}.
These calculations with massive \PQb quarks use parton distribution functions (PDFs) of the proton in the four-flavour scheme (\fourFS), where \PQb quarks are not part of the proton PDF.
In particular, in final states with gluon splittings, \gtobb, the \PQb quark mass is taken into account.
These calculations provide a description of \ttbb production based on the NLO MEs in the entire phase space where the \bbbar pair can be resolved as one or two \PQb quark jets.

Measurements of inclusive and differential \ttbb cross sections have previously been performed by the ATLAS and CMS Collaborations in \pp collisions at centre-of-mass energies of 7, 8, and 13\TeV in final states with zero, one, or two charged leptons ($\Pell=\Pe,\PGm$), using samples corresponding to integrated luminosities of up to 41.5\fbinv~\cite{Aad:2013tua, Aad:2015yja, Aaboud:2018eki, Khachatryan:2015mva, CMS:2014yxa, Sirunyan:2017snr, CMS:2019eih, CMS:2020grm, CMS:TOP-20-003}.
To date, \ttbb simulations have shown a tendency to underpredict the inclusive \ttbb cross sections.
Normalized differential distributions are generally in agreement with the predictions from event generators, within experimental and theoretical uncertainties, although the size of these uncertainties has not yet made it possible to definitively rule out or prefer specific modelling approaches.
The dominant uncertainties in previous studies were those related to \PQb tagging calibration, jet energy scale (JES), and the limited precision of the NLO calculations, \ie the choice of renormalization (\muR) and factorization scales (\muF).

This paper reports the measurement of inclusive and normalized differential cross sections of \ttbb production in four fiducial phase space regions in the lepton+jets channel.
The measurement uses \pp collision data recorded with the CMS detector at the LHC from 2016 to 2018 at $\sqrt{s}=13\TeV$, corresponding to an integrated luminosity of 138\fbinv.
Top quarks almost always decay into a \PW boson and a \PQb quark.
We consider events in which one \PW boson decays into a pair of quarks, and the other \PW boson decays into a charged lepton (electron or muon) and a neutrino.
Decays of the \PW boson into a tau lepton decaying into an electron or muon are also implicitly considered.
After hadronization of the quarks, and including the \PQb quarks not originating from the top quarks, the considered \ttbb events contain one electron or muon and at least five jets, of which three result from the hadronization of \PQb quarks (\PQb jets).

We measure inclusive cross sections and normalized differential cross sections, where the latter correspond to the ratios of the absolute differential cross sections to the inclusive ones.
Measurements are performed in different phase space regions, targeting distinct aspects of \ttbb production.
One phase space region targets fully resolved \ttbb events (labelled as \ttbb).
Another, more inclusive, phase space region requires only three \PQb jets, so as to select \ttbb events in which the pairs of additional \PQb quarks are reconstructed as a single jet, or in which one \PQb jet is outside of the detector acceptance (labelled as \ttb).
Finally, we measure cross sections aimed at the study of additional QCD radiation in \ttb or \ttbb events (labelled as \ttbj and \ttbbj, respectively), as these have been shown to be sensitive to the modelling of \ttbb production~\cite{Buccioni:2019plc}.

The observed distributions are unfolded to the particle level using binned maximum likelihood fits, whereby the data from muon and electron channels from all data-taking years are fitted simultaneously, and uncertainties are estimated using a nuisance parameter profiling procedure.
Differential cross sections are measured for several observables, using two complementary approaches for the event interpretation given below.

In the first approach, no attempt is made at directly identifying the additional \PQb jets in \ttbb events in either data or simulation.
The generator-level observables are defined using stable particles exclusively, with no reference to the simulated event history or the origin of \PQb jets from top quark decays or QCD radiation.
This ensures that the observable definitions at the generator and detector levels are highly consistent.
Furthermore, the distributions unfolded in this way can be compared with predictions from event generators which do not provide any information about the origin of \PQb quarks in the final state.
In the phase space region with four \PQb jets (\ttbb), observables are defined via the pair of \PQb jets with the smallest angular separation to target \PQb jets produced through the splitting of gluons into \bbbar.

In the second approach, the \PQb jets not originating from top quark decays are identified at the generator level using the simulated event history, and a multivariate algorithm is developed to identify the resulting reconstructed \PQb jets among all observed jets.
This approach is more accurate at identifying additional \PQb jets and thus these observables can be more sensitive to the modelling of additional heavy-quark production in \ttbar events, at the cost of restricting future reinterpretations of the results and introducing additional modelling assumptions about the parton history.

The unfolded results are compared to different predictions based on various NLO ME calculations, interfaced with different PS simulations.
Tabulated results are provided in the HEPData record for this analysis~\cite{hepdata}.

This paper is organized as follows.
Section~\ref{sec:cms} describes the CMS detector.
The event generation and detector simulation are detailed in Section~\ref{sec:mc}.
The reconstruction of electrons, muons, and jets and the identification of \PQb jets, as well as the corresponding definition of particle-level objects, are discussed in Section~\ref{sec:event_reco}.
Section~\ref{sec:sel_obs} describes the event selection and the definition of the measured observables at the generator and detector levels.
In Section~\ref{sec:signal_unf}, the extraction of the signal, the identification of additional \PQb jets using a multivariate algorithm, and the unfolding of the observables are presented.
The treatment of systematic uncertainties is detailed in Section~\ref{sec:systematics}, and the results, compared to several theoretical predictions, are presented in Section~\ref{sec:results}.
Finally, a summary of the results is provided in Section~\ref{sec:summary}.

\section{The CMS detector}
\label{sec:cms}

The central feature of the Compact Muon Solenoid (CMS) apparatus is a superconducting solenoid of 6\unit{m} internal diameter, providing a magnetic field of 3.8\unit{T}.
Within the solenoid volume are a silicon pixel and strip tracker, a lead tungstate crystal electromagnetic calorimeter (ECAL), and a brass and scintillator hadron calorimeter (HCAL), each composed of a barrel and two endcap sections.
Forward calorimeters extend the pseudorapidity ($\eta$) coverage provided by the barrel and endcap detectors.
Muons are measured in gas-ionization detectors embedded in the steel flux-return yoke outside the solenoid.
A more detailed description of the CMS detector, together with a definition of the coordinate system used and the relevant kinematic variables, can be found in Refs.~\cite{Chatrchyan:2008zzk, CMSTrackerGroup:2020edz}.

Events of interest are selected using a two-tiered trigger system.
The first level (L1), composed of custom hardware processors, uses information from the calorimeters and muon detectors to select events at a rate of around 100\unit{kHz} within a fixed latency of about 4\mus~\cite{CMS:2020cmk}.
The second level, known as the high-level trigger, consists of a farm of processors running a version of the full event reconstruction software optimized for fast processing, and reduces the event rate to around 1\unit{kHz} before data storage~\cite{Khachatryan:2016bia}.

\section{Simulated samples}
\label{sec:mc}

Samples of simulated events, produced with Monte Carlo (MC) event generators, are used in this analysis to estimate the contributions from background processes, model the correspondence between the observables at the generator and detector levels for unfolding, and compare the unfolded results with theoretical predictions.

For all simulated samples, additional \pp interactions in the same or neighbouring bunch crossings (pileup) are generated with \PYTHIA~(v8.240)~\cite{Sjostrand:2014zea} and overlapped with the simulated hard interactions to match the pileup multiplicity measured in data.
The detector response is modelled using a detailed simulation of the CMS detector, based on \GEANTfour~\cite{GEANT4:2002zbu}.

The response of the detector and event reconstruction of the \ttbb signal is modelled using a sample of \ttbb events generated using \POWHEGBOX~\cite{Jezo:2015aia} and \OPENLOOPS~\cite{Buccioni:2019sur}, referred to as the nominal \ttbb sample (or \ttbbPP), where the \ttbb MEs are calculated at NLO in QCD with massive \PQb quarks~\cite{Jezo:2018yaf}, and matched with \PYTHIA for parton showering and hadronization.
The \fourFS NNPDF3.1 next-to-NLO (NNLO) PDF set is used for the description of the proton structure.
The \PQb quark mass is set to 4.75\GeV and the \POWHEG damping parameter that regulates the damping of real emissions in the NLO calculation when matching to the PS is set to a value of $\hdamp=1.379\mtop$.
Dynamic \muF and \muR scales were chosen as $\muF=\HT/4$ and $\muR=\frac12 \prodttbb \mTi^{1/4}$, respectively, where $\HT = \sumttbbg \mTi$ and the transverse mass \mTidef, following Ref.~\cite{Buccioni:2019plc}.
The choice of renormalization scale is a geometric average of top and bottom quark transverse masses and is a natural choice for taking into account the widely separated energy scales of both particles.
The factorization scale is based on the maximum momentum of final state radiation that is still resummed into the PDF and hence is parameterized as the scalar sum of transverse top and bottom quark masses.

The sensitivity of the detector response to the modelling of the \ttbb process is evaluated using an alternative sample of \ttbb events, obtained from the inclusive \ttbar simulation with \POWHEG~(v2) matched to \PYTHIA (also referred to as \ttbarPP).
The \PQb quarks are assumed to be massless, and accordingly, five flavour scheme (\fiveFS) proton PDFs (also NNPDF3.1 NNLO) are used in the calculation.
In that sample, events with one additional \PQb quark are modelled by the ME at leading order (LO) in QCD, while any further \PQb quarks are generated by the PS.
In particular, in \gtobb splittings only the emission of the gluon off initial-state partons or the top quarks is described at the ME-level.
Hence, for a description of \ttbb, the modelling of the splitting itself is necessarily handled by the PS.
Since it has been shown that \gtobb splittings are the dominant mechanism for the production of additional \PQb jets with top quark pairs, both in \ttb and \ttbb events~\cite{Jezo:2018yaf}, this results in large uncertainties from the choice of the \muR scale used for the strong coupling constant in the final-state PS.
Conversely, in \ttbb samples generated in the \fourFS, the dominant uncertainty comes from the choice of \muR scale in the MEs, while uncertainties from the PS scale are smaller.
The \muF and \muR scales in the \ttbarPP sample were set to $\muF = \muR = \mTtop$.

In addition to the above, we consider several alternative predictions of \ttbb cross sections, to be compared with the measurements.
The generator settings used for the \ttbbPP and inclusive \ttbar simulation, as well as a number of alternative generator setups described below, which are used for comparison of results.

\begin{table}[!b]
\centering
\topcaption{%
    Generator settings for different modelling approaches of \ttbb production.
    The top quark mass value is set to $\mtop = 172.5\GeV$ for all generator setups, and for the generator setups using massive \PQb quarks, the \PQb quark mass value is set to $m_{\PQb} = 4.75\GeV$.
    In the scale settings, \HT corresponds to the scalar \mT sum, $\HT = \sumttbbg \mTi$, and \smash{\mTidef} is the transverse mass.
    For generators setups using \POWHEG the \hdamp value is specified. Other generator setups do not use this parameter and are marked with (\NA).
}
\label{tab:generatorsettings}
\renewcommand{\arraystretch}{1.2}
\cmsTable{\begin{tabular}{lllllll}
    Generator setup & Process/ME order & Generator/Shower & Tune & PDF set & \hdamp & Scales \\
    \hline
    \multirow{2}{*}{\ttbarPP} & \ttbar/ &
        \POWHEG v2/ & \multirow{2}{*}{CP5} & \fiveFS NNPDF3.1 & \multirow{2}{*}{1.379\mtop} &
        \multirow{2}{*}{$\muF = \muR = \mTtop$} \\
        & NLO & \PYTHIA 8.240 & & NNLO & & \\[\cmsTabSkip]
    \multirow{2}{*}{\ttbarPH} & \ttbar/ &
        \POWHEG v2/ & \multirow{2}{*}{CH3} & \fiveFS NNPDF3.1 & \multirow{2}{*}{1.379\mtop} &
        \multirow{2}{*}{$\muF = \muR = \mTtop$} \\
        & NLO & \HERWIG 7.13 & & NNLO & & \\[\cmsTabSkip]
    \multirow{2}{*}{\ttbbPP} & \ttbb/ &
        \POWHEGBOX/ & \multirow{2}{*}{CP5} & \fourFS NNPDF3.1 & \multirow{2}{*}{1.379\mtop} &
        $\smash{\muR=\frac12 \prodttbb \mTi^{1/4}}$, \\
        & NLO & \PYTHIA 8.240 & & NNLO as 0118 & &
        $\muF=\HT/4$ \\[\cmsTabSkip]
    \multirow{2}{*}{\ttbbSherpa} & \ttbb/ & \multirow{2}{*}{\SHERPA 2.2.4} &
        \multirow{2}{*}{\SHERPA} & \fourFS NNPDF3.0 & \multirow{2}{*}{\NA} &
        $\muR = \prodttbb \mTi^{1/4}$, \\
        & NLO & & & NNLO as 0118 & &
        $\muF = \HT/2$ \\[\cmsTabSkip]
    \multirow{2}{*}{\ttbbAMC} & \ttbb/ &
        \MGvATNLO v2.4.2/ & \multirow{2}{*}{CP5} & \fourFS NNPDF3.1 & \multirow{2}{*}{\NA} &
        \multirow{2}{*}{$\muF = \muR = \sum \mT$} \\
        & NLO & \PYTHIA 8.230 & & NNLO as 0118 & & \\[\cmsTabSkip]
    \multirow{3}{*}{\ttjetsAMC} & \multirow{2}{*}{\ttjets \fxfx/} &
        \multirow{2}{*}{\MGvATNLO v2.6.1/} & \multirow{3}{*}{CP5} & \multirow{2}{*}{\fiveFS NNPDF3.1} & \multirow{3}{*}{\NA} &
        $\muF = \muR = \sum \mT$, \\
        & \multirow{2}{*}{NLO [$\leq$2 jets]} & \multirow{2}{*}{\PYTHIA 8.240} & & \multirow{2}{*}{NNLO} & &
        $\textrm{qCut}=40\GeV$, \\
        & & & & & & $\textrm{qCutME}=20\GeV$ \\
\end{tabular}}
\end{table}

{\tolerance=800
The \POWHEG inclusive \ttbar generator is also interfaced with \HERWIG~(v7.13) for parton showering and hadronization~\cite{Bahr:2008pv, Bellm:2015jjp}, using the CH3 underlying event tune~\cite{CMS:2020dqt} (referred to as \ttbarPH).
Two other sets of simulated \ttbb events in the \fourFS at NLO in QCD were obtained using \MGvATNLO (referred to as \ttbbAMC), and \SHERPA~(v2.2.4)~\cite{Gleisberg:2008ta, Cascioli:2013era} with \OPENLOOPS (referred to as \ttbbSherpa).
The \ttbbAMC simulation is matched with \PYTHIA and uses \MadSpin to decay the top quarks; the \muF and \muR scales are set to the sums of the transverse masses \mT of all partons in the final state ($\sum \mT$).
In the \ttbbSherpa sample, the ME and PS are matched in the \MCATNLO scheme, the PDF set used is NNPDF3.0 NNLO in the \fourFS, and the scales are set to $\muF = \HT/2$ and $\muR = \prodttbb \mTi^{1/4}$.
Finally, \MGvATNLO is used to generate a sample of \ttjets events with up to two additional jets described at NLO QCD in the ME calculations, merged in the \fxfx scheme~\cite{Frederix:2012ps} and using massless \PQb quarks (referred to as \ttjetsAMC).
The \muF and \muR scales are set to $\muF = \muR = \sum \mT$, while the jet cutoff in the ME calculations is set to 20\GeV and the merging scale to 40\GeV.
The generator settings of all \ttbb simulation approaches are summarized in Table~\ref{tab:generatorsettings}.
\par}
Contributions to the \ttbb phase space include \PQb quarks produced in multiple parton interactions (MPI) as well as those produced in the matrix element or shower of the hard process itself.  For the \ttbarPP sample, this contribution is found to be 8--12\% depending on the phase space considered. The dedicated \ttbb simulations have a much smaller contribution ($<$2\% for \ttbbPP ) because they do not include contributions from \ttbar production in the primary parton scattering, where additional \PQb quarks come only from MPI.
 
The predictions from the exclusive \ttbb samples are normalized using the total \ttbb cross sections obtained at NLO in QCD from the corresponding generators, whereas the inclusive \ttbar and \ttjets samples are normalized using the total cross section for \ttbar production, $\sigma(\ttbar) = 833.9\unit{pb}$, computed at NNLO in QCD using \TOPpp~(v2.0)~\cite{Czakon:2011xx}, including soft-gluon resummations to next-to-next-to-leading logarithmic accuracy~\cite{Kidonakis:2013zqa}, and assuming a top quark mass of $\mtop = 172.5\GeV$.

The \ttbarPP sample is used for the estimation of the inclusive \ttbar background.
However, using the nominal \ttbb sample for the signal in conjunction with the inclusive \ttbar sample for the backgrounds would result in a double counting of the \ttbb contribution.
This overlap is avoided by removing from the inclusive sample events containing at least one \PQb jet not originating from a top quark (as defined below in Section~\ref{sec:event_reco:gen}), with $\pt > 20\GeV$ and $\abseta<2.4$.
Conversely, in the nominal \ttbb sample only the events with at least one additional \PQb jet are kept.
Henceforth the \ttB process is defined as the production of events passing these criteria. All simulated events in our measured fiducial phase spaces (see Section~\ref{sec:sel_obs:fidps}) are found to pass the \ttB selection criteria.
The remaining events in the inclusive \ttbar samples are categorized into a \ttC contribution, with events containing at least one charm (\PQc) jet (also with $\pt > 20\GeV$ and $\abseta < 2.4$) for which the simulation histories of the matched \PQc hadrons do not include any top quark, and a \ttlight contribution with all remaining events.

Simulated samples of minor backgrounds include single top quark production in the $t$ and $s$ channels, as well as \tW production (collectively Single \PQt); \ttW, \ttH, and \ttZ production (collectively \ttX); and the production of \PZ/$\PGg^\ast$ or \PW in association with jets (collectively \Vjets).
An overview of the simulation settings used for these backgrounds is given in Table~\ref{tab:generatorsettingsbkg}.

\begin{table}[!b]
\centering
\topcaption{%
    Generator settings for various minor background samples simulated with \POWHEG~\cite{Nason:2004rx, Frixione:2007vw, Alioli:2010xd, Frixione:2007nw, Alioli:2009je, Re:2010bp, Hartanto:2015uka} or \MGvATNLO~\cite{Alwall:2014hca}.
    The ``Group'' column refers to the grouping of processes in the maximum likelihood fits.
}
\label{tab:generatorsettingsbkg}
\renewcommand{\arraystretch}{1.1}
\cmsTable{\begin{tabular}{lllll}
    Process & Group & ME order & Generator &  Notes \\
    \hline
    \tW & Single \PQt & NLO & \POWHEG v2 & \\
    $t$ channel & Single \PQt & NLO & \POWHEG v2 & \MadSpin for heavy particle decays~\cite{Artoisenet:2012st} \\
    $s$ channel & Single \PQt & NLO & \MGvATNLO~v2.6.1 & \\[\cmsTabSkip]
    \ttH & \ttH & NLO & \POWHEG v2 & \\
    \ttZ & \ttV & NLO & \MGvATNLO~v2.6.1 & \MadSpin for heavy particle decays~\cite{Artoisenet:2012st} \\
    \multirow{2}{*}{\ttW} & \multirow{2}{*}{\ttV} & \multirow{2}{*}{NLO} & \multirow{2}{*}{\MGvATNLO~v2.6.1} & \fxfx merging up to 1 additional jet~\cite{Frederix:2012ps} \\[-1pt]
                          & & & & \MadSpin for heavy particle decays~\cite{Artoisenet:2012st} \\[\cmsTabSkip]
    \Wjets & \Vjets & LO & \MGvATNLO~v2.6.5 &  MLM merging up to 4 additional jets~\cite{Alwall:2007fs} \\
    \Zjets & \Vjets & LO & \MGvATNLO~v2.6.5 &  MLM merging up to 4 additional jets~\cite{Alwall:2007fs} \\
\end{tabular}}
\end{table}

For all of these minor background samples, the proton structure is described by the NNPDF3.1 set of NNLO PDFs~\cite{Ball:2014uwa}, and parton showering and hadronization are simulated with \PYTHIA, using the CP5 tune~\cite{CMS:GEN-17-001} for the underlying event description.
The value of the Higgs boson mass is assumed to be 125\GeV, while the top quark mass value is set to $\mtop = 172.5\GeV$.
In the \POWHEG samples, the \hdamp parameter is set to a value of $\hdamp=1.379\mtop$ as a part of the CP5 tune.

The cross sections for \ttbar, \Zjets, \Wjets, and single top quark production are obtained at NNLO in QCD~\cite{Czakon:2011xx, Kidonakis:2013zqa, Li:2012wna}.
Samples for \ttW, \ttZ, and \ttH production are normalized to predictions at NLO in QCD~\cite{Alwall:2014hca, Hartanto:2015uka}.

\section{Event reconstruction}
\label{sec:event_reco}

\subsection{Detector-level object reconstruction and identification}
\label{sec:event_reco:reco}

A particle-flow (PF) algorithm~\cite{Sirunyan:2017ulk} is applied to reconstruct and identify each individual particle in an event, with an optimized combination of information from the various elements of the CMS detector.
The primary vertex (PV) is taken to be the vertex corresponding to the hardest scattering in the event, evaluated using tracking information alone, as described in Section 9.4.1 of Ref.~\cite{CMS:TDR-15-02}.

The energy of electrons is determined using a multivariate algorithm from a combination of the electron momentum at the PV as determined by the tracker, the energy of the corresponding ECAL cluster, and the energy sum of all bremsstrahlung photons, obtained from the ECAL, spatially compatible with originating from the electron track~\cite{CMS:2020uim}.
Electrons are identified by placing requirements on the cluster shape in the ECAL, the track quality, and the compatibility between the track and the ECAL cluster.
Electrons from photon conversions are rejected.
The electrons used in this analysis are required to have $\abseta<2.5$ and $\pt > 29\GeV$ for data collected in 2016.
In data collected in 2017--2018, the \pt threshold is raised to 34\GeV, except for electrons with $\abseta<2.1$, for which the threshold is 30\GeV.
These requirements are chosen to be as low as possible so as to maximize the signal selection efficiency while staying above the trigger thresholds in the respective data-taking periods.
Electrons with a cluster pseudorapidity \absetaSC between 1.444 and 1.566 are not considered, in order to avoid the gap between the barrel and endcap ECAL sections.
The average identification efficiency for the primary electrons is about 70\%, including the isolation requirements described below.
Another looser set of requirements with 95\% selection efficiency is also considered, with $\pt > 15\GeV$ and $\abseta<2.5$, and relaxed identification criteria.
These define the ``veto'' electrons, which are used to reject events with more than one lepton.

The \pt of muons is obtained from the curvature of the corresponding track, combining information from the inner tracker and the outer muon detector system~\cite{Sirunyan:2018fpa}.
The muons are identified based on the quality of the combined track fit and on the number of hits in the different tracking detectors, with an efficiency of about 95\%.
Muons are required to have $\abseta<2.4$ and $\pt > 26$ (29)\GeV in 2016 (2017--2018), while ``veto'' muons are defined with $\pt > 15 \GeV$, $\abseta<2.4$, and looser identification criteria.

Electron or muon tracks are required to have a longitudinal and transverse distance to the PV smaller than 0.5--5\unit{mm}, depending on the lepton flavour and \abseta.
In order to suppress backgrounds from hadrons misidentified as leptons, or from leptons produced from hadrons decaying inside of jets, the leptons are required to be isolated from hadronic activity.
The lepton isolation is defined as the ratio between the scalar \pt sum of all PF candidates in a cone around the lepton excluding the lepton itself, and the lepton \pt.
The cone size is $\deltaR < 0.4$ (0.3) for electrons (muons).
The isolation is corrected by removing contributions from pileup~\cite{CMS:2020ebo}.
The maximum primary (veto) electron isolation varies between 0.03--0.08 (0.20--0.27), decreasing with \pt and increasing with \absetaSC.
For primary (veto) muons, the isolation is required to be $<$0.15 ($<$0.25).
Residual differences between lepton reconstruction, identification, and isolation efficiencies in data and simulation are corrected.
The efficiencies are measured as a function of lepton \pt and $\eta$ in data samples enriched in \ZtoLL events using a ``tag-and-probe'' method.
For electrons (muons), the corrections are between 1--5\% ($<$2\%) with uncertainties of less than 2 (1)\%~\cite{CMS:2020uim, Sirunyan:2018fpa}.

The energy of charged hadrons is determined from a combination of their momentum measured in the tracker and the matching ECAL and HCAL energy deposits, corrected for the response function of the calorimeters to hadronic showers.
The energy of neutral hadrons is obtained from the corresponding corrected ECAL and HCAL energies.
Hadronic jets are reconstructed by clustering PF candidates using the anti-\kt algorithm~\cite{Cacciari:2008gp}, as implemented in the \FASTJET package~\cite{Cacciari:2011ma}, with a distance parameter of $R=0.4$.
Jet momentum is determined as the vectorial sum of all particle momenta in the jet, and is found from simulation to be, on average, within 5--10\% of the true momentum over the whole \pt spectrum and detector acceptance.
The jet energy resolution (JER) amounts typically to 10 (15--20)\% at 100 (30)\GeV~\cite{Khachatryan:2016kdb}.
Pileup interactions can contribute additional tracks and calorimetric energy depositions, increasing the apparent jet momentum.
To mitigate this effect, tracks identified to be originating from pileup vertices are discarded and an offset correction is applied to correct for remaining contributions~\cite{CMS:2020ebo}.
Jet energy corrections are derived from simulation studies so that the average measured energy of jets becomes identical to that of jets at generator level.
In-situ measurements of the momentum balance in dijet, $\PGg{+}\text{jet}$, $\PZ{+}\text{jet}$, and multijet events are used to determine any residual differences between the JES in data and in simulation, and appropriate corrections are applied to data~\cite{Khachatryan:2016kdb}.
Additional selection criteria are applied to each jet to remove jets potentially falsely reconstructed, with dominant contributions from instrumental effects or reconstruction failures.
A multivariate algorithm is used to identify and remove jets likely originating from pileup interactions~\cite{CMS:2020ebo}.
Jets used in the analysis are required to have $\pt > 30\GeV$ and $\abseta<2.4$ and to be separated from the selected electron or muon by $\deltaR > 0.4$.

The \textsc{DeepJet} algorithm is used to identify \PQb jets~\cite{Sirunyan:2017ezt, Bols:2020bkb, CMS:DP-2023-005}.
This algorithm uses a deep neural network (DNN) discriminant to combine information from charged and neutral PF candidates clustered in the jet with features of secondary vertices within the jet.
Jets are labelled as \PQb tagged if they pass a ``medium'' working point of the discriminant, corresponding to an efficiency for correctly identifying \PQb jets in \ttbar events of 75--80\% and to a misidentification probability of about 15--17\% for \PQc jets and about 1.5--2\% for other jets.
We will also refer to jets failing the medium working point as light-tagged jets.
Another, more restrictive, (``tight'') \PQb tagging working point is also used, yielding a misidentification rate of 2.5--3.5 for \PQc jets and 0.15--0.25\% for other jets. This improved mistagging rate comes at the cost of reducing the \PQb jet identification efficiency to around 60\%.

The \PQb jet (mis)identification probabilities in the simulation are corrected to match the efficiencies measured in data at both the medium and tight working points~\cite{Sirunyan:2017ezt}.
The tagging efficiency in data for \PQb jets is obtained from a combination of five different measurements, three of which make use of a sample of multijet events enriched in \PQb jets, by requiring jets to contain a muon, while the remaining two use samples enriched in \ttbar events and containing respectively one or two isolated electrons or muons.
These three samples are statistically independent of each other and with the events used in this analysis and yield compatible measurements of the \PQb tagging efficiency.
The \PQb tagging efficiency correction factors are measured as functions of jet \pt and vary between 0.9--1.0 depending on the tagger working point, the data period, and the jet \pt, and have uncertainties of up to 10\%.
For \PQc jets, the misidentification efficiency in the simulation is corrected using the same correction factors as for \PQb jets, but with double their uncertainty.
This has been shown to cover the true \PQc jet misidentification probability~\cite{Sirunyan:2017ezt}.
For other jets, the misidentification probability is measured as a function of jet \pt in an inclusive sample of multijet events.
The corresponding simulation-to-data correction factors vary between 0.6--1.9, depending on the tagger working point, the data period, and the jet \pt.
Their uncertainties vary from 10--30\%.

\subsection{Particle-level object definitions}
\label{sec:event_reco:gen}

The fiducial phase space regions and observables for the inclusive and differential cross sections reported in this paper are defined based on the properties of stable final-state particles, with proper lifetimes $\tau_0 > 10\mm$.
Particles with $\abseta>5$ are not considered.
The definition of objects at the particle level, which follow closely those at the detector level introduced earlier, is described in the following.

Prompt particle-level electrons are selected with $\pt > 29\GeV$ and $\abseta<2.5$ and are ``dressed'' with photons from final-state radiation (FSR) by adding to their four-momentum the momentum of any photon within $\deltaR < 0.1$.
Prompt particle-level muons are required to have $\pt > 26\GeV$ and $\abseta<2.4$.
A similar FSR dressing procedure as for the electrons is applied to the muons.
Furthermore, ``veto'' electrons and muons are defined by relaxing the thresholds to $\pt > 15\GeV$, and are then used to reject events with more than one lepton.
The same dressing procedure as described above is also applied for the veto electrons and muons.

Jets are obtained by clustering all particles, excluding any neutrinos, using the anti-\kt algorithm with $R=0.4$.
They are required to have $\pt > 25\GeV$ and $\abseta<2.4$ and are not considered if they are within $\deltaR < 0.4$ of a prompt electron or muon.
At the particle level, the flavour of jets is defined unambiguously by rescaling the momenta of all generated \PQb hadrons to a negligible value and including them in the jet clustering procedure~\cite{Cacciari:2007fd}.
Jets thus matched to at least one \PQb hadron are considered as \PQb jets,
while all remaining jets are labelled as light jets.

For the purpose of defining observables for the second approach introduced in Section~\ref{sec:intro}, and the definition of the \ttB process, we also define as ``additional'' \PQb jets those particle-level \PQb jets for which the simulation histories of the matched \PQb hadrons do not include any top quark.
This definition is therefore not uniquely based on stable particles, since it refers to the simulated parton-level event content, which is specific to each generator and not provided by every generator.

\section{Event selection and definition of fiducial phase space regions and observables}
\label{sec:sel_obs}

\subsection{Event selection}
\label{sec:sel_obs:sel}

Data are collected using a set of triggers requiring the presence of one isolated electron or muon.
The lepton \pt thresholds applied by these single-lepton triggers vary between 27--32 (24--27)\GeV for electrons (muons), depending on the data-taking period.
In addition, triggers selecting events with one isolated electron and a scalar jet \pt sum (\HT) above 150\GeV are used, allowing the electron \pt threshold to be set to 28\GeV for $\abseta<2.1$.
The trigger selections used in collision data are also applied in the simulation, and residual differences between data and simulation are corrected.
For electron triggers, efficiencies are measured in a control region enriched with \ttbar events, collected using triggers requiring the presence of missing transverse momentum.
These present a negligible correlation with the triggers used in this analysis.
Muon trigger efficiencies are measured using the tag-and-probe technique in a control region enriched in \ZtoLL events.

Data from the different data-taking periods (2016, 2017, 2018) are analyzed separately and are only combined at the likelihood level, as will be explained in Section~\ref{sec:signal_unf:unfolding}.
Additionally, data collected in 2016 are split into two groups, the ``2016preVFP'' and ``2016postVFP'' eras, due to substantial changes in the detector conditions between them.
At the beginning of the 2016 period, saturation effects in the readout chips of the strip tracker led to lower signal-to-noise ratios and fewer hits on tracks.
This issue was mitigated by changing the feedback pre-amplifier bias voltage (VFP) during the 2016 data-taking run~\cite{CMS:DP-2020-045}.

\subsection{Fiducial phase space regions}
\label{sec:sel_obs:fidps}

The measured cross sections are derived from the fiducial phase space regions and observables described below.
The definitions at the particle and detector levels follow each other as closely as possible in order to minimize extrapolations outside of the detector acceptance.
Both at the particle and detector levels, events are first required to have exactly one primary electron or muon, and no additional ``veto'' electrons or ``veto'' muons, as defined in Section~\ref{sec:event_reco}.
At the particle level, events with electrons and muons are combined, and no requirement is placed on the decay channels of the top quarks, so that the fiducial phase space regions can also contain \ttbb events in which both top quarks decay leptonically, but one lepton is outside the detector acceptance and is not selected as a ``veto'' lepton.
Furthermore, electrons and muons produced indirectly through the decay of a tau lepton are included.
At the detector level, events with electrons or muons are classified into two distinct channels in this measurement.

Four different and partially overlapping phase space regions are then considered, each targeting different aspects of \ttbb production, as introduced in Section~\ref{sec:intro}.
The most inclusive selection considered, labelled as \fjtb (or \fjtbLONG) and targeting \ttb, requires the presence of at least five jets, of which at least three must be {\PQb}(-tagged) jets at particle (detector) level.
In a second selection, labelled as \sjfb (or \sjfbLONG) and targeting \ttbb, at least six jets are required, of which at least four are {\PQb}(-tagged) jets.
Two additional selections are defined which help target the properties of additional light jets produced in the event, targeting \ttbj and \ttbbj.
These are the \sjtbtl phase space (also labelled \sjtbtlLONG), requiring at least six jets, including at least three {\PQb}(-tagged) jets and at least three light(-tagged) jets; and the \sejfbtl phase space (also labelled \sejfbtlLONG), requiring at least seven jets, including at least four {\PQb}(-tagged) jets and at least three light(-tagged) jets.

\subsection{Observables}

As described in Section~\ref{sec:intro}, we consider two classes of observables for unfolding.
In the first class, all observables are defined using stable particles exclusively, without reference to any simulated event history.
This implies that no strict distinction is made between \PQb jets originating from the decay of top quarks, or from additional radiation.
In this way, there is a close correspondence between particle- and detector-level observables (see Sections~\ref{sec:event_reco:reco} and~\ref{sec:event_reco:gen}), and the definition of the observables is independent of any specific event generator.
In order to probe different features of \ttbb production, and in particular the properties of \PQb jets produced from QCD radiation, we distinguish different categories of objects using simple kinematic criteria.

In the \fjtb phase space, we focus the measurements on the \PQb jet with the third-largest \pt, which at the generator level is a true additional \PQb jet (not coming from top quark decays) in the nominal signal simulation in approximately 49\% of \fjtb events.
This identification allows for the study of the properties of the additional \PQb jets even in the case where they cannot be individually resolved.
In that phase space, we measure, each independently, the number of jets in the event (\Nj), the number of \PQb jets in the event (\Nb), the \pt and \abseta of the 3rd \PQb jet (\ptbN{3}, \absetabN{3}), the scalar \pt sum of all jets (\HT), and the scalar \pt sum of all \PQb jets (\HTb).

In the \sjfb phase space where there are at least four \PQb jets in the event, in order to measure properties that are expected to be sensitive to the modelling of gluon splittings to \bbbar, we define the ``extra'' \PQb jets as the pair of \PQb jets with the smallest angular separation, defined by
\begin{linenomath}
\begin{equation}
    \dRbb = \sqrt{ (\dPhibb)^2 + (\dEtabb)^2 },
\end{equation}
\end{linenomath}
where \dPhibb and \dEtabb are the angular separations of the pair of \PQb jets in azimuthal angle $\phi$ (in radians) and $\eta$, respectively.
We denote that pair by \bbextra in the following.
With this strategy, in the fiducial generator-level phase space, the two \bbextra jets correspond to true additional \PQb jets in $\sim$49\% of events in the \sjfb phase space.
Other choices, such as picking the pairs of jets with the 3rd and 4th largest \pt, result in smaller probabilities of correctly identifying the additional jets ($\sim$28\%).

In the \sjfb phase space, we measure, each independently, the number of jets in the event, the \pt and \abseta of the 3rd and 4th \PQb jets (\ptbN{3}, \ptbN{4}, \absetabN{3}, \absetabN{4}), the scalar jet \pt sum (\HT), and the scalar \PQb-jet \pt sum (\HTb).
Considering all possible pairs of distinct \PQb jets, we independently measure the average \dRbb over all pairings (\dRbbavg), and the largest invariant mass (\mbbmax).
For the \bbextra pair, we measure, each independently, the distance in $\eta$ between the two \PQb jets of the pair (\dRbbextra), the invariant mass of the pair (\mbbextra), the \pt and \abseta of the pair (\ptbbextra, \absetabbextra), and the \pt and \abseta of the leading (\bXN{1}) and sub-leading (\bXN{2}) \PQb jets in the pair (\ptbXN{1}, \ptbXN{2}, \absetabXN{1}, \absetabXN{2}).

In the \sjtbtl and \sejfbtl phase space regions, we target the properties of the additional light jets (see Sections~\ref{sec:event_reco:reco} and~\ref{sec:event_reco:gen}).
In each of these phase space regions, we measure the scalar \pt sum of the light jets in the event (\HTl).
We then remove from consideration the pair of light jets with the invariant mass closest to the \PW boson mass obtained from a fit to the two detector- or particle-level jets matched to the \PW boson decay in simulation.
We then measure, independently, the \pt of the leading remaining light jet (\ptljXN{1}), and the \dPhi between that light jet and the lowest \pt \PQb jet (\dPhiljb).
The latter variable probes the amount of recoil against additional QCD radiation in \ttbb events that is absorbed by the softest \PQb jet~\cite{Buccioni:2019plc}.
In the \sjtbtl and \sejfbtl phase space regions, the leading remaining light jet corresponds to a light jet not from top quark decays in $\sim$94\% of cases.
The softest \PQb jet is an additional \PQb jet in $\sim$50 (65)\% of cases in the \sjtbtl (\sejfbtl) phase space.

In the second class of observables, we focus on the \sjfb phase space and explicitly target the \PQb jets that do not originate from decaying top quarks.
At the particle level, these are the additional \PQb jets as defined in Section~\ref{sec:event_reco:gen}, labelled as \bbadd.
In case more than two additional \PQb jets are present, the two additional \PQb jets leading in \pt are selected.
At the detector level, the pair of \PQb-tagged jets most consistent with the true additional \PQb jets is identified using a DNN discriminant, described in Section~\ref{sec:signal_unf:dnn}.
For the \bbadd pair defined in this way, we measure the opening angle between the two \PQb jets of the pair (\dRbbadd), the invariant mass of the pair (\mbbadd), the \pt and \abseta of the pair (\ptbbadd, \absetabbadd), and the \pt and \abseta of the leading and subleading \PQb jets in the pair (\ptbAN{1}, \ptbAN{2}, \absetabAN{1}, \absetabAN{2}).

A summary of all observables is given in Table~\ref{tab:observables}.
In the measurement of each observable we also measure the inclusive cross section in the respective phase space (\sigmatot).

\begin{table}[!p]
\centering
\topcaption{%
    Description of all measured observables for each of the four fiducial phase space regions.
    Observables marked as (\checkmarkast) rely on the definition of additional \PQb jets, and do not fully correspond to the \sjfb fiducial phase space defined at the particle level, but also require the presence of \PQb jets without top (anti)quarks in their simulated history.
}
\label{tab:observables}
\renewcommand{\arraystretch}{1.2}
\cmsTable{\begin{tabular}{cclcccc}
    & \multicolumn{2}{c}{Observable}  & \fjtb & \sjfb & \sjtbtl & \sejfbtl \\ \hline
    & \sigmatot & Inclusive cross section   & \checkmark & \checkmark & \checkmark & \checkmark \\[\cmsTabSkip]
\multicolumn{3}{l}{Global observables} & & & & \\
    & \Nj & Jet multiplicity                & \checkmark & \checkmark &            &            \\
    & \Nb & \PQb jet multiplicity           & \checkmark &            &            &            \\
    & \HTj & Scalar \pt sum of all jets           & \checkmark & \checkmark &            &            \\
    & \HTb & Scalar \pt sum of all \PQb jets     & \checkmark & \checkmark &            &            \\
    & \HTl & Scalar \pt sum of all light jets    &            &            & \checkmark & \checkmark \\[\cmsTabSkip]
\multicolumn{3}{l}{Observables related to \PQb jets} & & & & \\
    & \ptbN{3} & \pt of third hardest \PQb jet          & \checkmark & \checkmark & & \\
    & \absetabN{3} & \abseta of third hardest \PQb jet  & \checkmark & \checkmark & & \\
    & \ptbN{4} & \pt of fourth hardest \PQb jet         &            & \checkmark & & \\
    & \absetabN{4} & \abseta of fourth hardest \PQb jet &            & \checkmark & & \\[\cmsTabSkip]
\multicolumn{3}{l}{Observables considering all pairs of \PQb jets (\bb)} & & & & \\
    & \dRbbavg & Average \deltaR of all \bb pairs           & & \checkmark & & \\
    & \mbbmax  & Highest invariant mass among all \bb pairs & & \checkmark & & \\[\cmsTabSkip]
\multicolumn{3}{l}{Observables related to the pair of \PQb jets closest in \deltaR (\bbextra)} & & & & \\
    & \ptbXN{1}      & \pt of leading extra \PQb jet         & & \checkmark & & \\
    & \absetabXN{1}  & \abseta of leading extra \PQb jet     & & \checkmark & & \\
    & \ptbXN{2}      & \pt of subleading extra \PQb jet      & & \checkmark & & \\
    & \absetabXN{2}  & \abseta of subleading extra \PQb jet  & & \checkmark & & \\
    & \dRbbextra     & \deltaR of \bbextra pair              & & \checkmark & & \\
    & \absetabbextra & \abseta of \bbextra pair              & & \checkmark & & \\
    & \mbbextra      & invariant mass of \bbextra pair       & & \checkmark & & \\
    & \ptbbextra     & \pt of \bbextra pair                  & & \checkmark & & \\[\cmsTabSkip]
\multicolumn{3}{l}{Observables related to the pair of \PQb jets not from \ttbar decay (\bbadd)} & & & & \\
    & \ptbAN{1}     & \pt of leading additional \PQb jet        & & \checkmarkast & & \\
    & \absetabAN{1} & \abseta of leading additional \PQb jet    & & \checkmarkast & & \\
    & \ptbAN{2}     & \pt of subleading additional \PQb jet     & & \checkmarkast & & \\
    & \absetabAN{2} & \abseta of subleading additional \PQb jet & & \checkmarkast & & \\
    & \dRbbadd      & \deltaR of \bbadd pair                    & & \checkmarkast & & \\
    & \absetabbadd  & \abseta of \bbadd pair                    & & \checkmarkast & & \\
    & \mbbadd       & invariant mass of \bbadd pair             & & \checkmarkast & & \\
    & \ptbbadd      & \pt of \bbadd pair                        & & \checkmarkast & & \\[\cmsTabSkip]
\multicolumn{3}{l}{Observables related to extra light jets} & & & & \\
    & \ptljXN{1} & \pt of leading extra light jet                        & & & \checkmark & \checkmark \\
    & \dPhiljb   & \dPhi of leading extra light jet and softest \PQb jet & & & \checkmark & \checkmark \\
\end{tabular}}
\end{table}

\section{Signal extraction and unfolding}
\label{sec:signal_unf}

\begin{figure}[!t]
\centering
\includegraphics[width=0.5\textwidth]{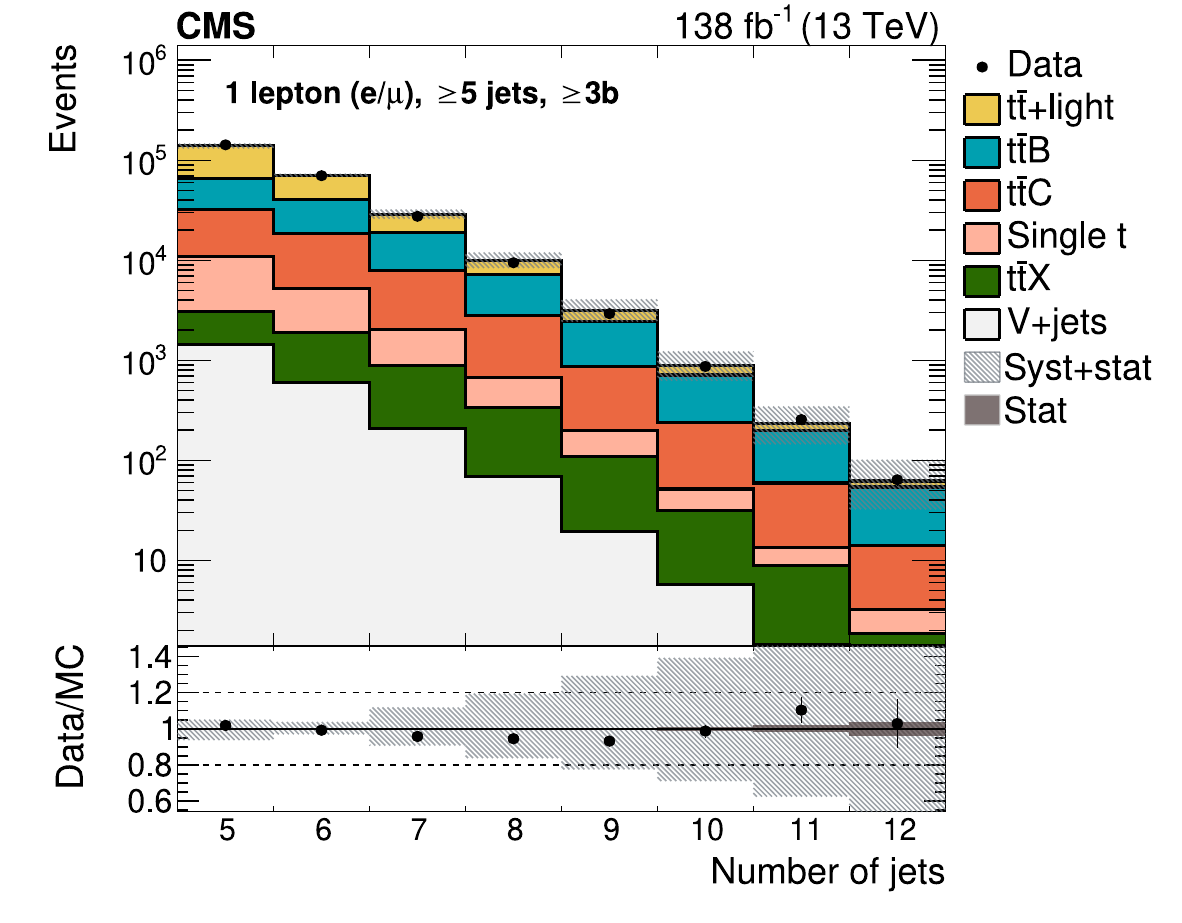}%
\includegraphics[width=0.5\textwidth]{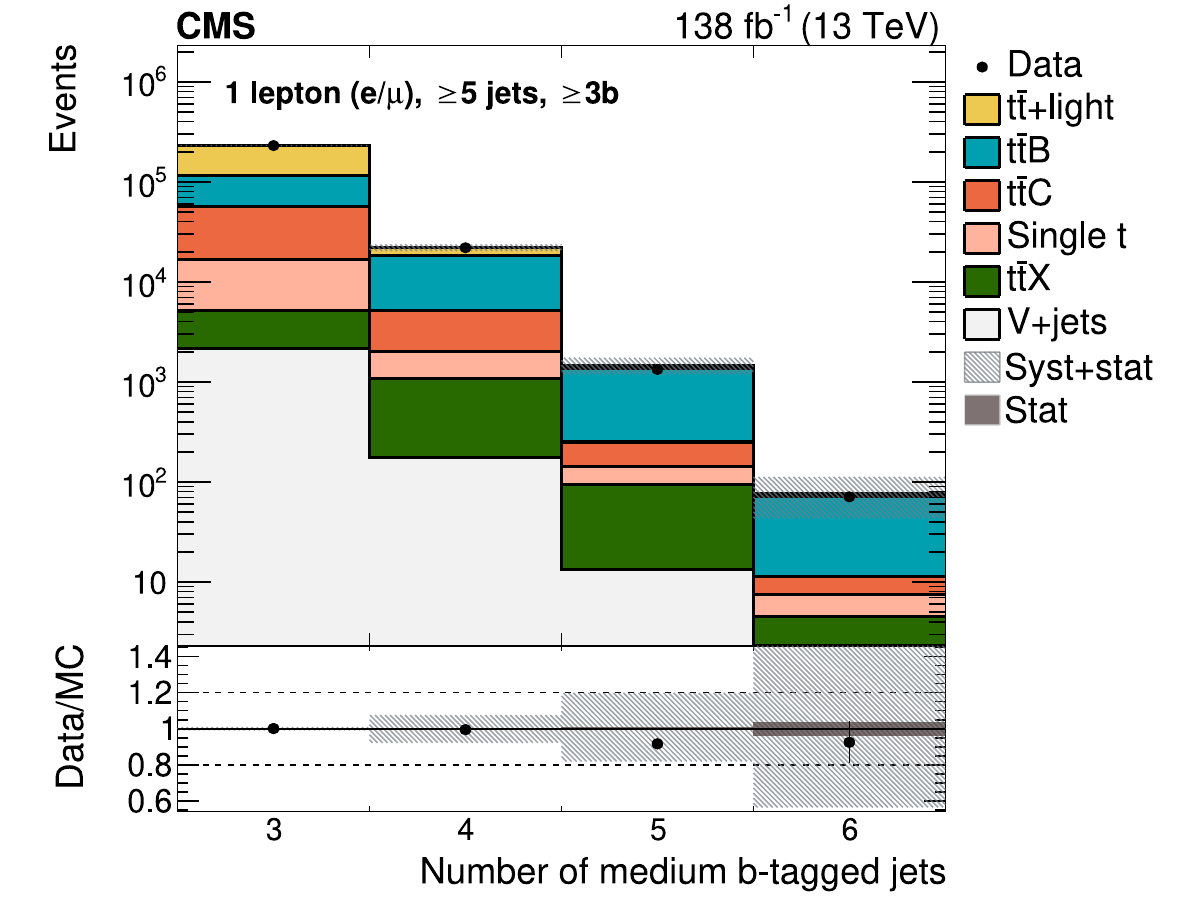}
\caption{%
    Jet (left) and \PQb-tagged jet (right) multiplicity with the \fjtbLONG selection prior to any fit, shown for both lepton channels and all data periods combined.
    For the purpose of visualisation, the contributions from simulation have been scaled by a common factor (0.98) to match the yield in data.
    The \ttX contribution includes the \ttH, \ttW, and \ttZ processes.
    The \ttB contribution includes the superset of fiducial and out of acceptance \ttbb events, following the definitions used for the measurements.
    The shaded bands include all a priori uncertainties described in Section~\ref{sec:systematics}, including the \ttB cross section uncertainty estimated from the nominal \ttbb simulation.
    Only effects on the shape of the distributions are considered.
    The last bins also contain the overflow.
}
\label{fig:signal_extraction:njets_bjets_5j3b}
\end{figure}

In the selected regions with at least three \PQb-tagged jets, the data are highly enriched in \ttjets events, which consist of about 30\% \ttB, 20\% \ttC, and 50\% \ttlight events, with minor background contributions, in descending order of importance, from single top quark production, \ttH, \Vjets, \ttZ, and \ttW processes.
Events from \ttH, \ttZ, or \ttW processes are referred to as a combined \ttX contribution if a distinction is not necessary.
In the other selected regions with at least four \PQb-tagged jets, the \ttB contribution is about two thirds of all events.
Background contamination from inclusive multijet production, where a jet or a nonisolated lepton is misidentified as a primary lepton, has been verified with simulated multijet events to be negligible at the level of the precision of this measurement ($\leq$1\% in the \fjtb phase space), and this background is therefore not considered further.
The prefit composition of the events as a function of number of jets and number of \PQb-tagged jets at the medium working point is shown in Figure~\ref{fig:signal_extraction:njets_bjets_5j3b}, with the \fjtb selection applied.
The events displayed in these distributions are the superset of events used for the measurements presented in Sec.~\ref{sec:results}.

For each observable, the cross section is measured using a dedicated binned maximum likelihood fit, which simultaneously yields the inclusive cross section in the corresponding fiducial phase space and the unfolded normalized differential cross section of the observable (Section~\ref{sec:signal_unf:unfolding}).
In order to improve the separation between the signal and background components, thereby better constraining the background contributions using the data, we use an ``ancillary'' variable that divides the detector-level selections into signal- and background-enriched categories, which are then fitted simultaneously.

\subsection{Ancillary variable}
\label{sec:signal_unf:anc_obs}

\begin{figure}[!t]
\centering
\includegraphics[width=0.5\textwidth]{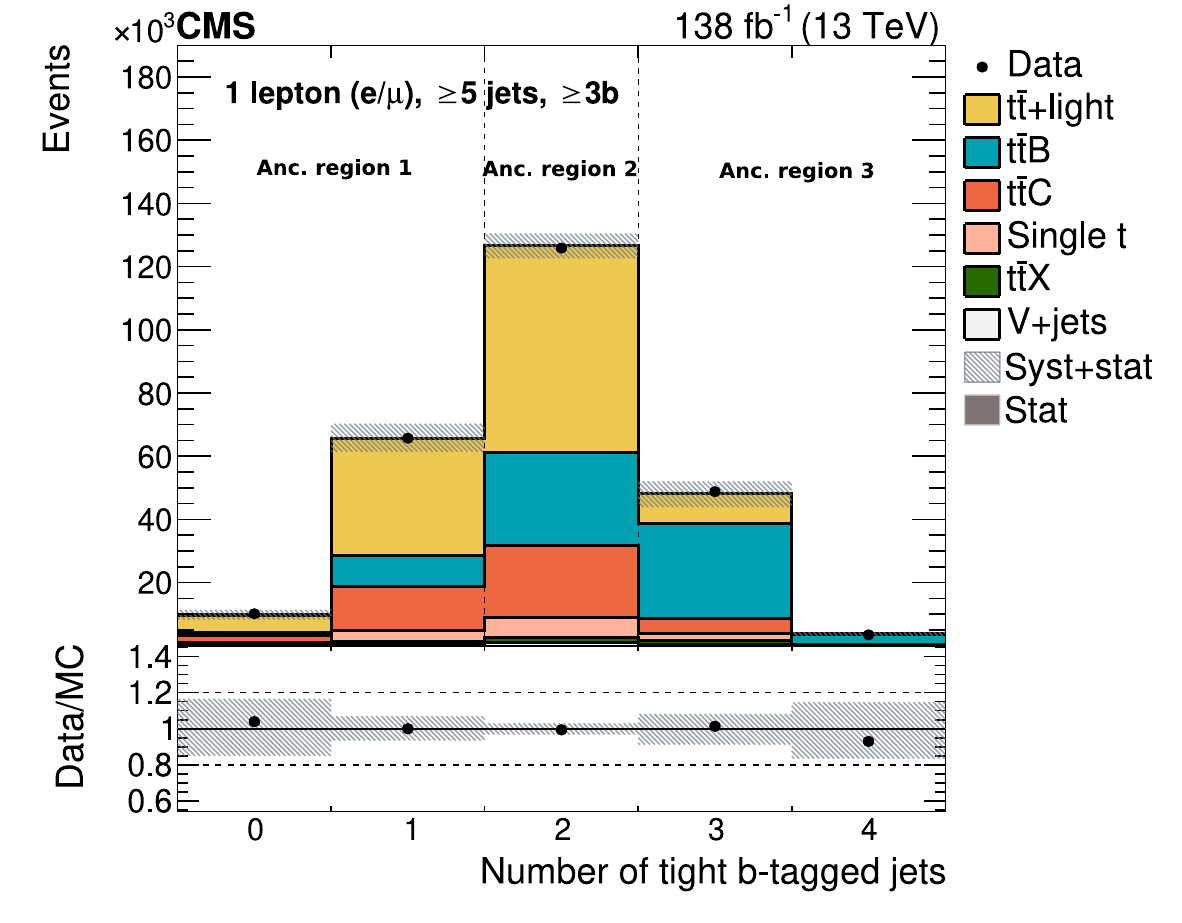}%
\includegraphics[width=0.5\textwidth]{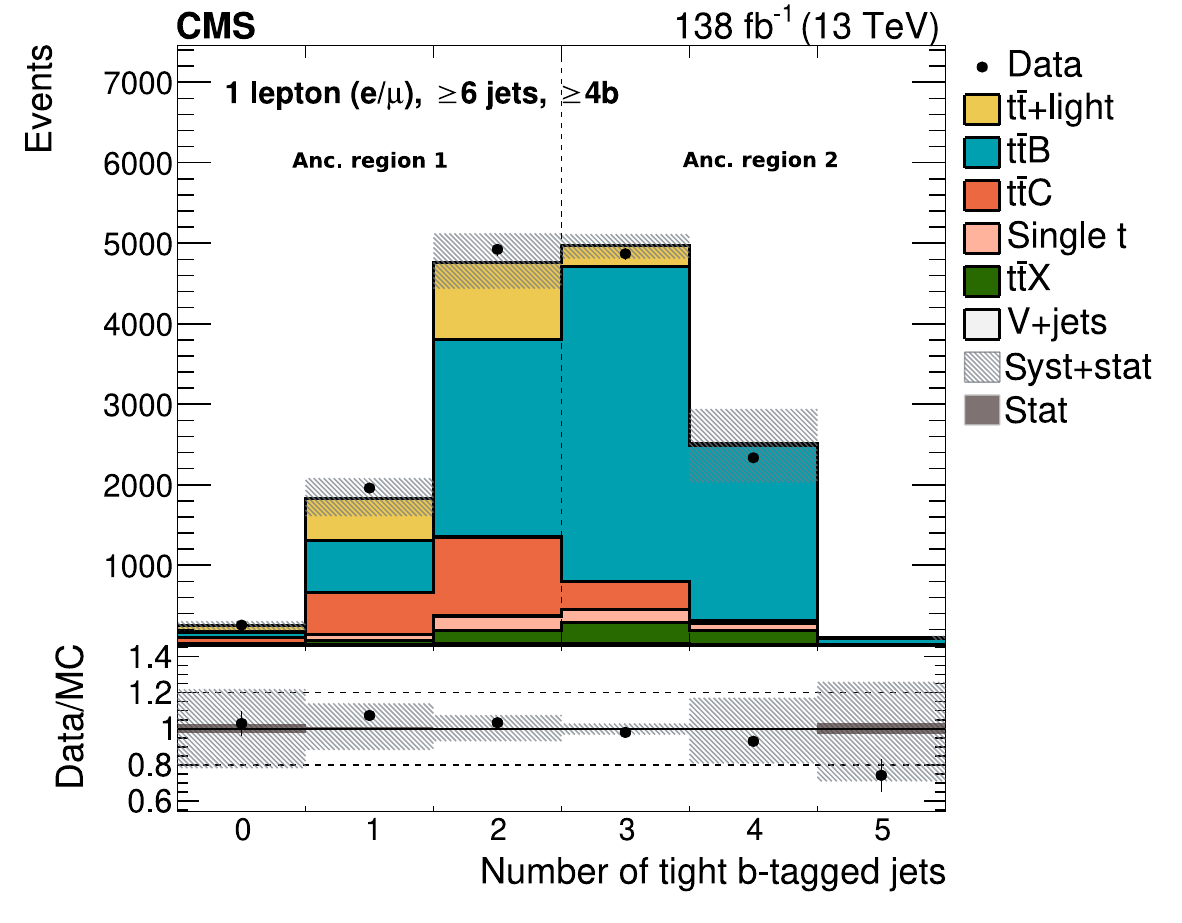}
\caption{%
    Number of jets \PQb tagged at the tight working point with the \fjtbLONG (left) and \sjfbLONG selections (right) prior to any fit, shown for all lepton channels and years combined.
    For the purpose of visualisation, the contributions from simulation have been scaled by a common factor (0.98 on the left, 0.95 on the right) to match the yield in data.
    The \ttB contribution includes the superset of fiducial and out of acceptance \ttbb events, following the definitions used for the measurements.
    The shaded bands include all uncertainties described in Section~\ref{sec:systematics}, including the \ttB cross section uncertainty estimated from the nominal \ttbb simulation.
    Only effects on the shape of the distributions are considered.
    The last bins also contain the overflow.
    The vertical dashed lines indicate the ancillary regions.
}
\label{fig:signal_extraction:auxvars}
\end{figure}

We use the number of jets passing the tight \PQb tagging working point as an ancillary variable.
The distributions of the number of tight \PQb-tagged jets with the \fjtb and \sjfb selections applied are shown in Fig.~\ref{fig:signal_extraction:auxvars}.
The regions with three or more tight \PQb-tagged jets are highly enriched in \ttB signal events, as defined in Section~\ref{sec:mc}, since they typically feature three or four true \PQb jets after fiducial selection, whereas the regions with 0--2 tight \PQb-tagged jets are dominated by \ttC or \ttlight backgrounds where \PQc or light jets have been misidentified as medium \PQb-tagged jets and, therefore, serve as control regions for these backgrounds.
In the fits of observables in the \fjtb and \sjtbtl phase space regions we use three ancillary regions containing zero or one, two, or three or more tight \PQb-tagged jets.
In the other phase space regions we only use two ancillary regions: one containing events with fewer than three tight \PQb-tagged jets, and another one containing events with at least three tight \PQb-tagged jets.
The use of ancillary variables is not only useful for obtaining signal-enriched regions but also allows for the constraint of \PQb tagging related uncertainties in the fit.

\subsection{Multivariate algorithm for jet assignment}
\label{sec:signal_unf:dnn}

For the eight observables in the \sjfb phase space with explicit identification of the additional \PQb jets, a multivariate algorithm based on a DNN is used to identify the pair of \PQb-tagged jets most consistent with the true additional \PQb jets as defined by the generator-level information.
The four \PQb-tagged jets in an event with the highest \pt, in the following referred to as candidate jets, are grouped into six permutations of candidate jet pairs, depending on their ranking in \pt.
Permutations within a pair are not treated separately.
A detector-level jet is considered correctly identified as an additional jet if it lies within an angular distance of $\deltaR < 0.4$ to an additional particle-level \PQb jet.

\begin{figure}[!ht]
\centering
\includegraphics[width=0.85\textwidth]{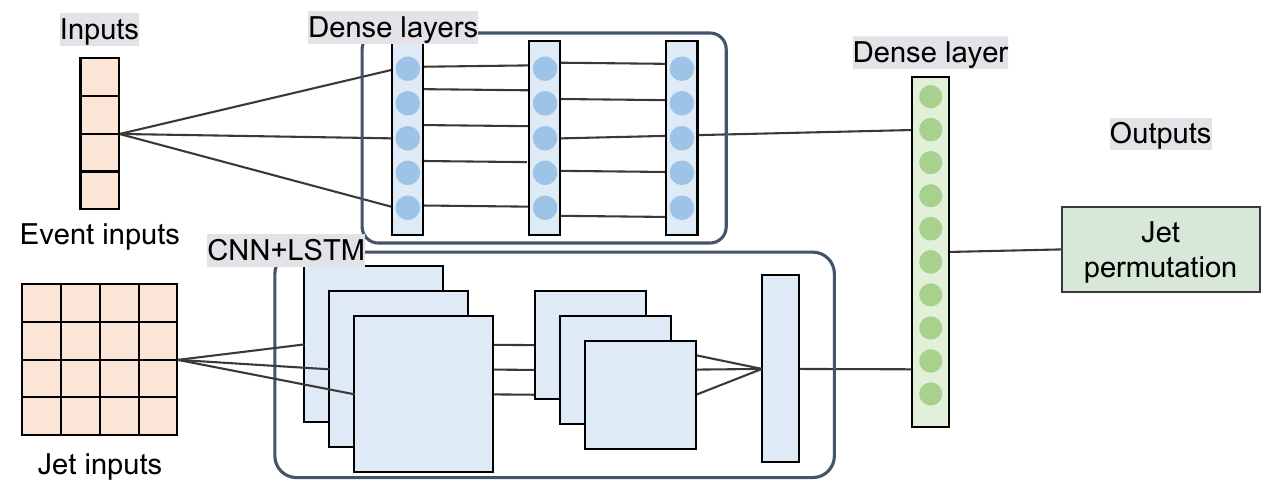}
\caption{%
    Structural representation of the neural network used for the assignment of the additional \PQb-jet pair.
    The neural network uses two sets of input variables:\ global event information is connected to three dense network layers, and jet-specific information is connected via convolutional network layers (CNN) and a long short-term memory (LSTM) cell.
    The input sequences are concatenated into one dense layer.
    The output layer consists of six nodes, each representing one of the six possible candidate jet combinations.
}
\label{fig:signal_extraction:dnn}
\end{figure}

An illustration of the DNN architecture is shown in Fig.~\ref{fig:signal_extraction:dnn}.
The DNN makes use of two sets of input variables, targeting jet-specific input information and global event information separately.

For the input variables targeting jet-specific information, \ie features of the four candidate jets, an automated feature engineering is performed for each jet using five convolutional neural network (CNN) layers~\cite{LeCun:1989nips} with filter matrices of size 1$\times$1~\cite{Bols:2020bkb}.
The size of the filter matrices is chosen such that they aggregate the different features, separately for each jet, into higher-level features of the consecutive layers.
These layers are followed by a long short-term memory (LSTM) cell~\cite{Hochreiter:1997yld}, where the independent jet features are combined into a sequence, allowing the network to learn from the correlations of the jet features.
The five features used for each candidate jet are the \pt, $\eta$, a flag indicating whether it passes the tight \PQb tagging working point, the angular separation (\deltaR) with the charged lepton, and the invariant mass with the charged lepton.
The 30 input variables targeting global event information include the properties of the six dijet combinations of the four candidate \PQb jets.
These input variables are connected via three dense network layers.
The inputs consist of the scalar \pt sum of the four candidate \PQb jets, the \pt, $\eta$, and $\phi$ of the charged lepton, the \dPhi, $\Delta\eta$, and invariant mass of the dijet combinations, the \deltaR of the dijet combinations and the charged lepton, and the jet and \PQb-tagged jet multiplicities.
Both input sequences are concatenated into one dense layer, which is connected to an output layer consisting of six nodes, each representing one of the six possible candidate jet combinations.
The pair of \PQb-tagged jets with the highest DNN output value per event is chosen as the additional \PQb-jet pair to be used further in the analysis.

{\tolerance=800
The training of the DNN is performed with simulated events passing the fiducial \sjfb phase space definition using the \textsc{keras}~\cite{Keras:2023git, Keras:2023webpage} package with a \textsc{TensorFlow}~\cite{Abadi:2016kic} backend and the \textsc{adam} optimizer~\cite{Kingma:2014vow}.
In the training, the categorical cross-entropy loss function is minimized.
To reduce biases from imbalances in the numbers of events for the different dijet categories, events are weighted in the training such that each category has the same number of weighted events.
Potential overfitting is mitigated using a dropout percentage of 10\%~\cite{Srivastava:2014jmlr} and an early-stopping procedure to stop the training if no decrease in the loss minimization has been achieved for 20 epochs (iterations over the training data set) on a set of spectator events not used for the training (validation data set).
In order not to bias the evaluation of the DNN, events from the \ttbarPP simulation passing the fiducial \sjfb definition are used in the training, while the evaluation for the measurement is performed with the nominal \ttbb sample.
It has been validated that the performance of the DNN is independent of the choice of which of those samples of simulated \ttbb events is used for the training or the evaluation.
This implies that the measurements of the DNN-based observables are not biased by the choice of signal model used for training the DNN.
\par}

The performance of the DNN is evaluated in the simulation based on the fraction of events in which it selects the correct pair of additional \PQb jets (accuracy).
The accuracy of the DNN is determined to be about 49\%, which represents a significant increase in identification accuracy compared to choosing the two \PQb jets closest in \deltaR (as in Section~\ref{sec:sel_obs}), which only yields an accuracy of about 41\%.

\subsection{Unfolding methodology}
\label{sec:signal_unf:unfolding}

The observed distributions are unfolded to the particle level by removing estimated background contributions and correcting for resolution, acceptance and efficiency effects of the detector and event reconstruction so that the measured distributions can be directly compared to theoretical predictions provided by MC event generators.

Unfolding is performed through a maximum likelihood fit by constructing, for each observable, a statistical model that links the distributions at the particle and detector levels.
The values of the particle-level cross sections which maximize the agreement between the predicted detector-level distributions and the observed data are determined from the fit.
In these models, freely floating parameters of interest determine the total cross section of the signal process in the corresponding phase space, as well as the normalized differential cross section of the signal process in discrete bins of the considered observable.
The models are constructed using the simulated signal and background samples described in Section~\ref{sec:mc} and include nuisance parameters that model the effect of systematic uncertainties in the signal and background predictions.
The fit is performed by minimizing the negative-log-likelihood (NLL) of the data with respect to the parameters of the model. The calculation of the NLL and its minimization are performed using \textsc{smoofit}~\cite{Smoofit:2022git, Smoofit:2022webpage}, itself relying on the \textsc{jax} package~\cite{JAX:2018git, JAX:2018webpage} for automatic differentiation of the NLL function, providing fast and numerically stable evaluations of the NLL gradients and of the second derivative (Hessian) matrix.
Both the central values and confidence intervals for the absolute and normalized differential cross sections are extracted from the fit.

The electron and muon channels as well as the four data-taking eras (2016preVFP, 2016postVFP, 2017, and 2018) and the ancillary variable regions are combined at the likelihood level.
The likelihood used to measure the unfolded distributions can be written as
\begin{linenomath}
\begin{equation}
    L(\vecmu, \vecalpha) = \left[ \prod_{e,i} \Poi \left( D_{e,i} \bigg\rvert S_{e,i}(\vecmu, \vecalpha) + \sum_{p \in \text{bkg.}} N^p_{e,i}(\vecalpha) \right) \right] \Nalpha,
    \label{eq:unfolding:likelihood}
\end{equation}
\end{linenomath}
where the parameters \vecmu are freely-floating parameters of interest, \vecalpha are profiled nuisance parameters used to model systematic uncertainties, $D_{e,i}$ are the observed yields in data-taking era $e$ and detector-level bin $i$ (including the lepton channels and ancillary regions), $N^p_{e,i}$ are the predicted yields of background process $p$ in era $e$ and bin $i$, $S_{e,i}$ are the predicted signal yields in era $e$ and bin $i$, $\Poi(d|\nu)$ is the Poisson probability mass function for counts $d$ with mean $\nu$, and \Nalpha is the Gaussian constraint term (with mean of zero and width of one) of the nuisance parameters.

We denote by $M^e_{ij}$ the expected number of signal events in the fiducial phase space in the detector-level bin $i$ and generator-level bin $j$ for the data-taking era $e$.
For every generator-level bin there are, therefore, up to
\begin{linenomath}
\begin{equation}
    (\text{4 eras}) \times (\text{2 channels}) \times (\text{2 or 3 anc. regions}) = \text{16 or 24 detector-level bins},
\end{equation}
\end{linenomath}
depending on the phase space.
In addition, the detector-level binning is chosen to be finer than the generator-level binning, with every bin at the detector level being about half the width of the generator-level bins (except for discrete observables such as the number of jets or \PQb jets, and the observables in the \sejfbtl region where the data yields are the lowest).
This improves the separation between signal and background.
Including other control regions in the fit or using a finer binning at the detector level has not been found to provide any significant further improvements in the sensitivity of the measurements.
The expected event yields in era $e$ can be written as $\smash[b]{M^e_{ij} = \lumi_e \sigmazj K^e_{ij}}$, where $\smash[b]{\lumi_e}$ is the integrated luminosity in era $e$, \sigmazj is the prefit cross section in bin $j$ estimated using the nominal \ttbb sample, and $K^e_{ij}$ are response matrices, \ie the probability for a simulated event in era $e$ and generator-level bin $j$ to be reconstructed and selected in detector-level bin $i$.
The total expected signal yields $S$ in Eq.~\eqref{eq:unfolding:likelihood} are computed as functions of the parameters of interest as
\begin{linenomath}
\begin{equation}
    S_{e,i}(\vecmu, \vecalpha) = \mufid \sum_{j=1}^n \mu_j M^e_{ij}(\vecalpha),
\end{equation}
\end{linenomath}
where $\mufid = \sigmatot / \sigmazfid$ is the signal-strength modifier for the inclusive cross section and $\mu_j$ are the parameters varying the fraction of signal events in each generator-level bin $j$.
To preserve unity, the yields in the last generator-level bin $n$ are not scaled independently, but as a function of the other bins as
\begin{linenomath}
\begin{equation}
    \mu_n(\mu_1 \dots \mu_{n-1}) = \frac{1}{F_n} \left( 1 - \sum_{i=1}^{n-1} \mu_i F_i  \right).
    \label{eq:unfolding:norm_constraint}
\end{equation}
\end{linenomath}
where $\smash[b]{F_j = \sigmazj / \sigmazfid = \sigmazj / \sum_{i=1}^n \sigma^0_i}$ is the \textit{a priori} fractional cross section in bin $j$.
In this way, the measured inclusive cross section is directly obtained as $\hat{\sigma}_{\text{fid}} = \hat{\mu}_{\text{fid}} \sigmazfid$.
Furthermore, the measured normalized differential cross section in bin $j$ is extracted as $1/\hat{\sigma}_{\text{fid}}\,\ddinline{\hat{\sigma}_j}{X} = \hat{\mu}_j F_j / w_j$, where $w_j$ is the width of generator-level bin $j$.
Thus, the yields in the last bin $n$ are constrained through Eq.~\eqref{eq:unfolding:norm_constraint} to conserve the total signal normalization for a given value of \mufid.
The covariance matrix of all fit parameters is obtained from the inverse Hessian of the NLL at the minimum and is used to compute confidence intervals on the measured cross sections.
We have verified that the resulting intervals are equivalent to those obtained by finding the level crossings of the profiled NLL, and we have validated the frequentist coverage properties of the confidence intervals using pseudo-experiments.

\begin{figure}[!p]
\centering
\includegraphics[width=0.65\textwidth]{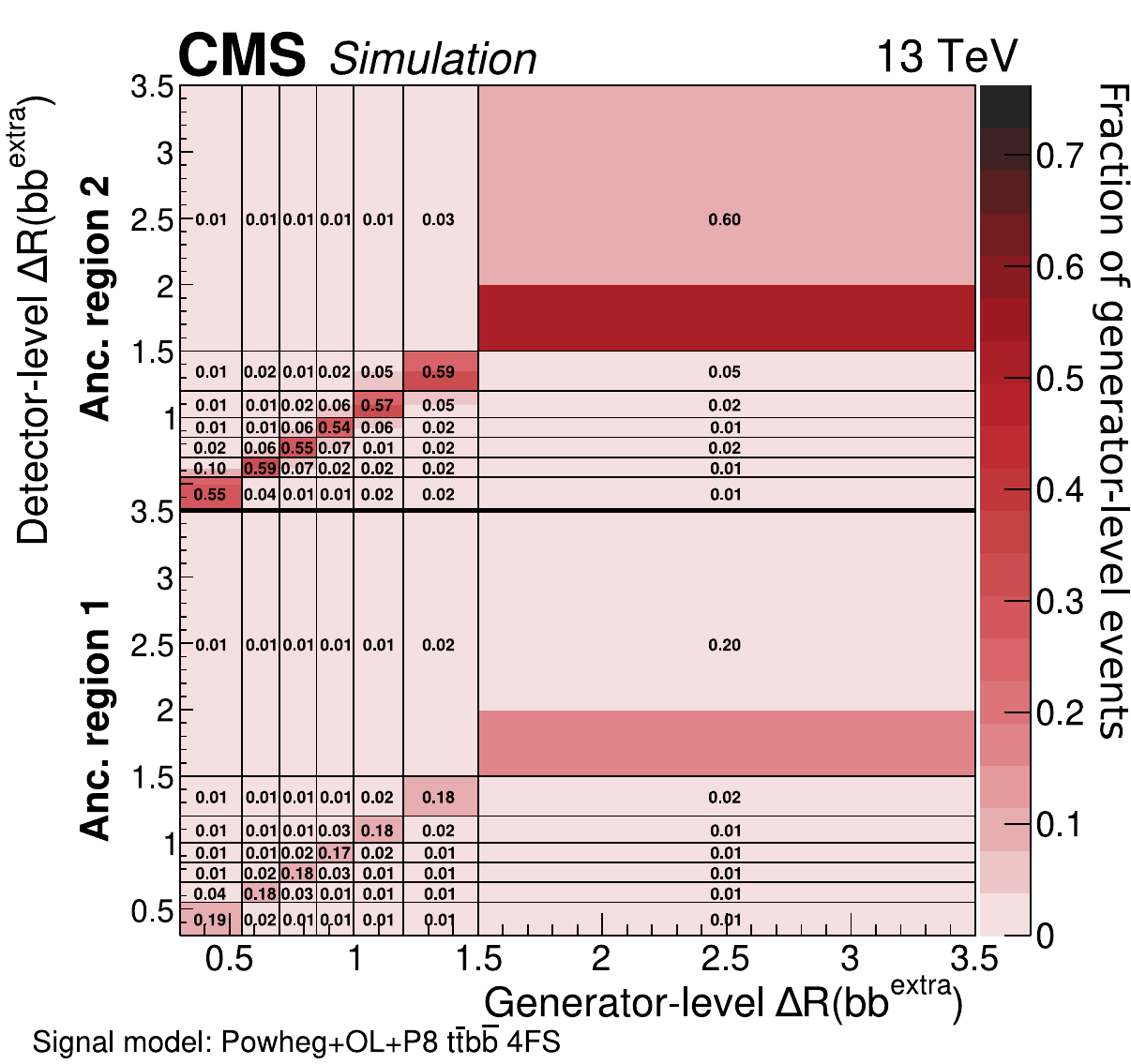} \\
\includegraphics[width=0.65\textwidth]{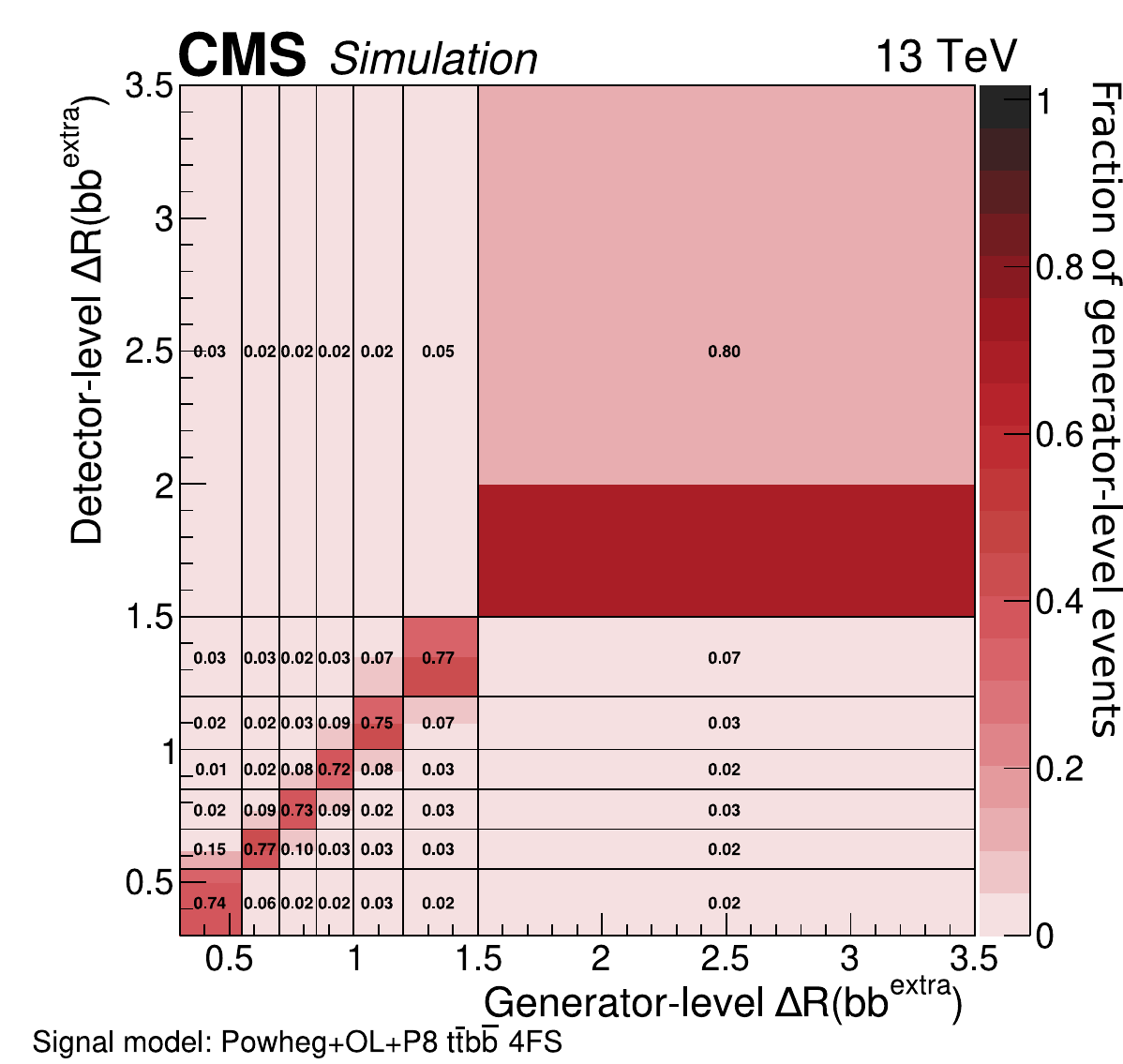}
\caption{%
    Response matrix for \dRbbextra in the \sjfbLONG phase space. The $x$ ($y$) axes show the generator- (detector-)level observables. The upper figure includes the ancillary variable, unrolled on the same axis as the detector-level observable, so that the binning of the detector-level observable, stacked vertically, is repeated twice. For the lower figure, the ancillary variables are projected out to more easily visualize the correspondence between true and reconstructed values. The coloured bins show the finer binning used at reconstructed level (bins split in two), while the numbers show the values one would obtain when using the same binning at the generator and detector level.
}
\label{fig:unfolding:migmat}
\end{figure}

Figure~\ref{fig:unfolding:migmat} shows the response matrix for an example observable, both with and without ancillary variables.
For the purpose of visualization, the response matrices shown are averaged across eras and lepton channels and are normalized so that the values sum to 100\% across each column (generator-level bin).
Most response matrices are highly diagonal, which can be explained by the close correspondence between the particle- and detector-level definitions of the observables.
Compared to these purely particle-level based observables, there is considerably more migration for the observables based on the additional \PQb jets, which are defined via the simulated history at the particle level but only identified with limited accuracy by the DNN at the detector level.

Despite a careful definition of the fiducial phase space at the generator and detector level, a fraction of the events selected at the detector level consists of simulated \ttB events that do not pass the corresponding generator-level fiducial phase space definitions.
There is an inherent ambiguity on how to treat this contribution, labelled as ``out of acceptance'' (OOA).
While technically part of the \ttbb process being measured, these events are outside of the fiducial phase space and should not be included in the measured cross sections.
Therefore, we treat this OOA contribution as a background and assign to it theoretical systematic uncertainties estimated using the nominal \ttbb simulation.
This OOA contribution constitutes about 40\% of selected \ttB events in the \sjfb and \sejfbtl regions and reduces to about 20 and 12\% in the more inclusive \sjtbtl and \fjtb regions, respectively.

Two distinct mechanisms explain the presence of OOA events in the detector-level selection.
First, mismeasurements of the energy or direction of leptons or jets can cause events that are outside the generator-level phase space to pass the detector-level selection.
This class of events features a similar topology to the fiducial signal events.
Second, the misidentification of light or \PQc jets as \PQb jets can lead to events with only two or three \PQb jets within the fiducial acceptance to be selected in the regions requiring the presence of three or four \PQb-tagged jets, respectively.
This latter mechanism is dominant in the two regions with four \PQb jets, \sjfb and \sejfbtl, and explains why the OOA contribution is larger in these regions, while the fractions of OOA events are reduced in the signal-enriched ancillary regions.
However, \ttB events with only three \PQb jets selected within the generator-level fiducial volume feature a distinct topology from those with four \PQb-tagged jets, since they might either contain a soft (low-\pt) or forward (high-\abseta) \PQb jet, or a \PQb jet resulting from the hadronization of a collinear \bbbar quark pair.
Due to this difference, the two OOA configurations are treated as separate background sources in the \sjfb and \sejfbtl regions and are labelled as \ttb OOA and \ttbb OOA, respectively, based on the presence of one or two ``additional'' \PQb jets (in the sense of Section~\ref{sec:event_reco:gen}).

For some of the unfolded distributions, there is no well-defined upper bound on the measured observable.
In order to include all observed events in the measurement, while keeping the upper edge of the unfolded distributions at a reasonable value, we normalize the differential cross sections in these bins by the width of the bin as it appears in the histogram, while at the same time including any overflow events, both in the data and in the simulation, into these last bins.

{\tolerance=1000
In order to assess the presence of biases due to the choice of the nominal signal model (\ttbbPP), bias tests were performed with the \ttbarPP simulation.
These tests consisted of unfolding pseudo-data generated using signal predictions from the \ttbarPP simulation, while using response matrices constructed from the \ttbbPP samples, thereby verifying whether the differences in signal modelling and response matrices resulted in any biases in the measurements.
These studies showed that the unfolded results were compatible with the expectations of the \ttbarPP signal prediction within systematic uncertainties, supporting the validity of using the response matrices of the \ttbbPP model for unbiased unfolding.
\par}

\section{Systematic uncertainties}
\label{sec:systematics}

Systematic uncertainties are evaluated by appropriate variations of the signal and background simulations.
The uncertainty sources may affect background yields and distributions, as well as the selection efficiency and the kinematic distributions of the signals.
These uncertainties are taken into account via nuisance parameters in the likelihood fits.

In certain cases where the statistical uncertainty of a systematic variation is comparable to or larger than the size of the systematic variation or there is significant bin-to-bin variation, the systematic templates need to be smoothed.
This is done via a lowess-based smoothing algorithm, which constrains the shape differences between the upward and downward fluctuations if they have a shape effect (smoothing) or, if not, converts them to rate-only effects.
If, on the other hand, the systematic uncertainty in question is determined not to have a shape or rate effect, then it is removed altogether.

A summary of all systematic uncertainties is given in Table~\ref{tab:systematics} and in the following sections, grouped by experimental uncertainties in Section~\ref{sec:systematics:exp} and modelling uncertainties in Section~\ref{sec:systematics:theory}.

\begin{table}[!t]
\centering
\topcaption{%
    Summary of the systematic uncertainty sources in the inclusive and differential \ttbb cross section measurements.
    The first column lists the source of the uncertainty.
    The second (third) column indicates the treatment of correlations of the uncertainties between different data-taking periods (processes), where \fullcorr means fully correlated, \partcorr means partially correlated (\ie contains sub-sources that are either fully correlated or uncorrelated), \nocorr means uncorrelated, and {\NA} means not applicable.
}
\begin{tabular}{clcc}
    & Source    & Corr. (period) & Corr. (process) \\
    \hline
    \multirow{9}{*}{\rotatebox{90}{Experimental}}
    & Integrated luminosity                         & \partcorr & \fullcorr  \\
    & Pileup reweighting                            & \fullcorr & \fullcorr   \\
    & Electron reconstruction and identification    & \fullcorr & \fullcorr   \\
    & Muon reconstruction and identification        & \fullcorr & \fullcorr   \\
    & Trigger efficiencies                          & \nocorr   & \fullcorr   \\
    & L1 prefiring                                  & \fullcorr & \fullcorr   \\
    & JES                                           & \partcorr & \fullcorr   \\
    & JER                                           & \nocorr   & \fullcorr   \\
    & \PQb tagging                                  & \partcorr & \fullcorr   \\[10pt]
    \multirow{11}{*}{\rotatebox{90}{Theoretical}}
    & \muR scale                        & \fullcorr & \partcorr   \\
    & \muF scale                        & \fullcorr & \partcorr   \\
    & Top quark \pt modelling           & \fullcorr & \fullcorr   \\
    & PDF                               & \fullcorr & \fullcorr   \\
    & PS scales: ISR                    & \fullcorr & \nocorr     \\
    & PS scales: FSR                    & \fullcorr & \nocorr     \\
    & ME-PS matching (\hdamp)           & \fullcorr & \partcorr   \\
    & Underlying-event tune             & \fullcorr & \fullcorr   \\
    & Colour reconnection               & \fullcorr & \fullcorr   \\
    & \PQb quark fragmentation          & \fullcorr & \fullcorr   \\
    & Inclusive \ttC cross section      & \fullcorr & \NA         \\
\end{tabular}
\label{tab:systematics}
\end{table}

\subsection{Experimental uncertainties}
\label{sec:systematics:exp}

\begin{description}
\item[Integrated luminosity:]
The integrated luminosities of the data-taking periods are individually measured with uncertainties of 1.2, 2.3, and 2.5\% for 2016, 2017, and 2018 data-taking periods, respectively~\cite{CMS:2021xjt, CMS:LUM-17-004, CMS:LUM-18-002}.
The uncertainty in the integrated luminosity of the combined data set is 1.6\% when taking into account the correlations between the periods.

\item[Pileup reweighting:]
The prediction of the number of pileup interactions in simulation is performed by assuming a total inelastic \pp cross section of 69.2\unit{mb}.
Changes in the assumed pileup multiplicity are estimated by varying the total inelastic cross section by $\pm$4.6\%~\cite{CMS:2020ebo}.
This uncertainty is treated as fully correlated between the data-taking periods.

\item[Lepton reconstruction and identification:]
Separate uncertainties are assigned to the corrections applied to the reconstruction and identification efficiencies for electrons in the simulation.
Similarly, the muon tracking, identification, and isolation efficiencies are estimated and the associated uncertainties are propagated to signal and background distributions used in the fits.
These uncertainties are taken to be fully correlated across the data-taking periods.

\item[Trigger efficiencies:]
The trigger efficiency scale factors are varied within their uncertainties, separately for electron and muon triggers and for each data-taking period.

\item[L1 prefiring:]
During the 2016--2017 data-taking periods, a gradual shift in the timing of the inputs of the ECAL L1 trigger in the forward endcap region ($\abseta > 2.4$) led to a specific inefficiency, known as ``prefiring''.
A similar effect is present for the muon system due to its limited time resolution, most pronounced in 2016, but also impacting data collected in 2017--2018.
Corrections of this effect are applied to simulated events, and 20\% of the corrections are assigned as the associated uncertainties.

\item[Jet energy scale and resolution:]
Uncertainties in the determination of the JES are taken into account by shifting the jet momenta in the simulation up and down, separately for several sources of uncertainty such as the overall energy scale, differences in flavour response, and residual differences between energy scale measurements.
Some of these sources are treated separately per data-taking period, while some are correlated for all periods.
Uncertainties in the JER are evaluated by increasing or decreasing the variation of jet energies between the reconstructed and particle levels, or by smearing the measured jet energy in case no matching particle-level jet could be found~\cite{Khachatryan:2016kdb}.
This uncertainty is uncorrelated between data-taking periods.

\item[\PQb tagging:]
Differences in the \PQb tagging efficiency between data and simulation are corrected by applying correction factors to simulated events, derived as a function of jet \pt and \abseta.
Systematic uncertainties are considered separately for light jets and \PQb/\PQc jets.
For \PQb and \PQc jets, nine different sources of uncertainties from the measurements of the correction factors are considered~\cite{Sirunyan:2017ezt}.
Included amongst these uncertainties are effects from variations in \PQb fragmentation and gluon to \bbbar splitting.
Scale factors and their uncertainties on the \PQb tagging efficiency and mistag rates are split based on the flavour of the intiating quark, but do not discriminate based on the origin of the \PQb quark as coming, for example, from a gluon splitting or top quark decay.
Statistical uncertainties in the scale factor measurements are treated independently for the medium and tight \PQb tagging working points, and for the four data-taking periods.
All other sources of uncertainties are correlated between the data-taking periods and the \PQb tagging working points.
\end{description}

\subsection{Modelling uncertainties}
\label{sec:systematics:theory}

This analysis is affected by uncertainties in the modelling of the background processes and the migration matrices linking the measured particle-level observables with the detector-level observables.
Variations applied to the signal process are defined such that the predicted yields at generator level remain constant for all values of the corresponding nuisance parameters, independently for each bin of the generator-level distributions.
This procedure ensures that the modelling uncertainties only have an effect on the signal selection efficiency, due to shape variations within each generator-level bin.
All modelling uncertainties are correlated between the data-taking eras.
Some uncertainties are not correlated between processes.
In the phase space regions where the OOA contribution is separated into \ttb OOA and \ttbb OOA contributions, the two components can have different correlations with the signal.
However, by repeating the fits with different correlation assumptions for the signal and the OOA processes, it has been verified that the results were not sensitive to this choice.
While most modelling uncertainties are assumed to be correlated between ancillary regions, various alternative correlation assumptions were tested for the scale uncertainties, with no significant effect on the result.

\begin{description}
{\tolerance=800
\item[Renormalization and factorization scales:]
Uncertainties covering the choice of \muR and \muF scales in the ME generators are considered by shifting the scales independently up and down by a factor of two.
These uncertainties are treated separately for each process but correlated for \ttlight and \ttC, as these contributions are estimated using the same \ttbarPP simulation.
Where separate \ttbb OOA and \ttb OOA contributions are considered, the uncertainty in the former contribution is correlated with the signal while the latter taken as uncorrelated.
For the \ttB OOA processes, shape and normalization effects of the \muR scale are decorrelated, and both are uncorrelated with the signal.
\par}

\item[Top quark \pt modelling:]
Because of discrepancies between the observed and simulated \pt spectrum of top quarks in \ttbar events, simulated events are reweighted to match the top quark \pt distribution predicted at NNLO~\cite{Czakon:2017wor}.
This procedure improves the agreement of the simulated predictions with the data.
We consider as uncertainty in this reweighting the full effect of the reweighting itself.
The uncertainty is applied to all \ttbar subprocesses (including the signal and all OOA processes), but not to single top quark or \ttX production.

\item[PDF:]
The uncertainties in the PDFs are evaluated by using replicas of the NNPDF set~\cite{Ball:2014uwa}.
For the \fourFS set used for the \ttbb signal simulation, uncertainties are estimated using the root-mean-square of all residuals in the predictions obtained using the PDF replicas, as these are defined by sampling from the covariance matrix of the PDF fit.
For the \fiveFS set used in all the other simulated samples, the replicas correspond to the leading eigenvectors of the PDF fit covariance matrix, and the uncertainties are obtained as the quadratic sum of the residuals in the predictions across the replicas.
This uncertainty is treated as correlated for all processes.
An additional uncertainty from the value of the strong coupling constant, \alpS, used in the PDF, is also included.

\item[PS scales:]
The uncertainty from the choice of scale at which the strong coupling constant is evaluated in the PS is estimated by varying the scale up and down by a factor of two, independently for initial-state radiation (ISR) and final state radiation (FSR).
These uncertainties are considered as uncorrelated for all processes.
Accordingly, the \ttB OOA processes is considered uncorrelated with the signal.
Where separate \ttbb OOA and \ttb OOA contributions are considered, the \ttbb OOA contribution is correlated with the signal, and \ttb OOA is treated separately.

\item[ME-PS matching (\hdamp):]
In the \POWHEG generator, the scale that separates the phase space of the first QCD emission into soft and hard parts is controlled by the \hdamp parameter.
The nominal value of the \hdamp parameter in the CP5 tune~\cite{CMS:GEN-17-001} is $\smash[b]{\hdamp = 1.379\mtop}$ and the uncertainties are estimated with varied values of $\smash[b]{\hdamp = 2.305\mtop}$ and \smash[b]{0.874\mtop} for the \ttbar and \ttbb samples.
This uncertainty is treated as correlated for \ttlight and \ttC processes but decorrelated from the \ttB contributions. This uncertainty is only applied to \ttbar subprocesses, which does not include single top or \ttX production.

 \item[Underlying event tune:]
The simulation of the underlying event is based on the CP5 tune of the \PYTHIA event generator, and uncertainties are estimated via varied tune settings~\cite{CMS:GEN-17-001}, applied to the \ttbar samples.
This uncertainty is only applied to \ttB, \ttlight, and \ttC processes and is treated as correlated.

 \item[Colour reconnection:]
 The default colour reconnection model in the \PYTHIA PS simulation is replaced by three alternative models~\cite{Argyropoulos:2014zoa, Christiansen:2015yqa}.
 These uncertainties are treated as correlated for \ttB, \ttlight, and \ttC processes. This uncertainty is only applied to \ttbar subprocesses, which does not include single top or \ttX production.

\item[\PQb quark fragmentation:]
We include a theoretical uncertainty for the fragmentation of the \PQb quarks into hadrons.
The fragmentation of the initiating parton into observable hadrons is subject to modelling uncertainties.
We estimate this effect by varying simultaneously the generator- and detector-level momenta of \PQb jets up and down by 1\% in the simulated samples.
This is consistent with the effect from reasonable variations of the fragmentation functions on the \PQb jet \pt distribution in Ref.~\cite{ATLAS:2021agf}.
Residual effects on the efficiency to identify \PQb jets, due to a possible mismodelling of the fragmentation of \PQb quarks, are already accounted for by the calibration of the \PQb tagging efficiency in the simulation.

\item[Inclusive \ttC cross section:]
For the \ttC process an additional 20\% normalization uncertainty is applied, corresponding to the precision of the inclusive \ttC cross section measurement by CMS~\cite{CMS:TOP-20-003}.
This measurement found that the \ttC cross section agreed with the \ttbarPP prediction.
\end{description}

No additional cross section uncertainties are considered for the backgrounds since the uncertainties listed above already result in variations in the predicted event yields that cover the uncertainties in the theoretical cross sections used to normalize the background contributions.

The finite size of the simulated MC samples is taken into account as a systematic uncertainty, following a method similar to the one proposed by Barlow and Beeston~\cite{Barlow:1993dm, Conway:2011in}.
For every bin of the detector-level distributions, a single Gaussian-constrained nuisance parameter varies the predicted yields, summed over all processes including the signal, within their statistical uncertainty.

The impact of a group of $k$ nuisance parameters $\alpha_{n_1} \dots \alpha_{n_k}$ on the parameter of interest $p$ is computed as
\begin{linenomath}
\begin{equation}
    I_p = \sqrt{\sum_{i,j=1}^k C_{p n_i} \widetilde{C}^{-1}_{n_i n_j} C_{n_j p}},
    \label{eq:combined_impacts}
\end{equation}
\end{linenomath}
where $C$ is the covariance matrix between all parameters, and $\widetilde{C}$ is the covariance matrix restricted to the parameters $n_1 \dots n_k$.
In this way, the effects of a set of nuisance parameters are combined while taking into account their correlation in the fit.
The total systematic uncertainties are calculated by considering all nuisance parameters in the sum in Eq.~\eqref{eq:combined_impacts}, and the statistical uncertainties as the difference in quadrature between the total uncertainties and the total systematic uncertainties.

\begin{table}[!p]
\centering
\topcaption{%
    Contributions of the considered sources of uncertainty to the total uncertainty in the inclusive cross sections.
    For each group of uncertainty sources, the impacts of the corresponding nuisance parameters on the total cross section are combined, taking into account their correlation in the fit.
    The numbers show relative uncertainties (in \%).
    The statistical uncertainty is obtained as the difference, in quadrature, between the total uncertainty and the sum of all systematic uncertainties.
}
\label{tab:impact:total_xs}
\renewcommand{\arraystretch}{1.1}
\begin{tabular}{lcccc}
                               & \multicolumn{4}{c}{Relative uncertainty (\%)} \\
Uncertainty source             &  \fjtb  & \sjtbtl & \sjfb & \sejfbtl \\
\hline
Integrated luminosity          &   1.6 &   1.6 &   2.0 &   1.8 \\
Pileup reweighting             &   0.2 &   0.8 &   0.4 &   0.5 \\
Lepton and trigger             &   1.1 &   0.9 &   1.9 &   1.8 \\
JES, JER                       &   2.1 &   1.6 &   3.5 &   5.7 \\
\PQb tagging                   &   4.5 &   3.9 &   7.0 &   9.1 \\[\cmsTabSkip]
\muR and \muF scales           &   2.8 &   6.8 &   8.2 &  12 \\
Top quark \pt modelling        &   0.3 &   1.0 &   0.6 &   1.3 \\
PDF                            &   0.2 &   0.7 &   1.0 &   1.9 \\
PS scales                      &   2.8 &   2.7 &   2.4 &   1.5 \\
ME-PS matching (\hdamp)        &   0.4 &   0.9 &   1.3 &   2.8 \\
Underlying event               &   0.4 &   \makebox[0pt][r]{$<$}0.1 &   0.4 &   0.4 \\
Colour reconnection            &   1.1 &   1.5 &   1.9 &   4.5 \\
\PQb quark fragmentation       &   0.3 &   0.4 &   0.4 &   0.4 \\
Inclusive \ttC cross section   &   0.5 &   0.3 &   1.9 &   2.6 \\[\cmsTabSkip]
MC statistical                 &   0.8 &   1.6 &   2.4 &   2.8 \\[\cmsTabSkip]
Total systematic uncertainty   &   6.0 &   8.7 &  13 &  17 \\
Statistical uncertainty        &   0.6 &   1.2 &   2.2 &   3.3 \\[\cmsTabSkip]
Total uncertainty              &   6.0 &   8.8 &  13 &  17 \\
\end{tabular}
\end{table}

Table~\ref{tab:impact:total_xs} shows the contributions of various sources of uncertainties to the total uncertainties in the inclusive cross sections, obtained from the combined impacts $I_{\mufid}$ on the parameters scaling the inclusive cross sections in the fits of the representative observables discussed in Section~\ref{sec:results:tot_xs}.
For these fits, the contributions of the 20 single nuisance parameters with the largest contributions to the uncertainty in the inclusive cross section can be found in Appendix~\ref{app:fidXS_impacts}.
Figure~\ref{fig:syst:impacts} shows the effect of the sources of uncertainty on the normalized differential cross section measurements for the \HT of \PQb jets in the \fjtb phase space.
For the four representative observables, corresponding figures are shown in Appendix~\ref{app:diffXS_impacts}.
The uncertainties in the inclusive cross sections are dominated by systematic sources, while the precision in the differential measurements is mainly limited by the statistical uncertainty in the data since the rate-based effects of many systematic uncertainties cancel out in the ratio between the absolute differential cross sections and the inclusive cross sections.
The leading systematic uncertainties originate from the calibration of the \PQb tagging and of the JES, the choice of \muR scale in the signal \ttbb and background \ttbar processes, and, for the differential measurements only, the finite number of simulated events.
Previous measurements of the inclusive \ttbb cross section by the ATLAS~\cite{Aaboud:2018eki} and CMS~\cite{Sirunyan:2017snr, CMS:2019eih, CMS:2020grm} Collaborations had the same leading sources of systematic uncertainty.

\begin{figure}[!ht]
\centering
\includegraphics[width=0.9\textwidth]{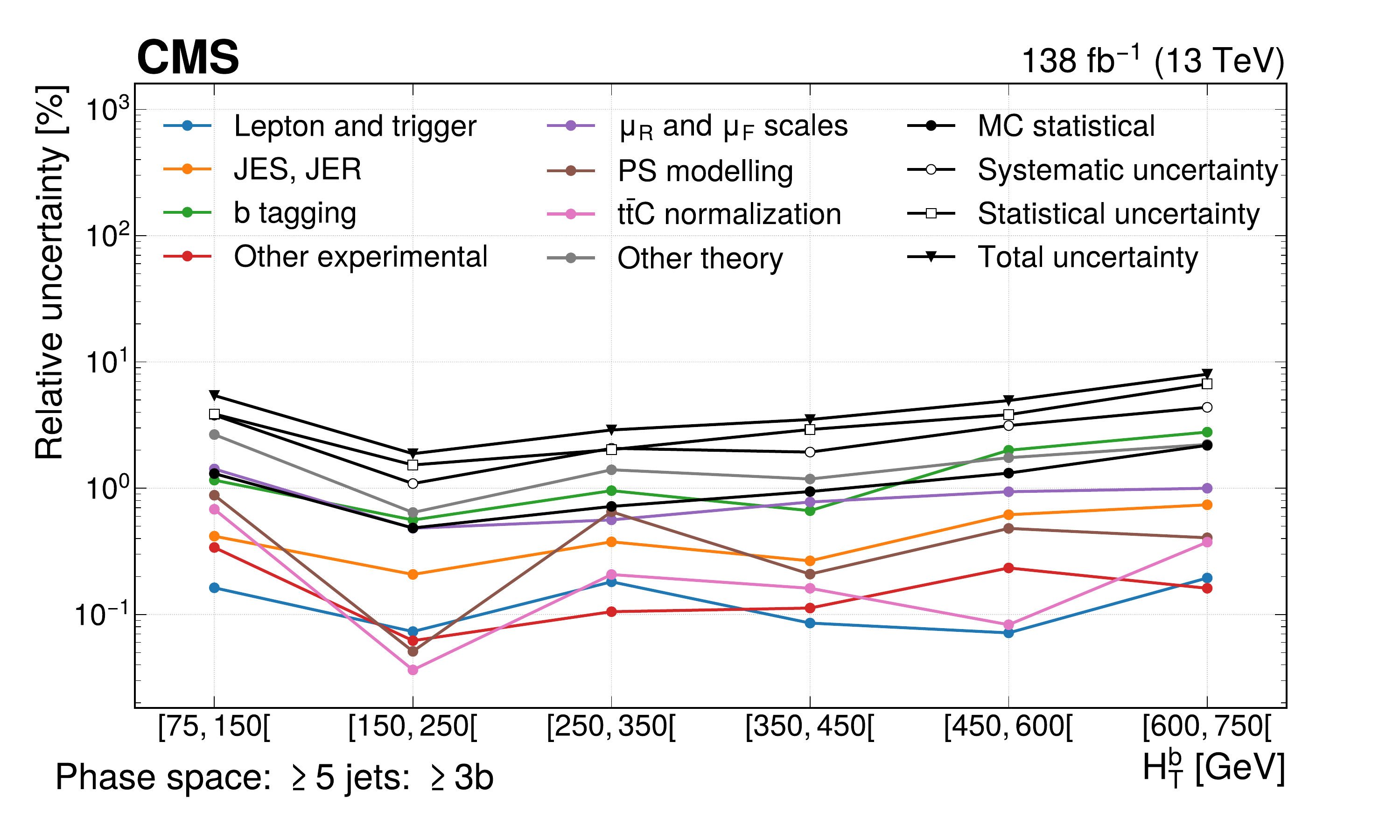}
\caption{%
    Effect of the considered sources of uncertainties on the measurement of the normalized differential cross section of the \HT of \PQb jets in the \fjtbLONG phase space, obtained by combining the impacts of associated nuisance parameters according to Eq.~\eqref{eq:unfolding:likelihood}.
    The ordering of the various sources is similar for other observables and in the other phase space regions.
    The last bin of the distribution is not shown, since it has no associated parameter of interest but is constrained by the other bins as described in Section~\ref{sec:signal_unf:unfolding}.
    The category ``other theory'' includes \PQb quark fragmentation, top quark \pt modelling, PDF, \hdamp, colour reconnection, and underlying event uncertainties.
    The category ``other experimental'' includes pileup and the integrated luminosity uncertainties.
}
\label{fig:syst:impacts}
\end{figure}

The nuisance parameter associated with the normalization of the \ttC process is not found to be constrained significantly beyond its prefit expectation and also shows no significant deviation from its expected value, which is consistent with the results of Ref.~\cite{CMS:TOP-20-003}.
The correlation of that nuisance parameter with the inclusive \ttbb cross section in the fits to different observables is below 5 (20)\% in the phase space regions with at least three (four) \PQb jets.
\section{Results \label{sec:results}}

The results are obtained with the statistical procedures described in Section~\ref{sec:signal_unf:unfolding}.
Inclusive and differential cross section results are presented in Sections~\ref{sec:results:tot_xs} and~\ref{sec:results:diff_xs}, respectively.

\subsection{Inclusive cross sections}
\label{sec:results:tot_xs}

\begin{figure}[!p]
\centering
\includegraphics[width=\textwidth]{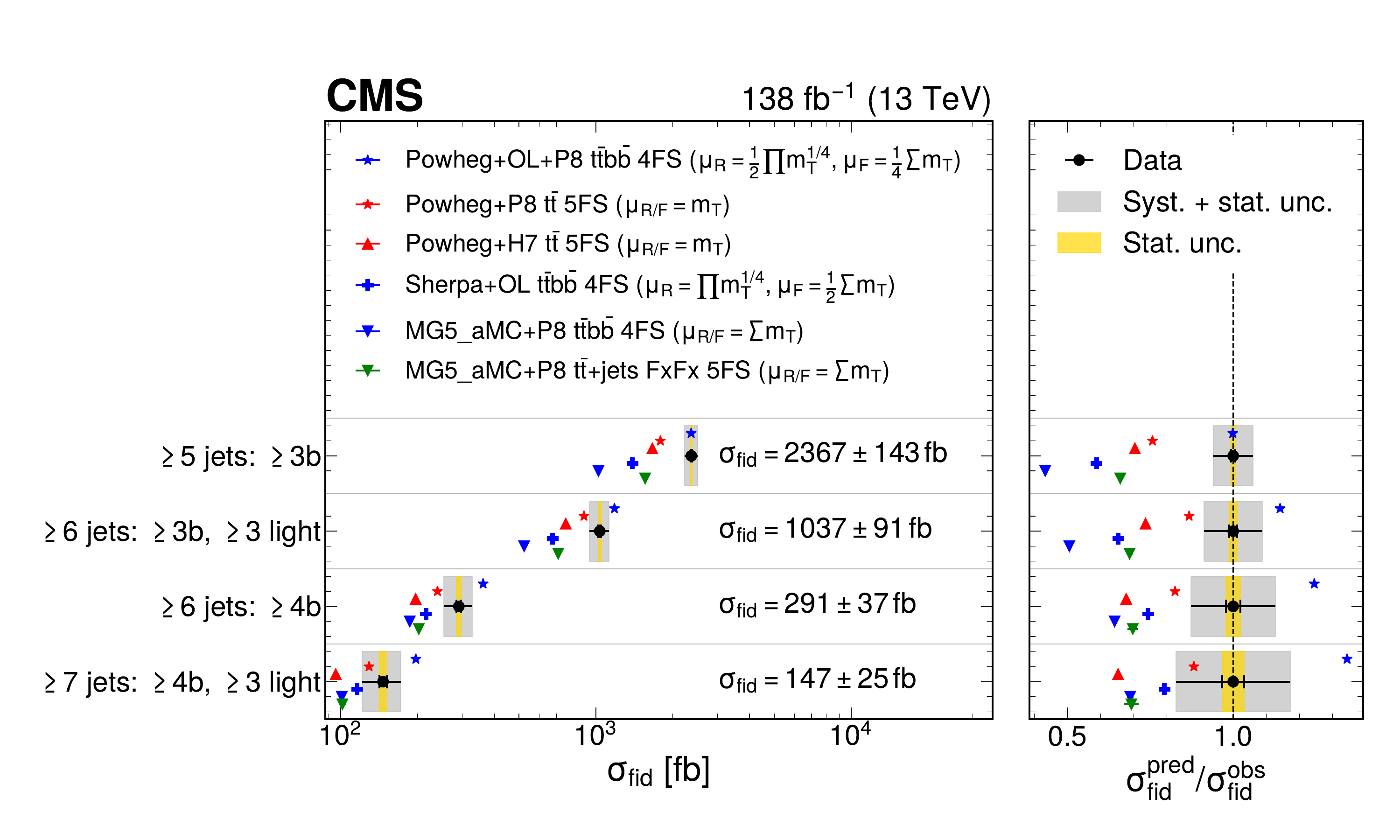}
\caption{%
    Measured inclusive cross sections for each considered phase space, compared to predictions from different \ttbb simulation approaches shown as coloured symbols.
    The predictions include uncertainties (horizontal bars) due to the limited number of simulated events.
    The blue colour is reserved for models using massive \PQb quarks and NLO QCD \ttbb MEs, while red is used for the inclusive \ttbar generators at NLO in QCD with massless \PQb quarks.
    The right panel shows the ratios between the predicted and measured cross sections, with the black bars showing the relative uncertainties in the measurements.
}
\label{fig:results:tot_xs}
\end{figure}

\begin{figure}[!p]
\centering
\includegraphics[width=0.9\textwidth]{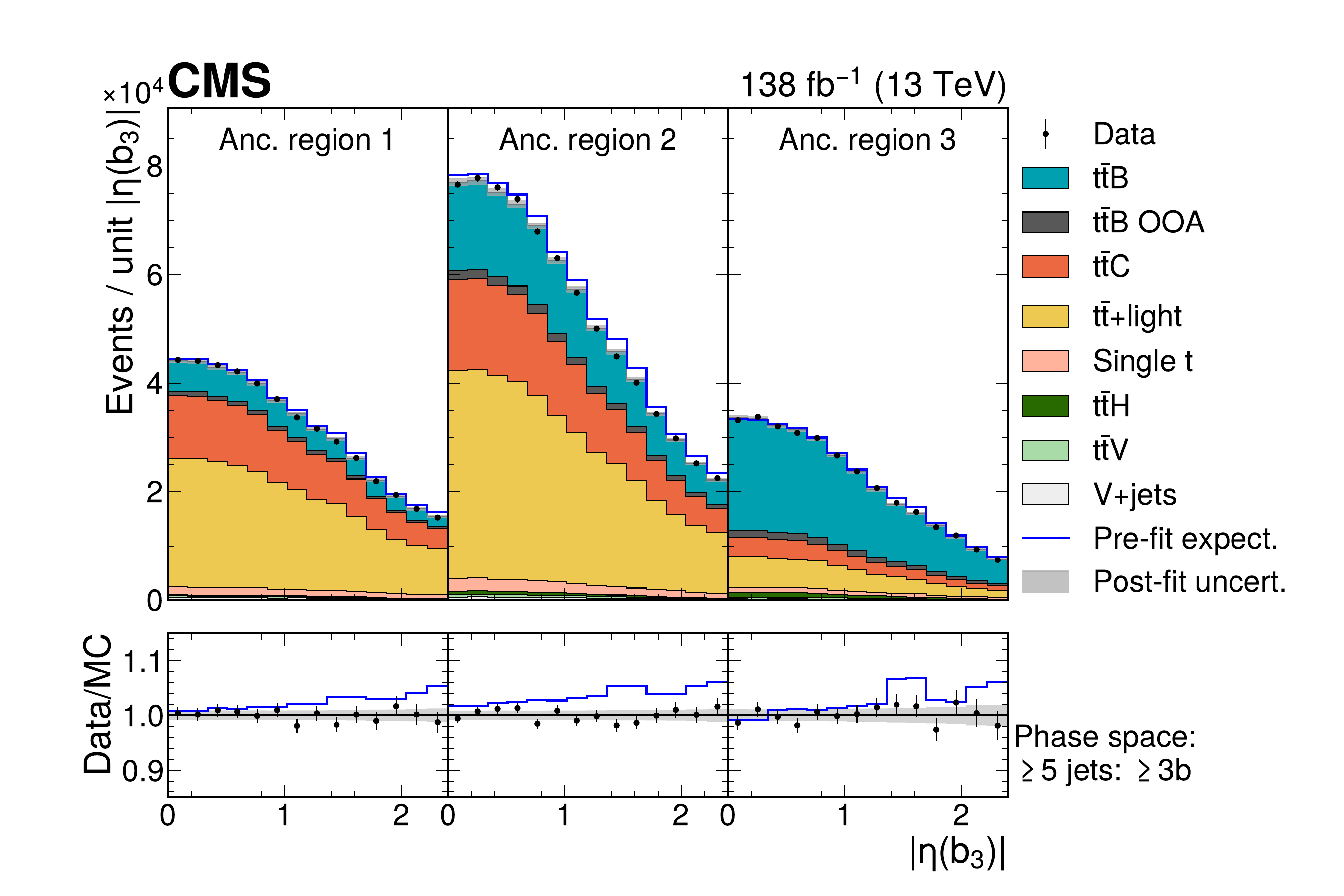}
\includegraphics[width=0.9\textwidth]{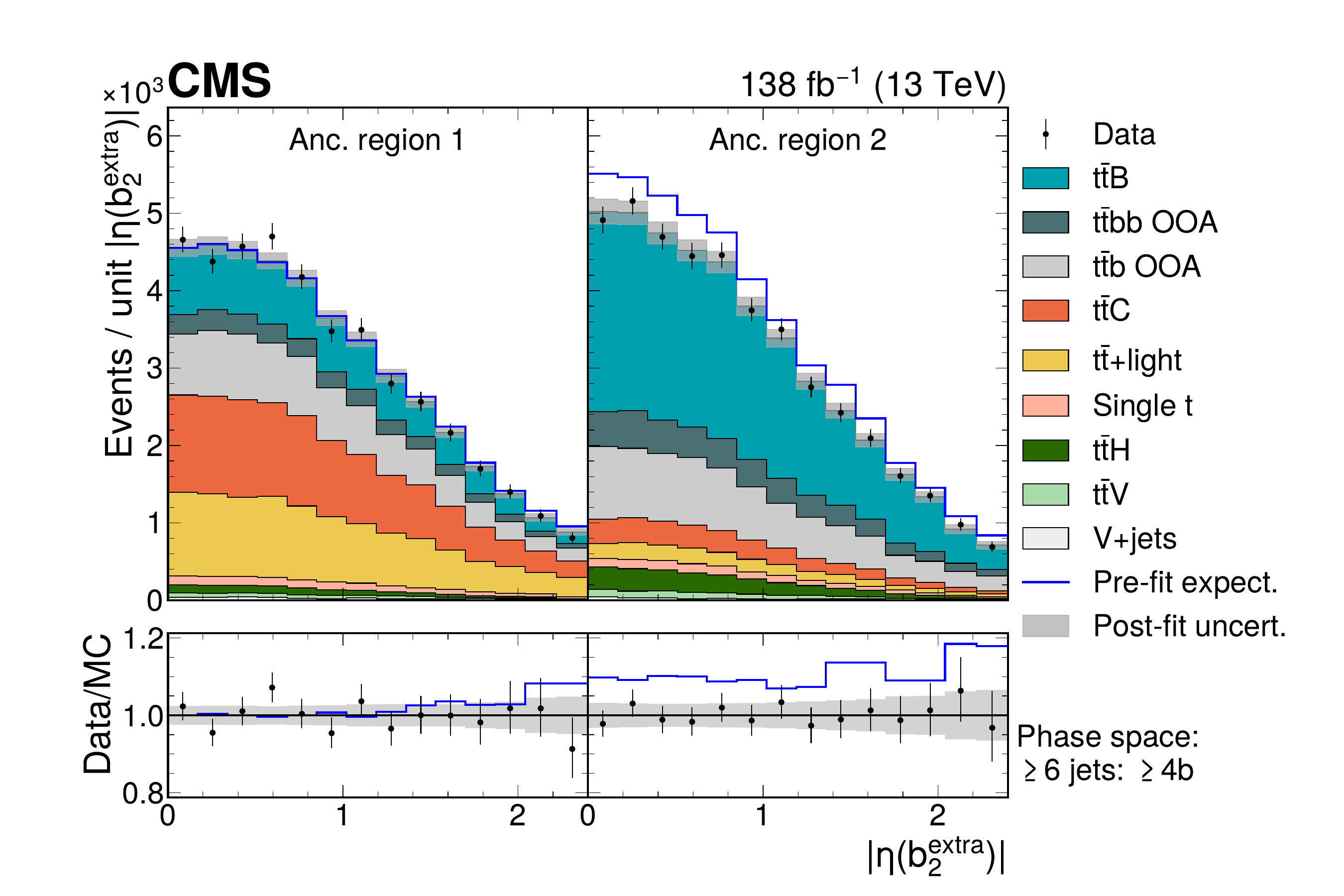}
\caption{%
    The \abseta of the third-hardest \PQb jet in \pt (\absetabN{3}) in the \fjtbLONG phase space (upper) and the \abseta of the subleading additional \PQb jet (\absetabXN{2}) in the \sjfbLONG phase space (lower) after the fit to data, shown for both lepton channels and all data periods combined.
    The distributions are shown separately for each ancillary region, as defined in Section~\ref{sec:signal_unf:anc_obs}.
    In the \fjtbLONG (\sjfbLONG) phase space the ancillary regions are defined as $\leq$2, 2, and $\geq$3 ($\leq$3 and $\geq$3) tight \PQb-tagged jets.
    The shaded bands include all uncertainties described in Section~\ref{sec:systematics} after profiling the nuisance parameters in the fit, estimated by sampling the predicted yields from the fit covariance matrix.
    The blue line shows the sum of the predicted yields for all processes before the fit to data, using the nominal \ttbb samples and its corresponding cross section for the signal.
    In the ratio panel the expected yields before the fit to data are shown relative to the predicted yields after the fit to data.
    The last bins contain the overflow.
}
\label{fig:results:post_fit_5j_3b_6j_4b}
\end{figure}

\begin{figure}[!p]
\centering
\includegraphics[width=0.9\textwidth]{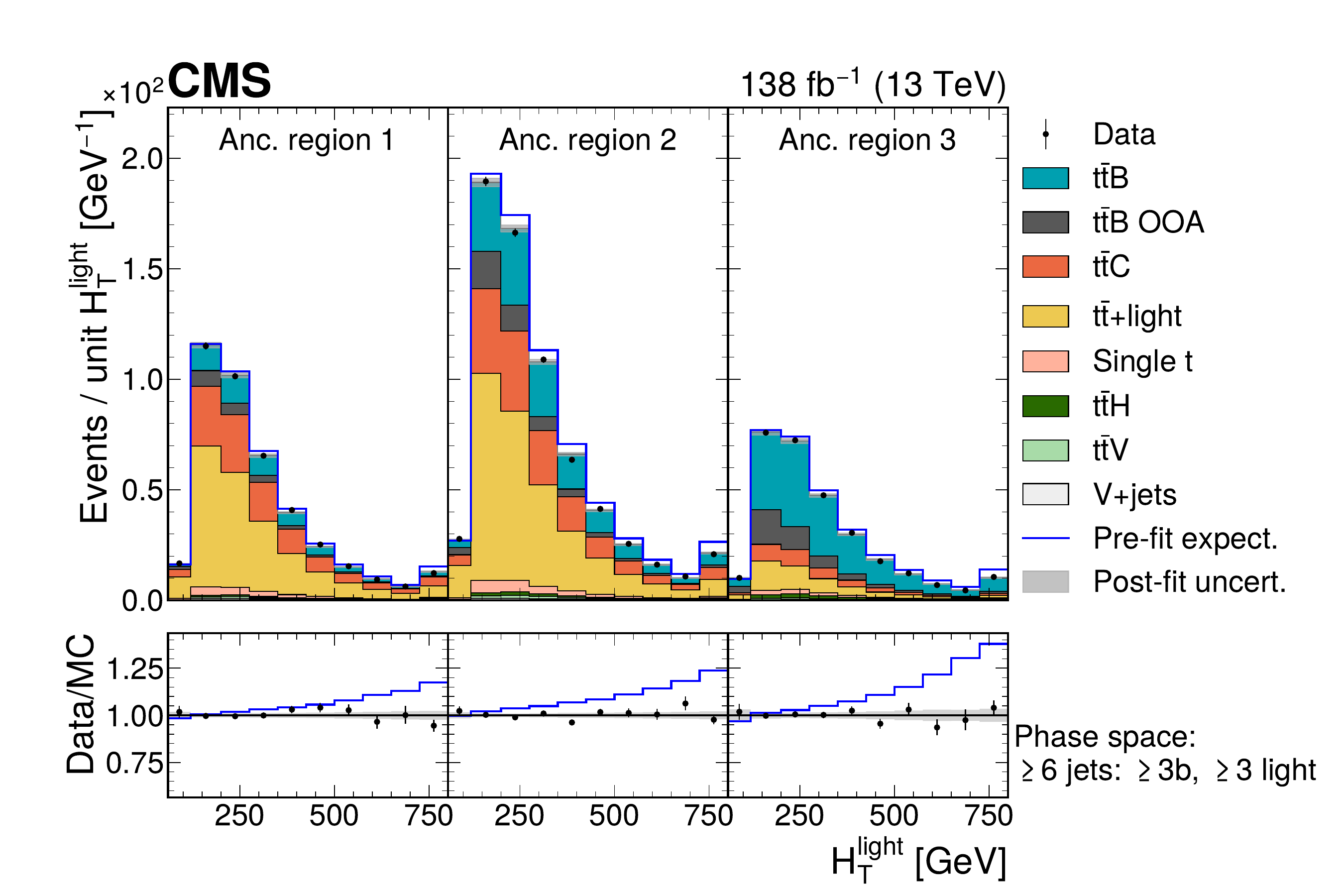}
\includegraphics[width=0.9\textwidth]{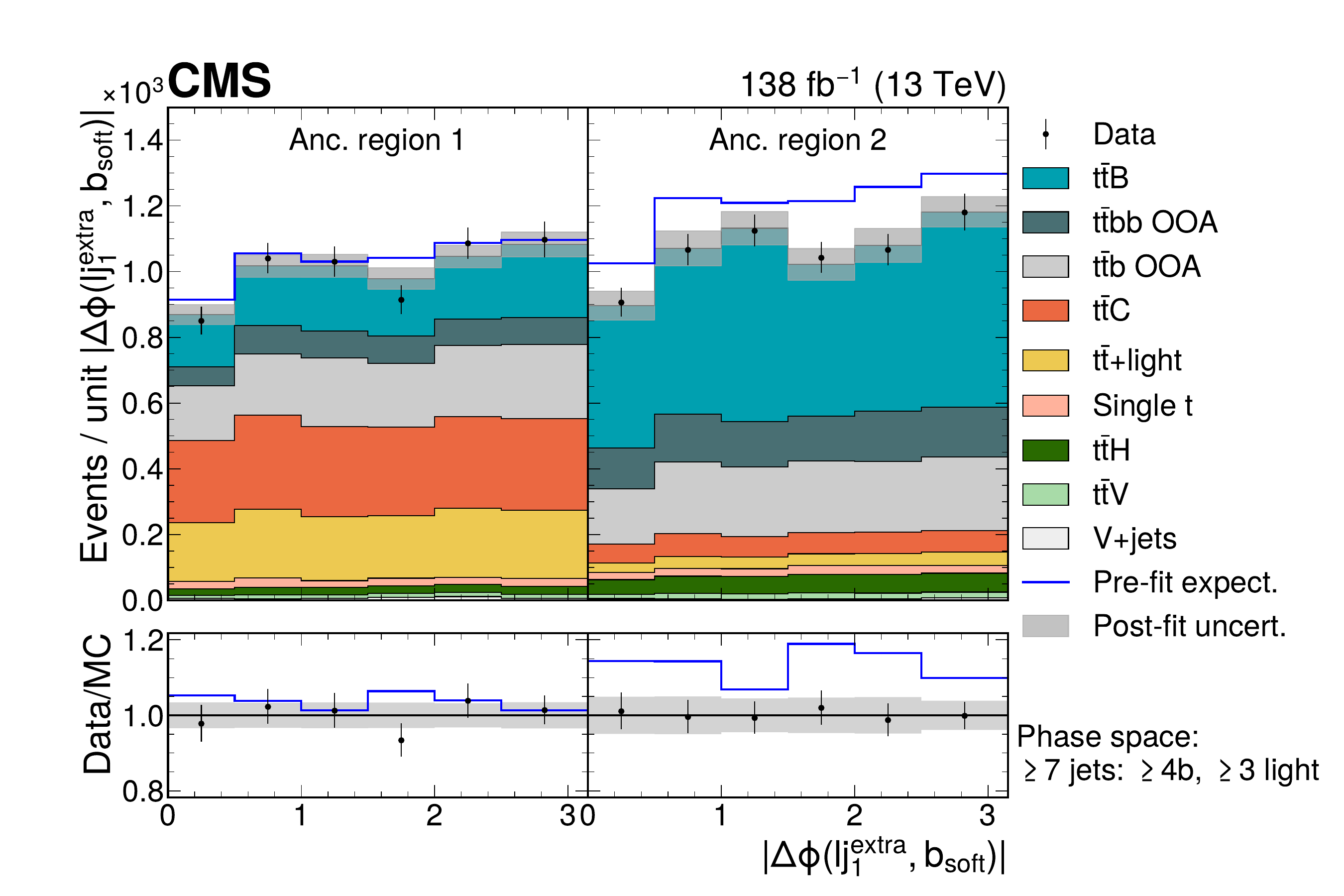}
\caption{%
    The \HT of all light jets in the \sjtbtlLONG phase space (upper) and the azimuthal angle between the hardest remaining light jet and the softest \PQb jet (\dPhiljb) in the \sejfbtlLONG phase space (lower) after the fit to data, shown for both lepton channels and all data periods combined.
    The distributions are shown separately for each ancillary region, as defined in Section~\ref{sec:signal_unf:anc_obs}.
    In the \sjtbtlLONG (\sejfbtlLONG) phase space the ancillary regions are defined as $\leq$2, 2, and $\geq$3 ($\leq$3 and $\geq$3) tight \PQb-tagged jets.
    The shaded bands include all uncertainties described in Section~\ref{sec:systematics} after profiling the nuisance parameters in the fit, estimated by sampling the predicted yields from the fit covariance matrix.
    The blue line shows the sum of the predicted yields for all processes before the fit to data, using the nominal \ttbb samples and its corresponding cross section for the signal.
    In the ratio panel the expected yields before the fit to data are shown relative to the predicted yields after the fit to data.
    The last bins contain the overflow.
}
\label{fig:results:post_fit_6j_3b_3l_7j_4b_3l}
\end{figure}

The inclusive \ttbb cross sections, measured in each of the fiducial phase space regions, are shown in Fig.~\ref{fig:results:tot_xs} and listed in Table~\ref{tab:results:tot_xs}, along with the predictions obtained from the \ttbb generator setups described in Section~\ref{sec:mc}.
In each phase space, the representative observable for which we report the measured inclusive cross section is that for which the measured value is closest to the mean of all measured inclusive cross sections in order to provide a value representative of the ensemble of measurements.
These observables are \absetabN{3} in the \fjtb phase space, \HTl in the \sjtbtl phase space, \absetabXN{2} in the \sjfb phase space, and \dPhiljb in the \sejfbtl phase space.
Their observed distributions at the detector level, combined for both lepton channels and all data-taking periods, are shown in Figs.~\ref{fig:results:post_fit_5j_3b_6j_4b} and~\ref{fig:results:post_fit_6j_3b_3l_7j_4b_3l} after the fits to data.
The correlations between the parameters \vecmu in the fit of \absetabN{3} in the \fjtb phase space are shown in Fig.~\ref{fig:supp:correlations_53:3}. For the other representative observables, corresponding figures are shown in Appendix~\ref{app:correlations}.
The inclusive cross section results for all observables are provided in the HEPData record for this analysis~\cite{hepdata}.

\begin{table}[!p]
\centering
\topcaption{%
    Measured and predicted inclusive cross sections in the four considered phase space regions (in fb).
}
\label{tab:results:tot_xs}
\renewcommand{\arraystretch}{1.2}
\cmsTable{\begin{tabular}{lr@{}l@{}lr@{}l@{}lr@{}l@{}lr@{}l@{}l}
    Fiducial phase space & \multicolumn{3}{c}{\fjtb} & \multicolumn{3}{c}{\sjtbtl} & \multicolumn{3}{c}{\sjfb} & \multicolumn{3}{c}{\sejfbtl} \\[\cmsTabSkip]
    \hline
    Measured cross section
    & 2367 &       &       & 1037 &      &       & 291 &       &       & 147 &       &       \\[-2pt]
    & & ${}\pm142$ & \syst & & ${}\pm90$ & \syst & & ${}\pm36$ & \syst & & ${}\pm24$ & \syst \\[-2pt]
    & & ${}\pm14$  & \stat & & ${}\pm12$ & \stat & & ${}\pm6$  & \stat & & ${}\pm5$  & \stat \\[\cmsTabSkip]
    \ttbbPP         & 2361 & & & 1183 & & & 361 & & & 197 & & \\
    \multicolumn{1}{r}{\muR variation} & & $+1161$ & /$-737$ & & $+826$ & /$-433$ & & $+183$ & /$-113$ & & $+121$ & /$-67$ \\
    \multicolumn{1}{r}{\muF variation} & & $+126$ & /$-100$ & & $+97$ & /$-78$ & & $+23$ & /$-18$ & & $+16$ & /$-13$ \\
    \ttbarPP        & 1791 & & &  899 & & & 240 & & & 129 & & \\
    \ttbarPH        & 1665 & & &  762 & & & 197 & & &  95 & & \\
    \ttbbSherpa     & 1391 & & &  677 & & & 216 & & & 116 & & \\
    \ttbbAMC        & 1024 & & &  524 & & & 187 & & & 101 & & \\
    \ttjetsAMC      & 1560 & & &  712 & & & 203 & & & 101 & & \\
\end{tabular}}
\end{table}

\begin{figure}[!t]
\centering
\includegraphics[width=0.85\textwidth]{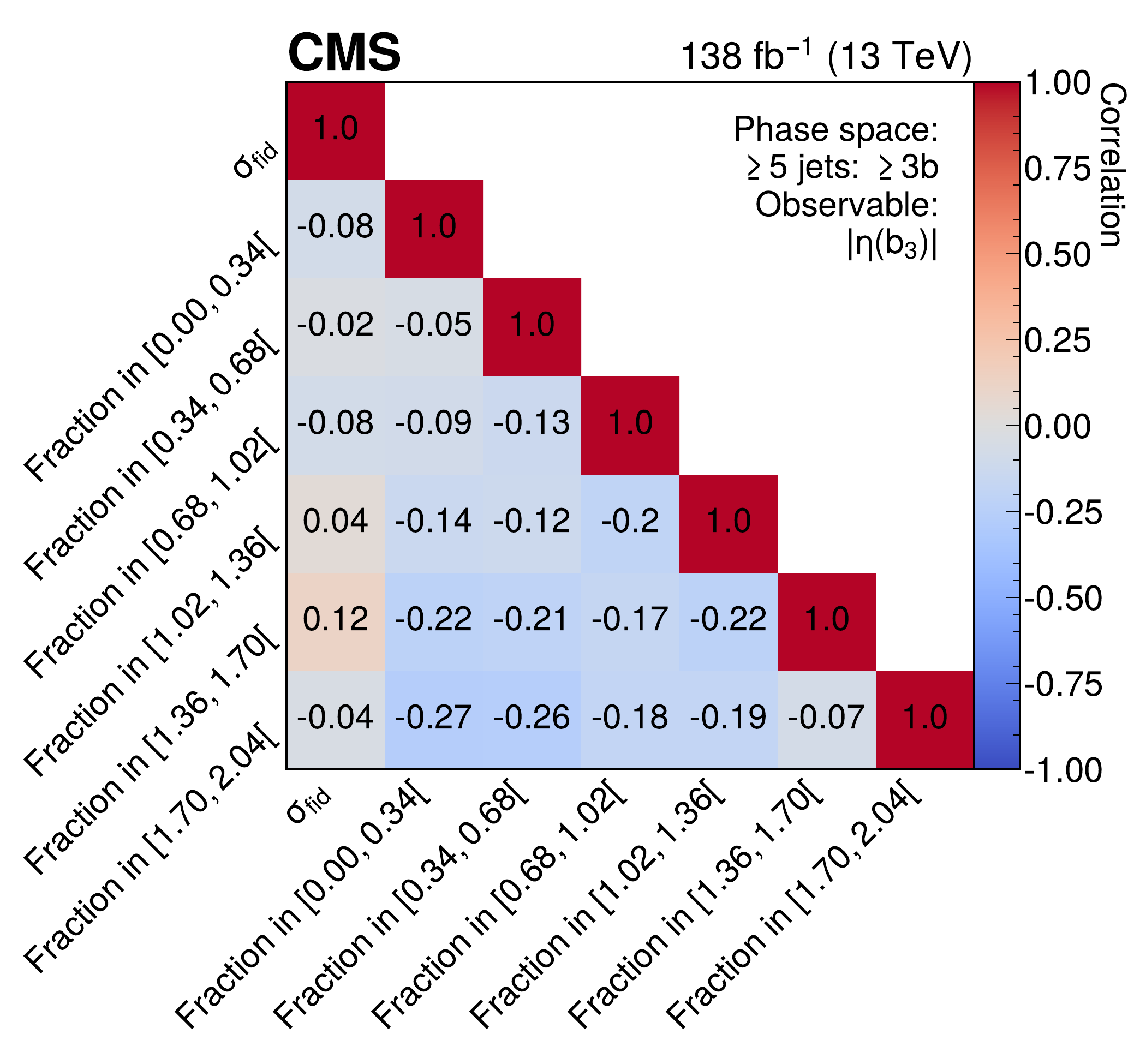}
\caption{%
    Correlations between the parameters of interest \vecmu in the fit for \absetabN{3} in the \fjtbLONG phase space.
}
\label{fig:supp:correlations_53:3}
\end{figure}

The measured cross sections in all phase space regions are larger than the theoretical predictions, as was also observed in previous measurements of the inclusive \ttbb cross section (with somewhat different fiducial definitions) by the CMS~\cite{Sirunyan:2017snr, CMS:2019eih, CMS:2020grm} and ATLAS~\cite{Aaboud:2018eki} Collaborations, with the notable exception of the \ttbbPP generator setup.
The choice of reduced central \muR and \muF scales in the \ttbbPP sample, justified by the previous measurements and by studies of fixed-order NLO QCD corrections to the \ttbbj process~\cite{Buccioni:2019plc}, results in significantly larger cross sections in all phase space regions.
This prediction agrees well with the measurement in the \fjtb phase space, but overestimates the cross section by 10--35\% in the other phase space regions.
Nevertheless, the \ttbbPP predictions agree with the measurements when considering their \muR scale uncertainties of about 50\%, estimated by varying the \muR scale by a factor of two.
When considering \muR scale uncertainties in both the ME and in the PS, the \ttbarPP predictions agree with the measurements in the phase space regions targeting additional light radiation in \ttb and \ttbb events.
The cross sections based on \ttbb matrix element calculations are expected to increase if \ttbar events with additional \PQb quarks produced in MPI are included. For the \ttbbPP prediction, this increase is estimated using the \ttbarPP simulation to be 6--10\% depending on the phase space.

The \ttbbSherpa predictions are between 20 and 50\% lower than the measured cross sections. This can mostly be attributed to the \muR and \muF scale choices in this \ttbbSherpa setup, which are a factor of two higher than the scale choices of the \ttbbPP simulation. With common scale settings between the \ttbbSherpa and \ttbbPP simulation approaches, compatible inclusive and differential cross section predictions are achieved in the studies presented in Ref.~\cite{Ferencz:2023fso}.
Increased cross section predictions could similarly be achieved in the other simulations by lowering the \muR and \muF scale settings.

\subsection{Normalized differential cross sections}
\label{sec:results:diff_xs}

The normalized differential cross section is measured in four different fiducial phase space regions for 29 observables that use exclusively stable-particle level information without reference to any simulated event history, and eight observables targeting explicitly the \PQb jets that do not originate from decaying top quarks.
For each observable, customized bin sizes are chosen, depending on the resolution of the observables and the statistical uncertainty in the measured event yields.

\begin{figure}[!p]
\centering
\includegraphics[width=0.5\textwidth]{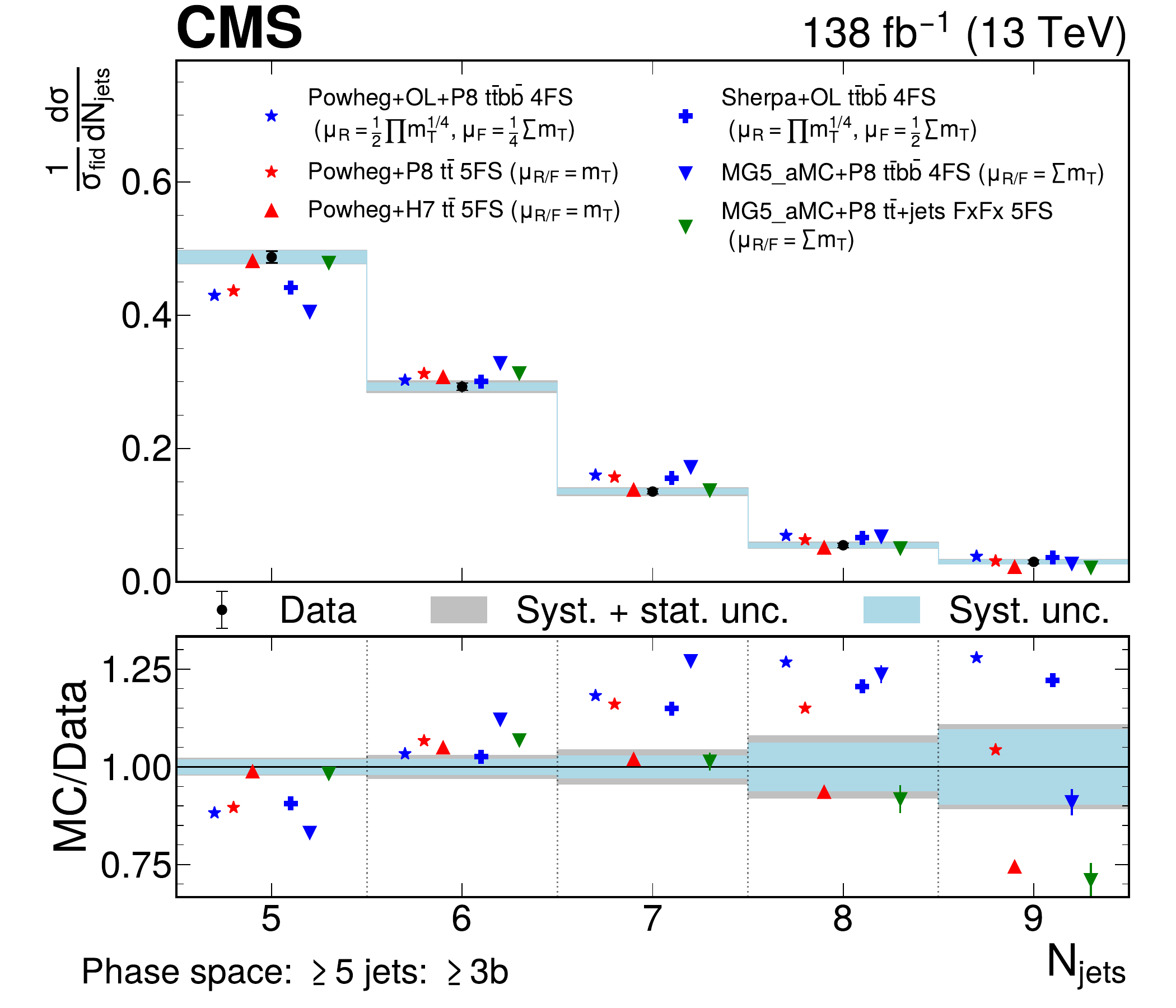}%
\includegraphics[width=0.5\textwidth]{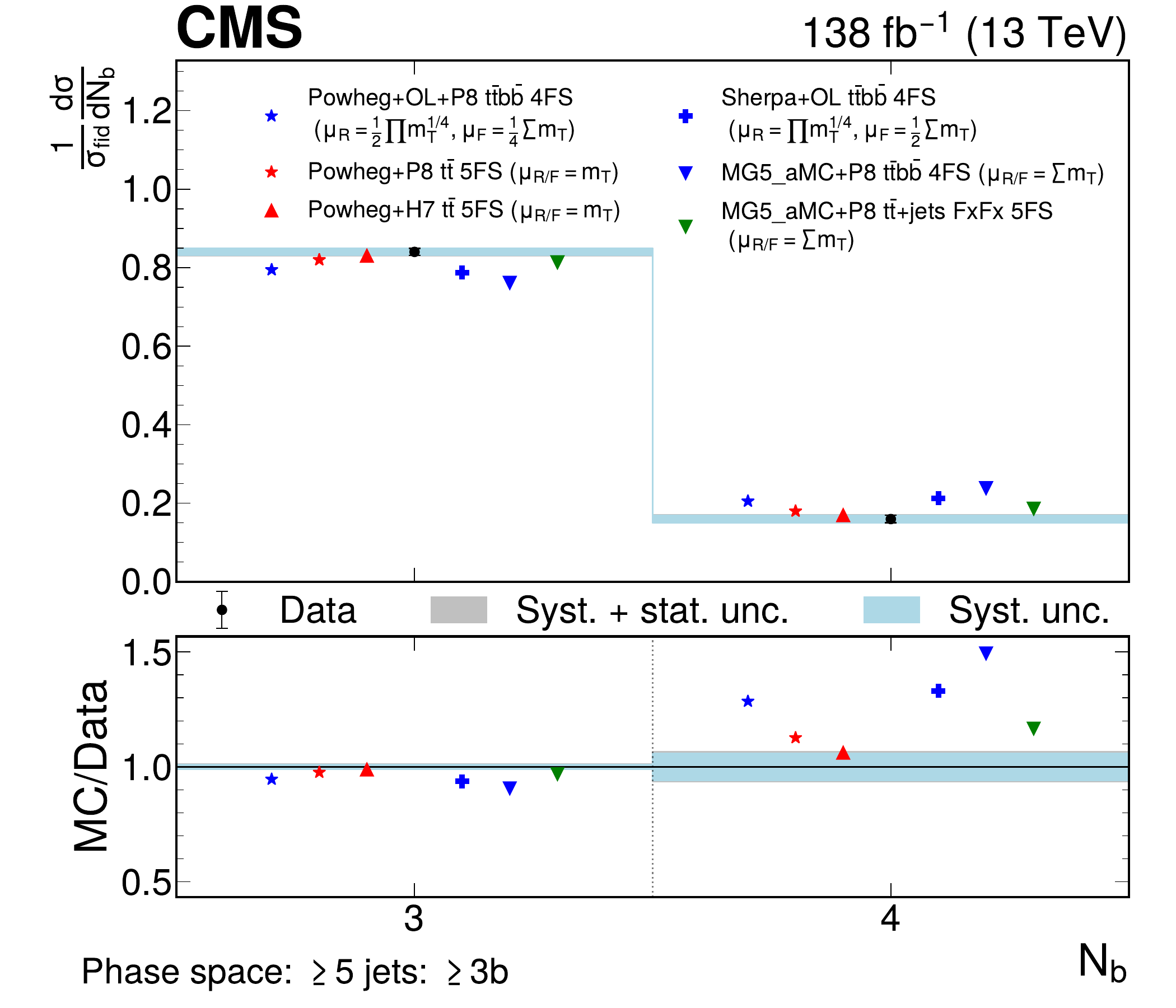} \\
\includegraphics[width=0.5\textwidth]{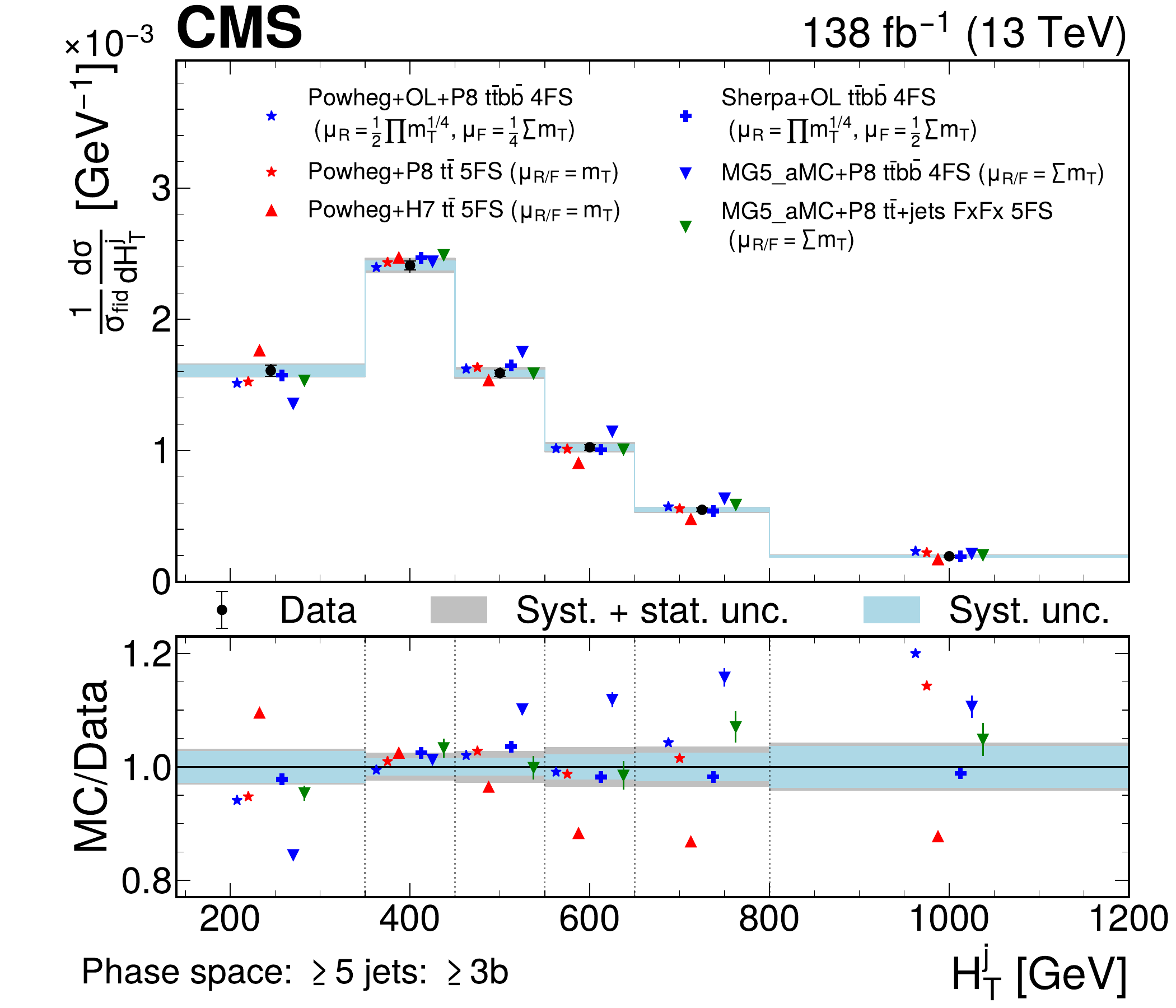}%
\includegraphics[width=0.5\textwidth]{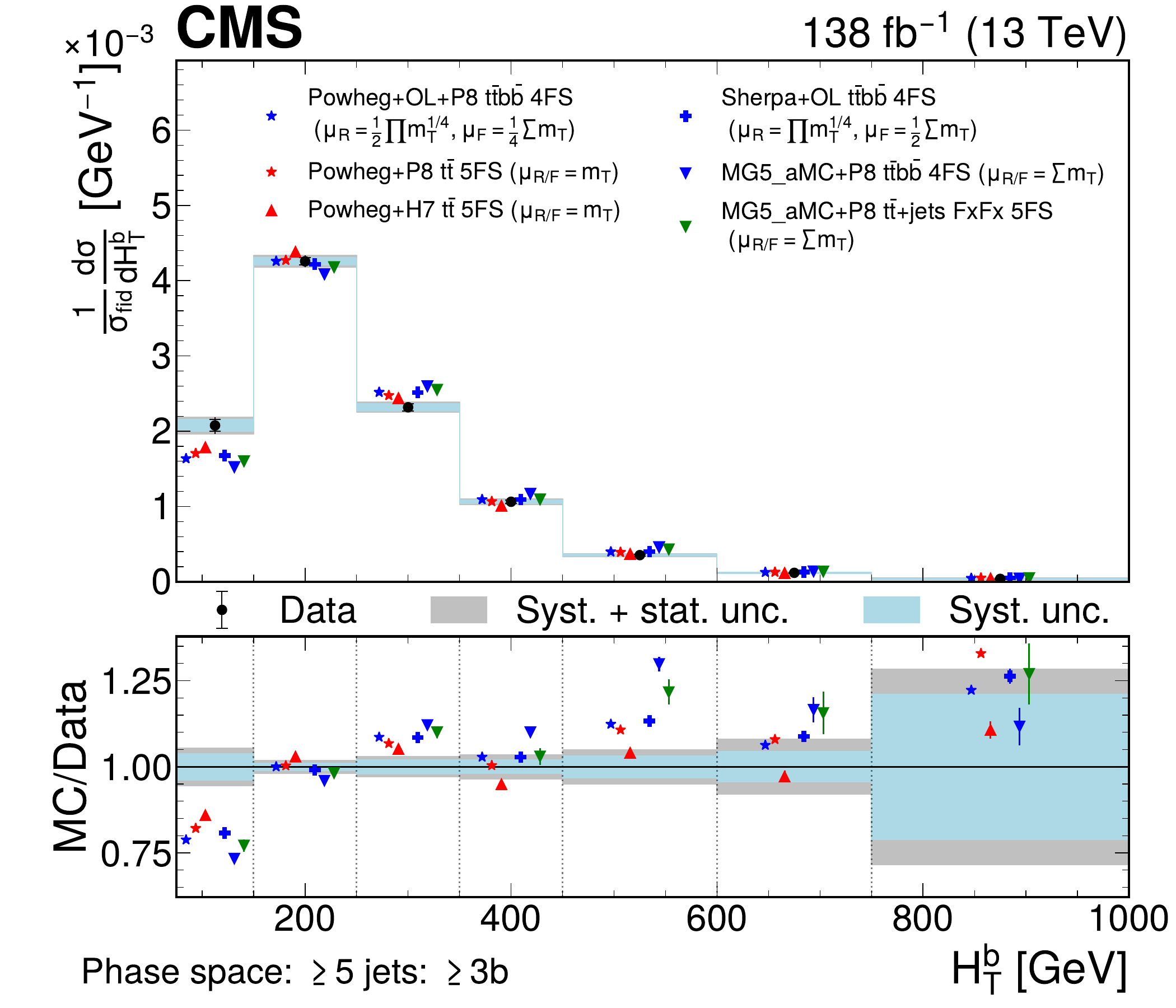} \\
\includegraphics[width=0.5\textwidth]{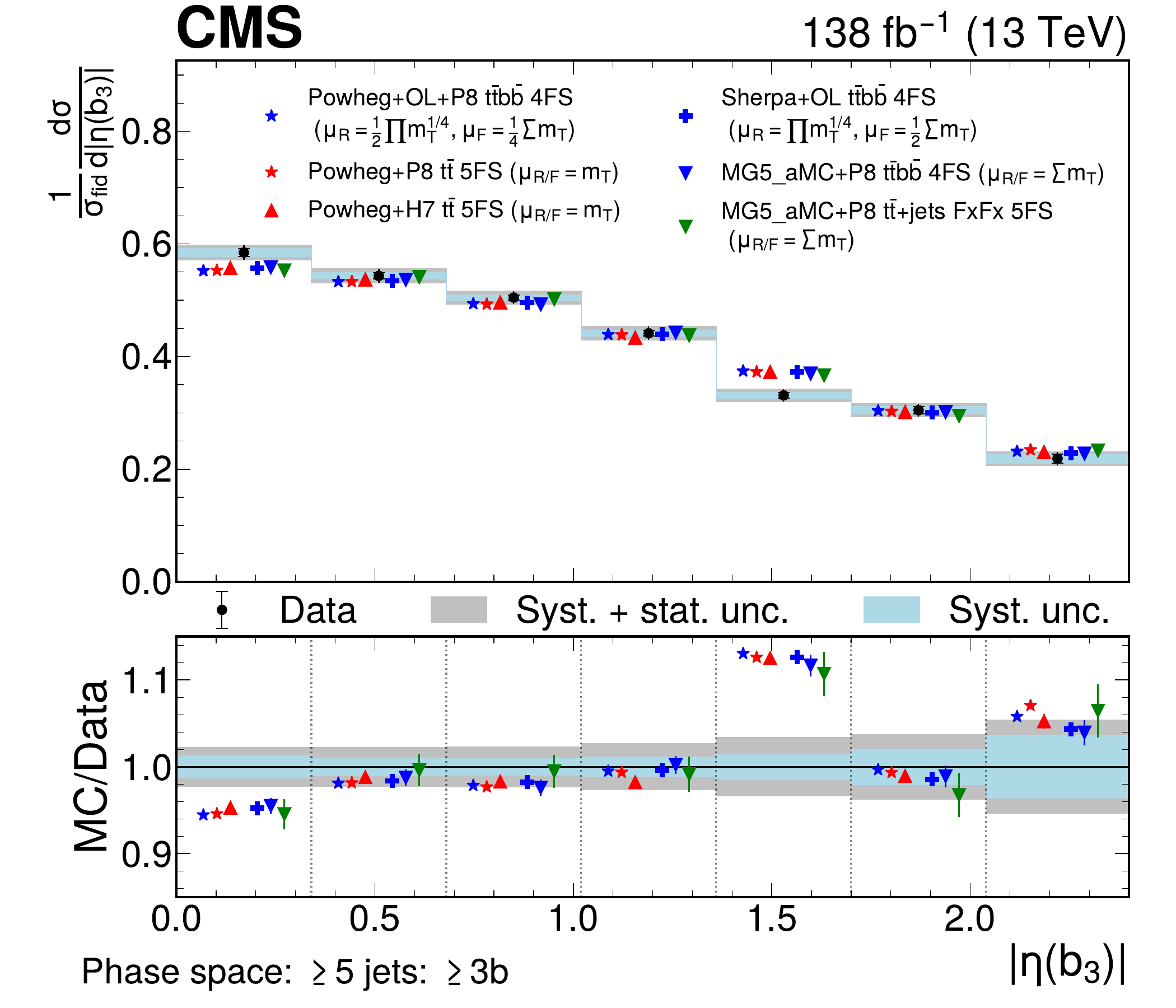}%
\includegraphics[width=0.5\textwidth]{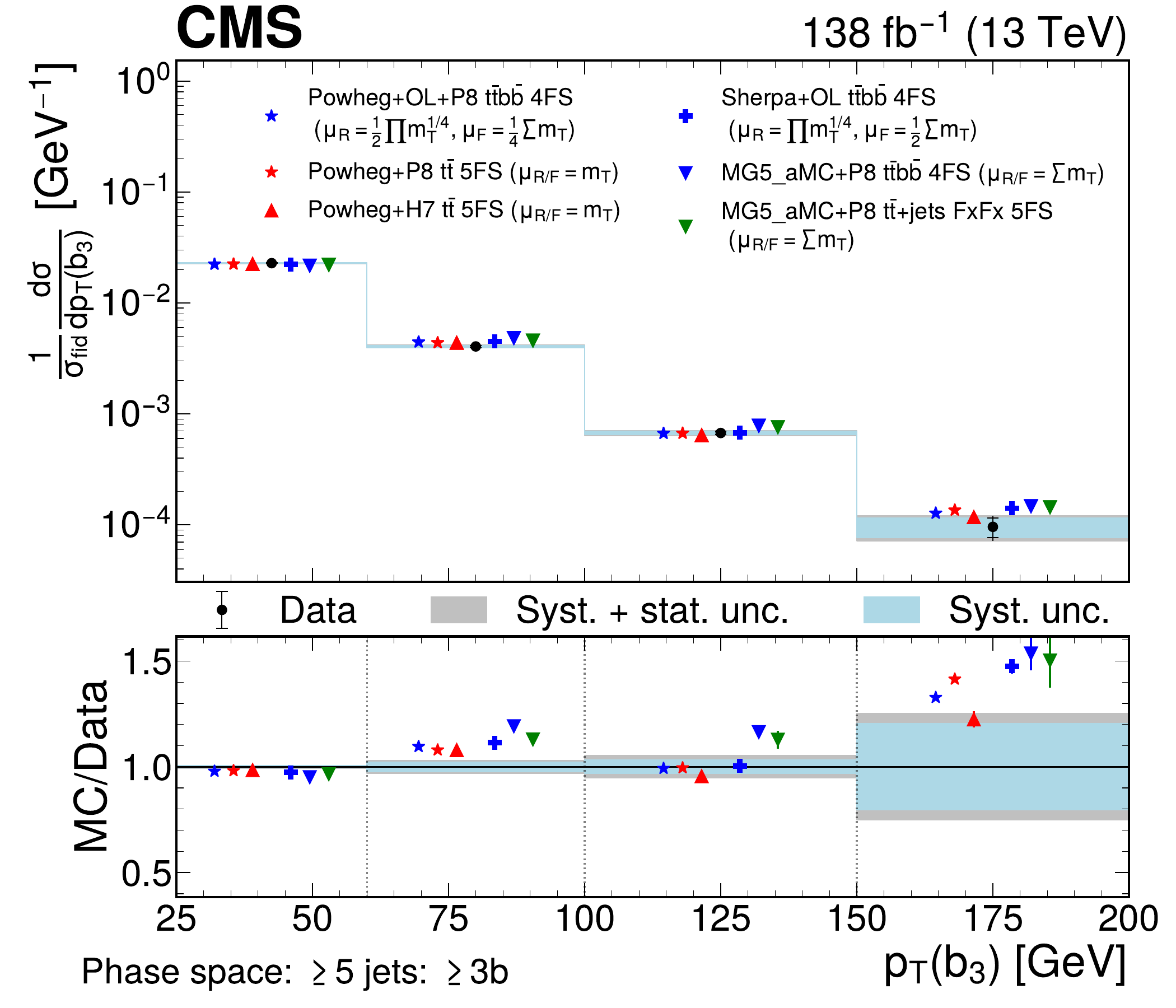}
\caption{%
    Predicted and observed normalized differential cross sections in the \fjtbLONG fiducial phase space, for the inclusive jet multiplicity (upper left), the \PQb jet multiplicity (upper right), the inclusive jet \HT (middle left, \HTj), the \HT of \PQb jets (middle right, \HTb), the \abseta of the third \PQb jet (lower left), and the \pt of the third \PQb jet (lower right).
    The data are represented by points, with inner (outer) vertical bars indicating the systematic (total) uncertainties, also represented as blue (grey) bands.
    Cross section prediction obtained at the particle level from different simulation approaches are shown, including their statistical uncertainties, as coloured symbols with different shapes.
    For \Nj, \Nb, \HTj, \HTb, and \ptbN{3}, the last bins contain the overflow.
}
\label{fig:results:5j3b}
\end{figure}

\begin{figure}[!p]
\centering
\includegraphics[width=0.5\textwidth]{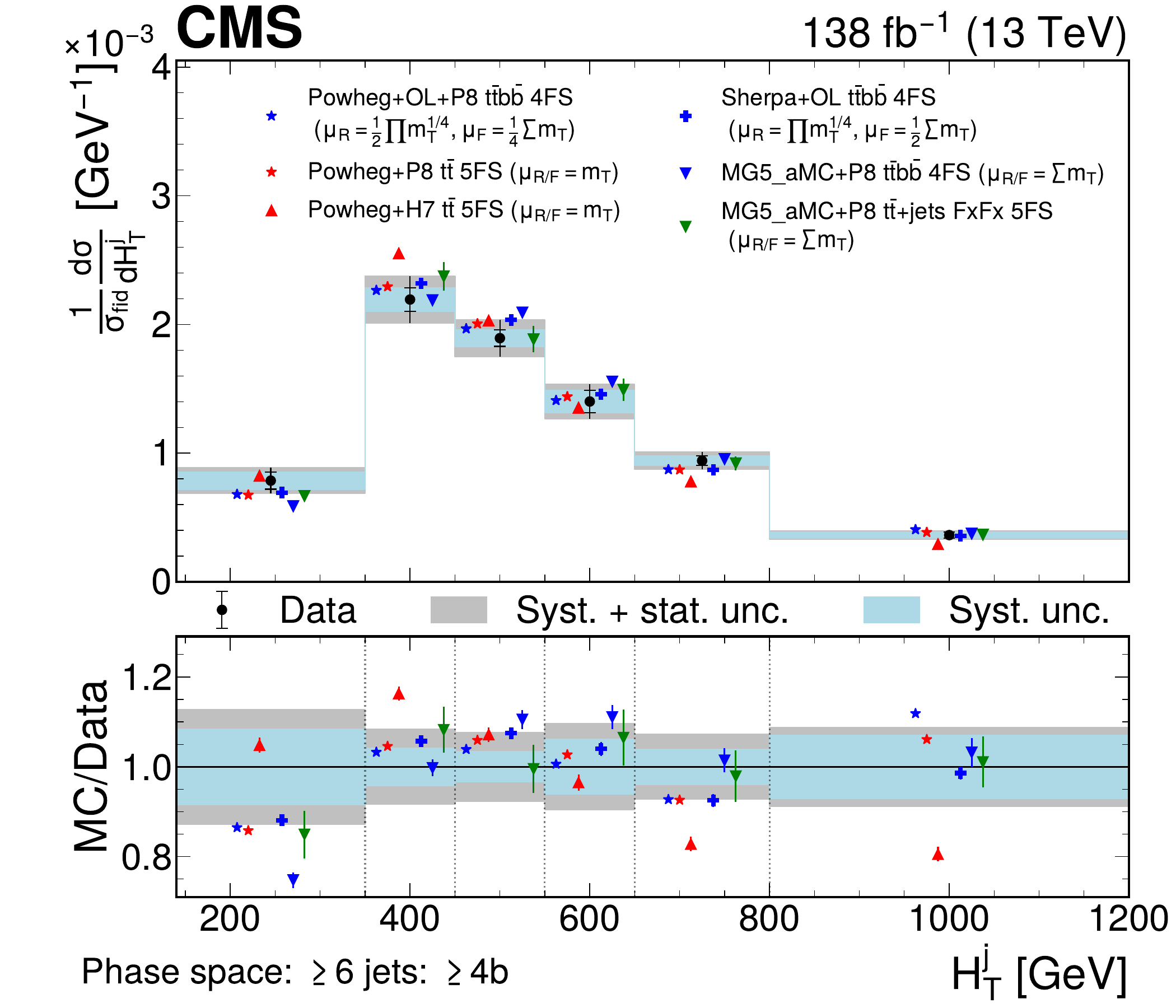}%
\includegraphics[width=0.5\textwidth]{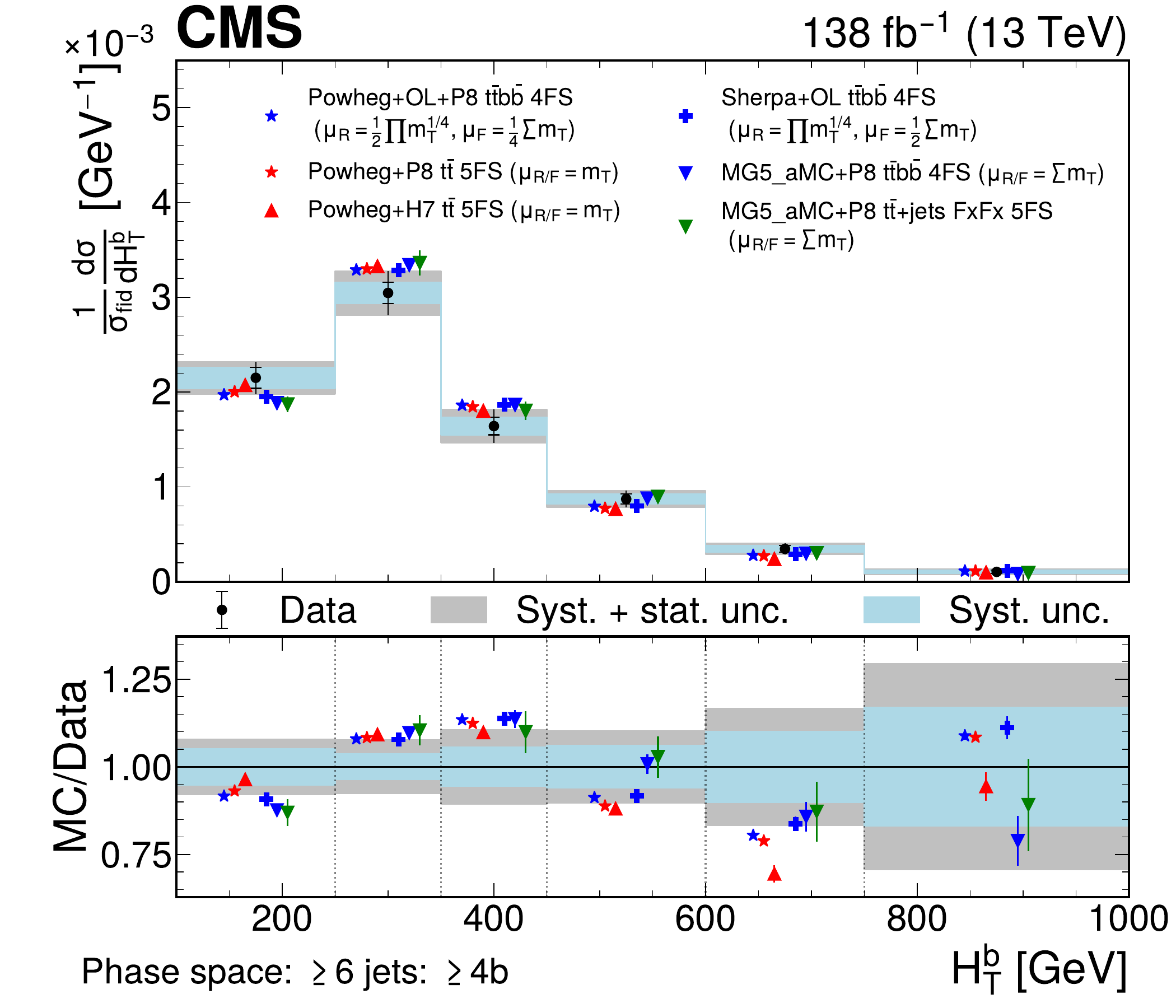} \\
\includegraphics[width=0.5\textwidth]{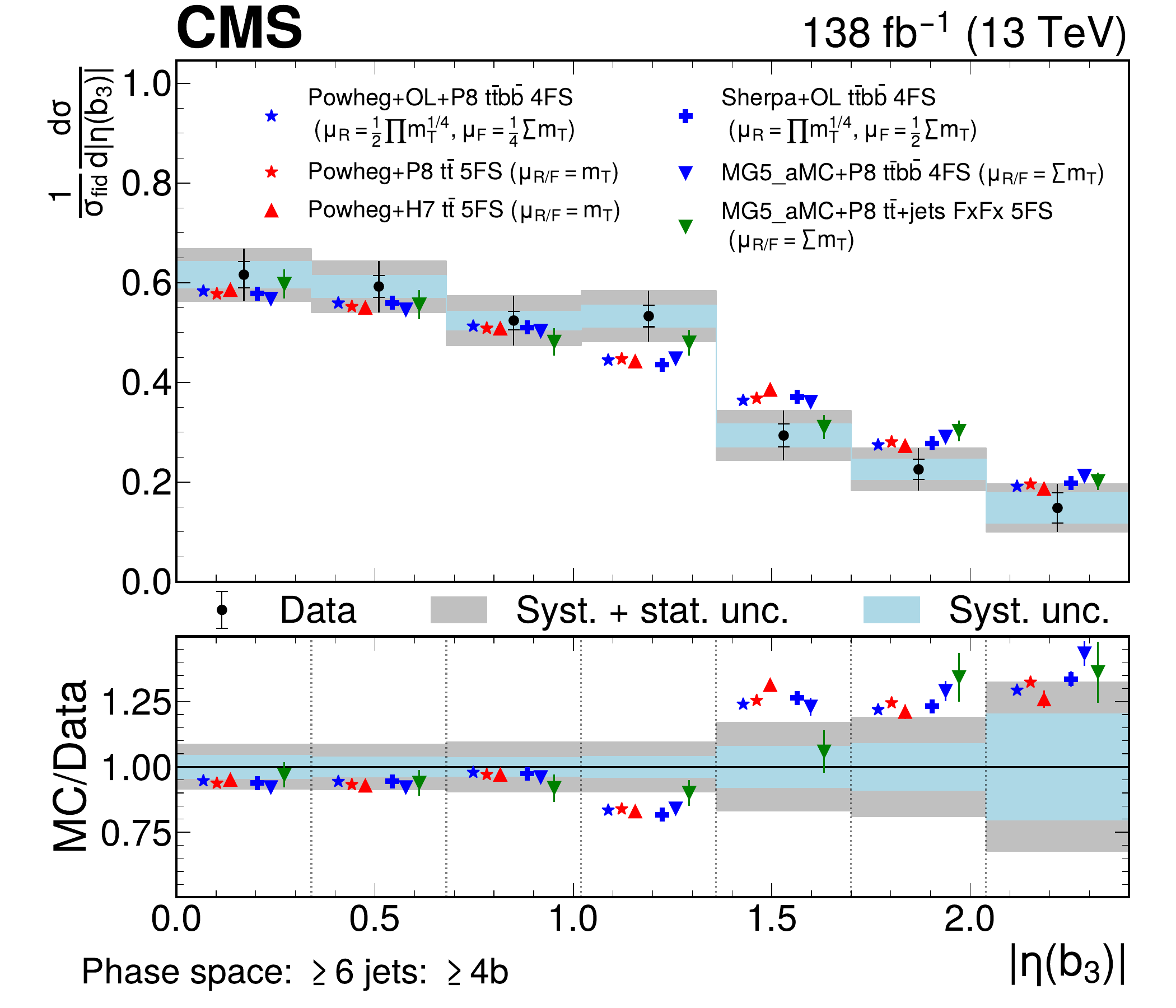}%
\includegraphics[width=0.5\textwidth]{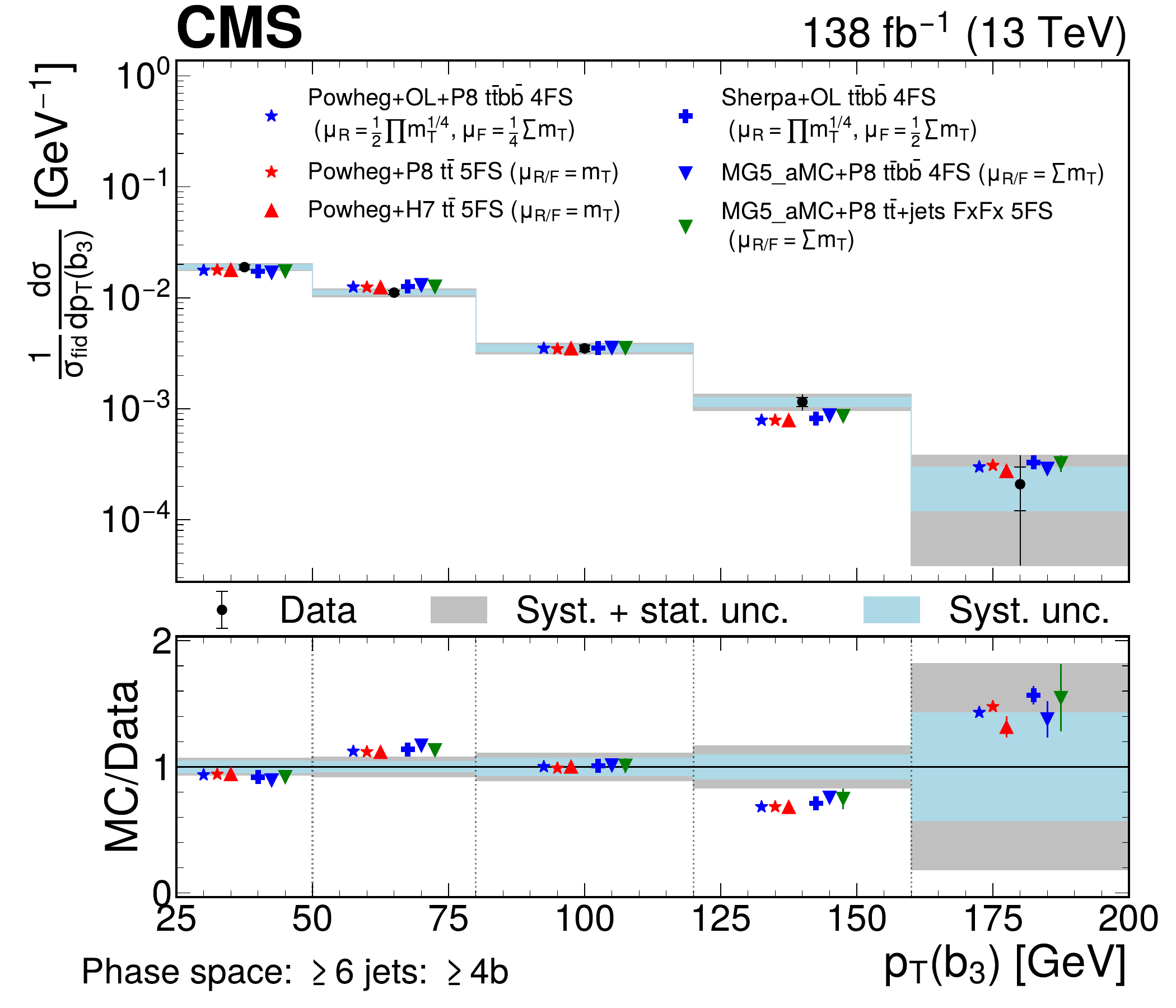} \\
\includegraphics[width=0.5\textwidth]{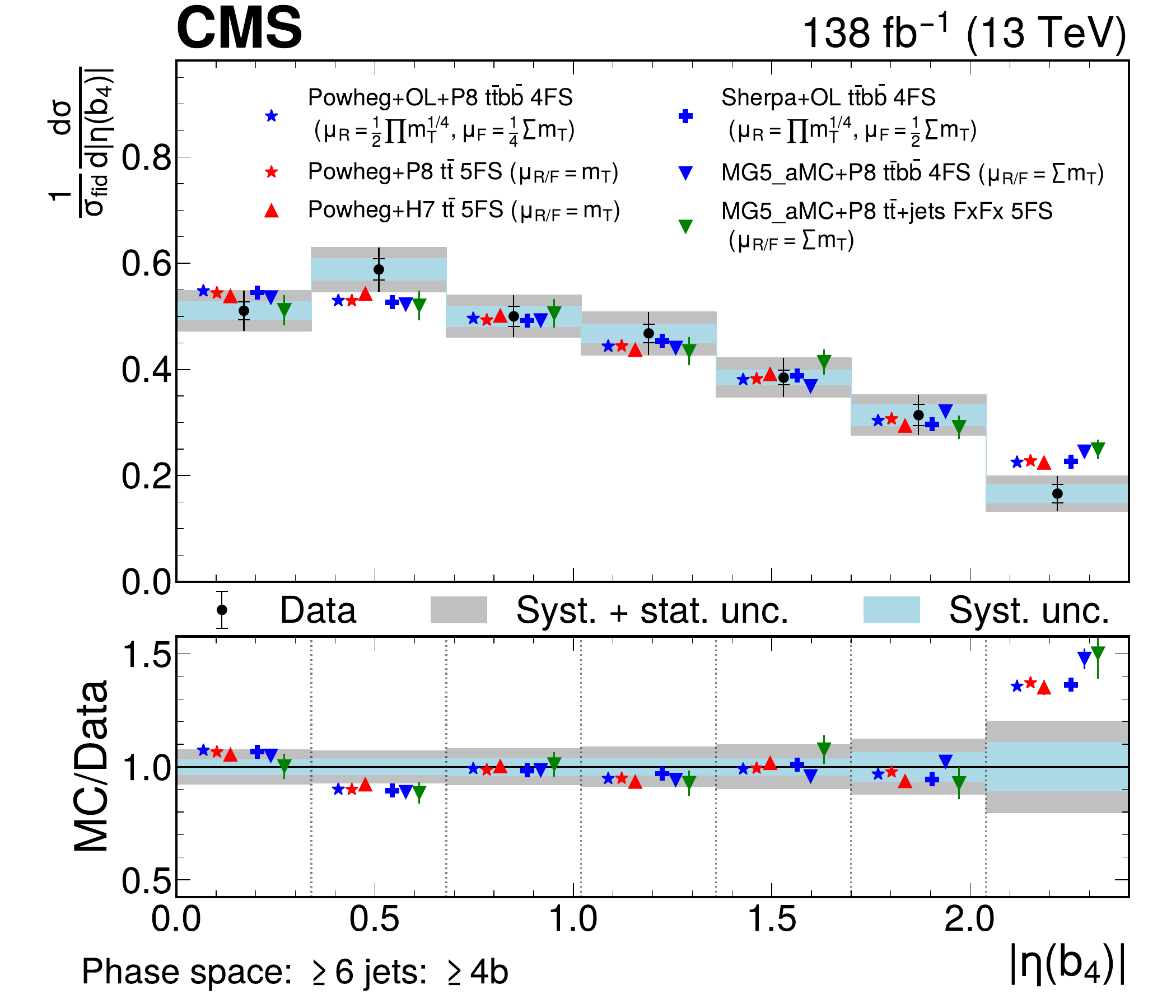}%
\includegraphics[width=0.5\textwidth]{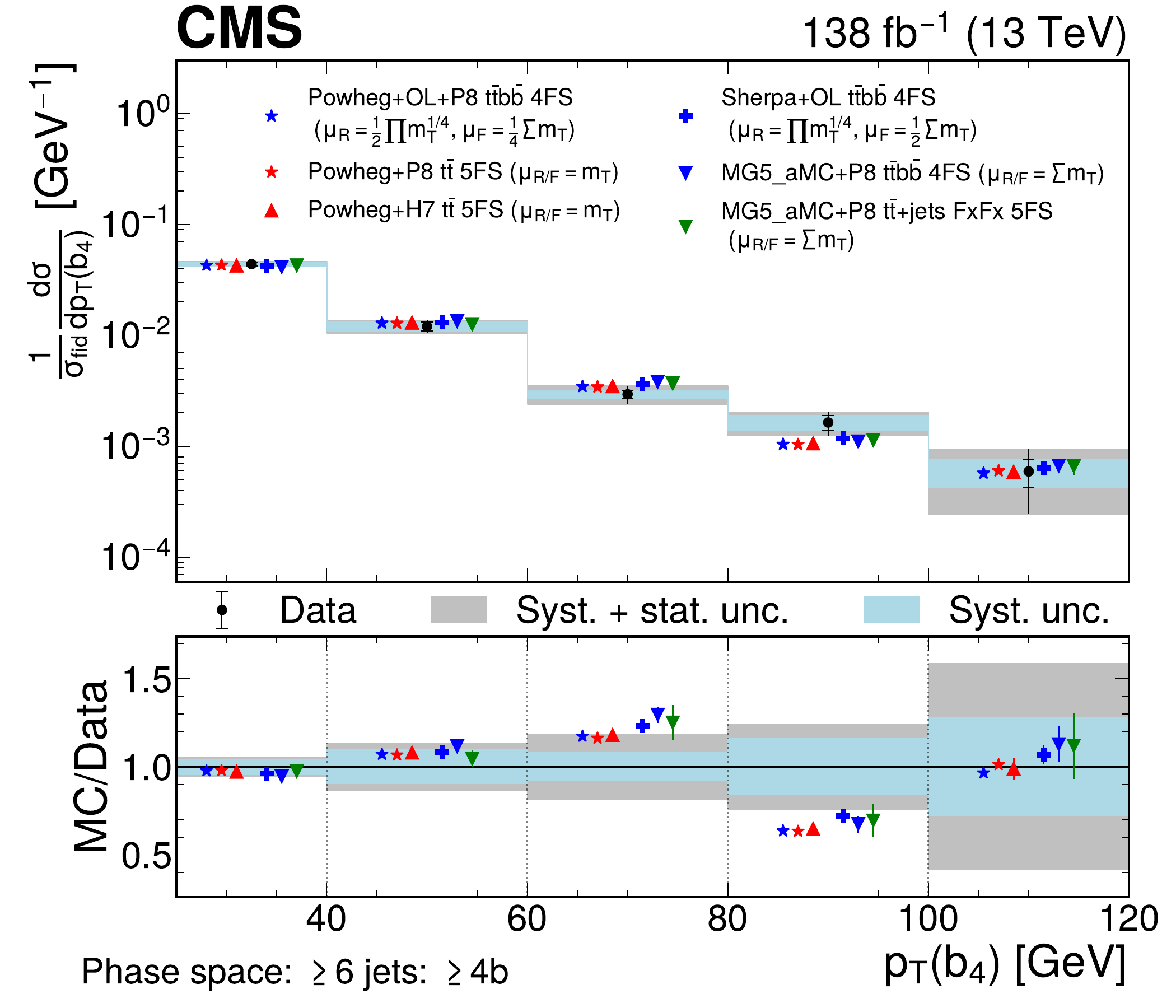}
\caption{%
    Predicted and observed normalized differential cross sections in the \sjfbLONG fiducial phase space, for the inclusive jet \HT (upper left), the \HT of \PQb jets (upper right), the \abseta of the third \PQb jet (middle left), the \pt of the third \PQb jet (middle right), the \abseta of the fourth \PQb jet (lower left), and the \pt of the fourth \PQb jet (lower right).
    The data are represented by points, with inner (outer) vertical bars indicating the systematic (total) uncertainties, also represented as blue (grey) bands.
    Cross section predictions obtained at the particle level from different simulation approaches are shown, including their statistical uncertainties, as coloured symbols with different shapes.
    For \HT and \pt, the last bins contain the overflow.
}
\label{fig:results:6j4b:1}
\end{figure}

\begin{figure}[!p]
\centering
\includegraphics[width=0.5\textwidth]{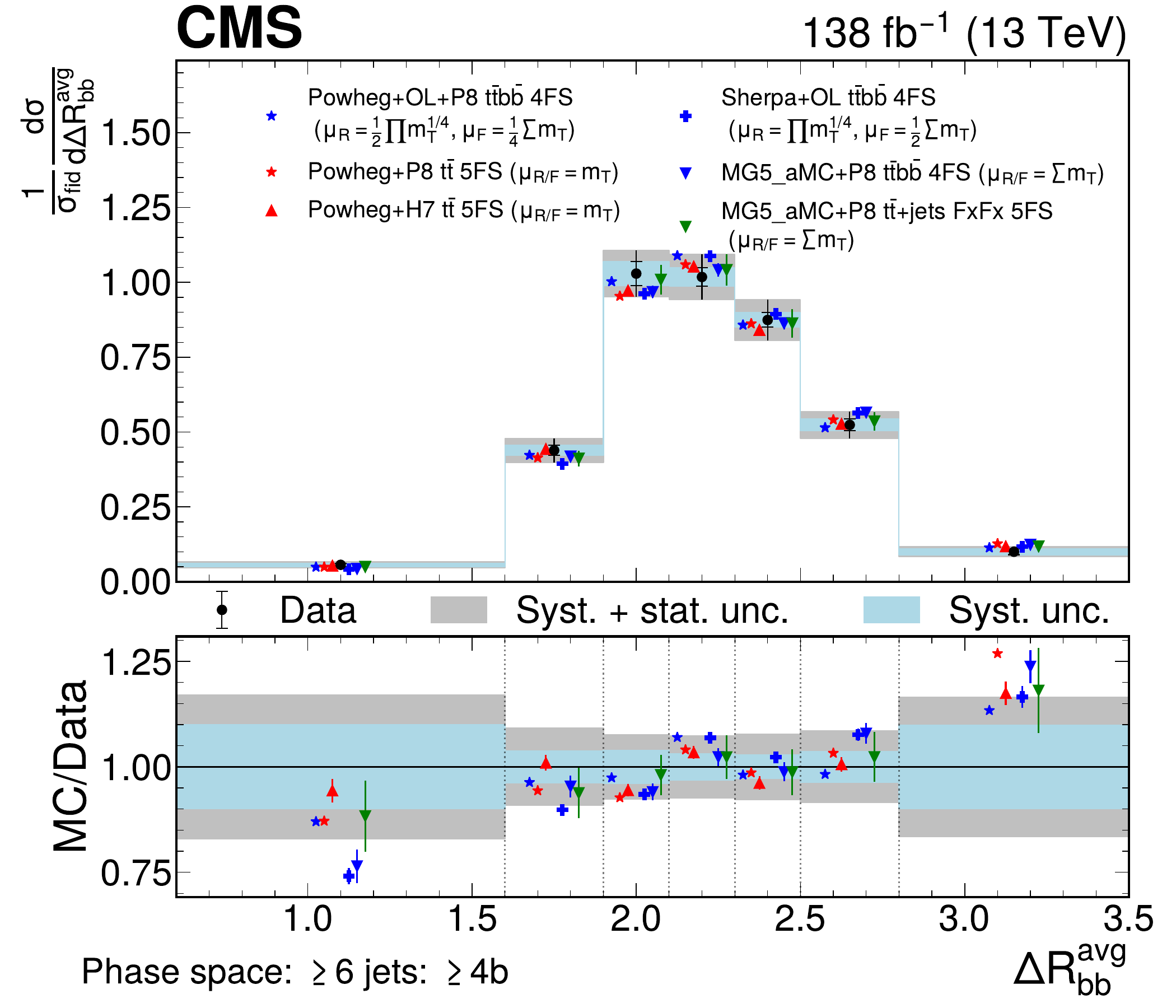}%
\includegraphics[width=0.5\textwidth]{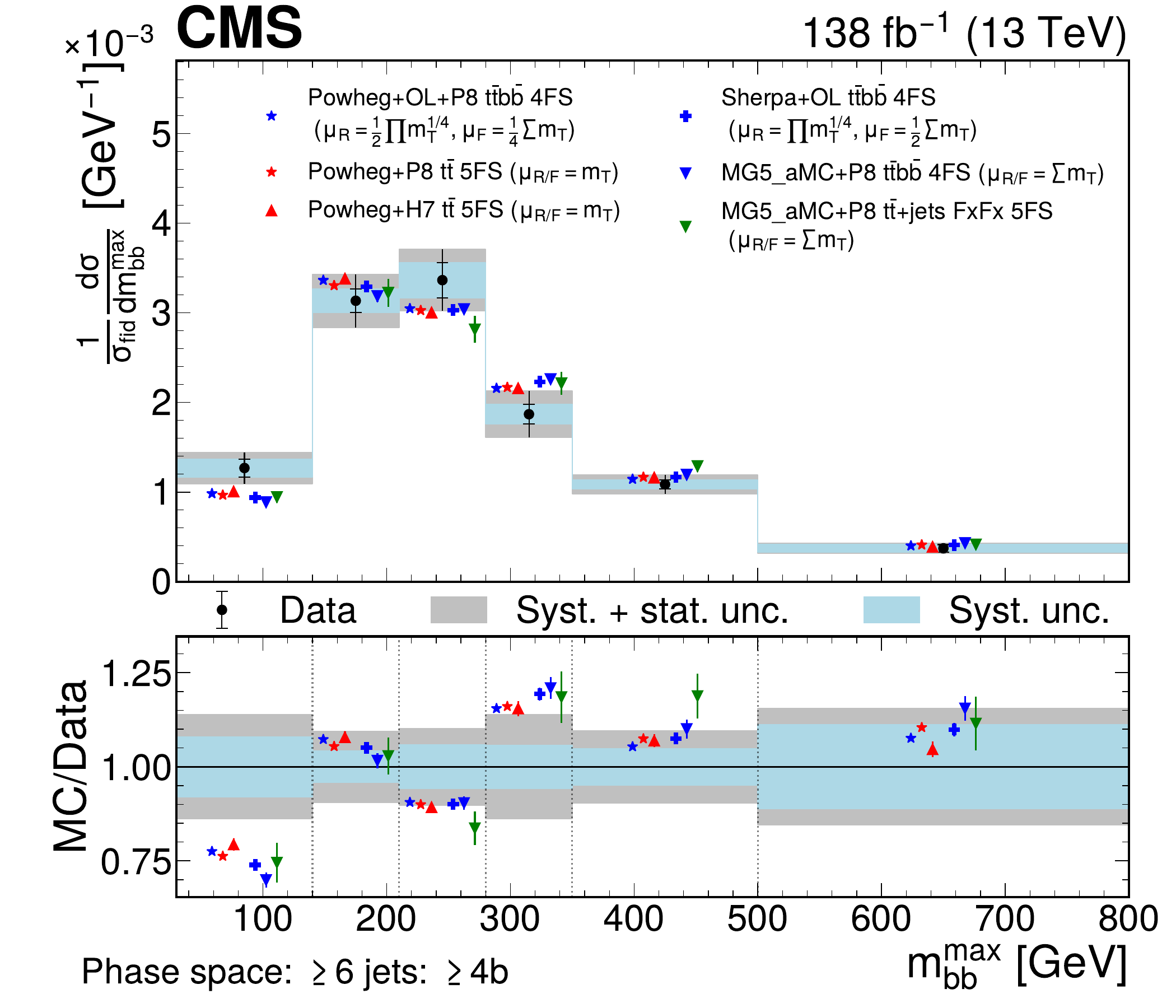} \\
\includegraphics[width=0.5\textwidth]{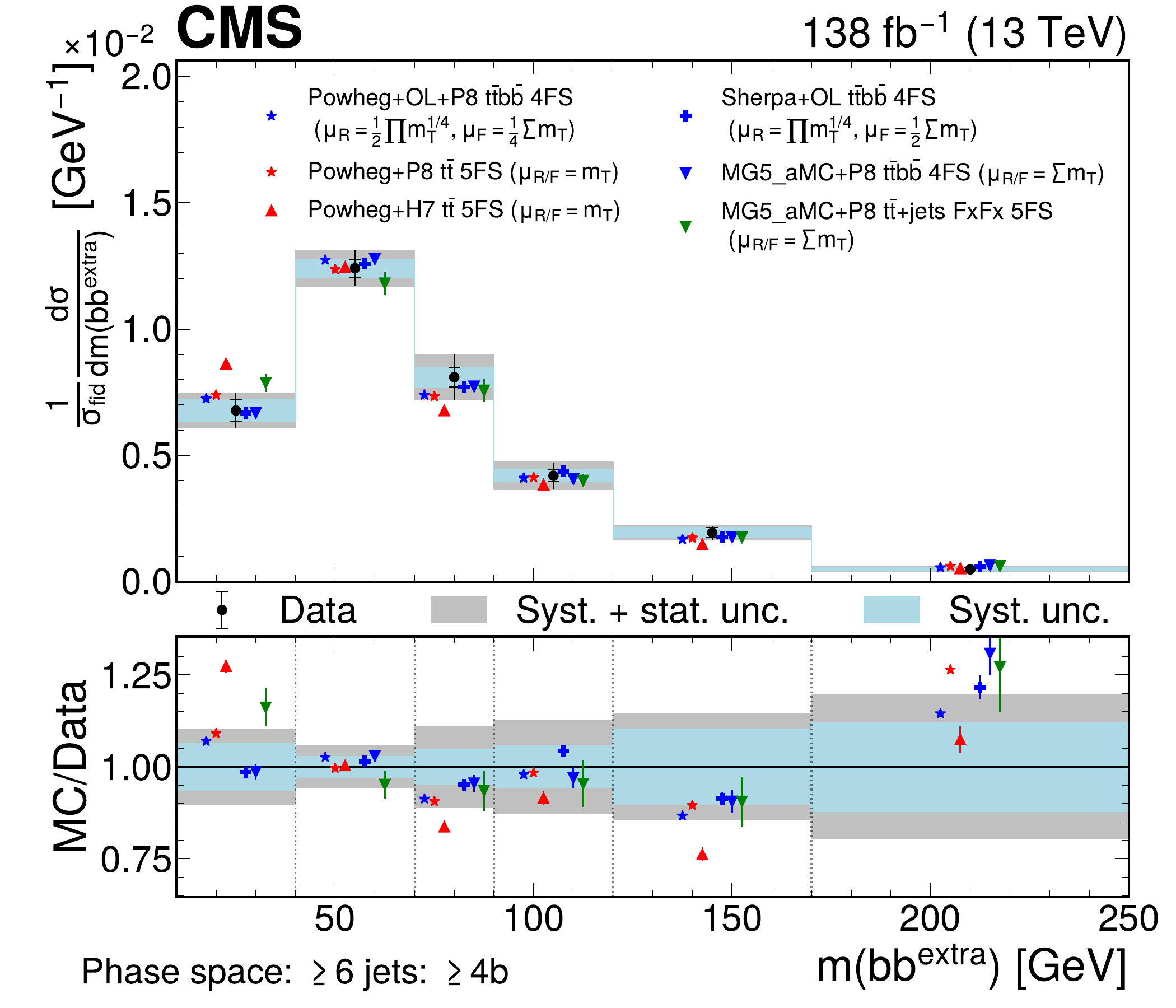}%
\includegraphics[width=0.5\textwidth]{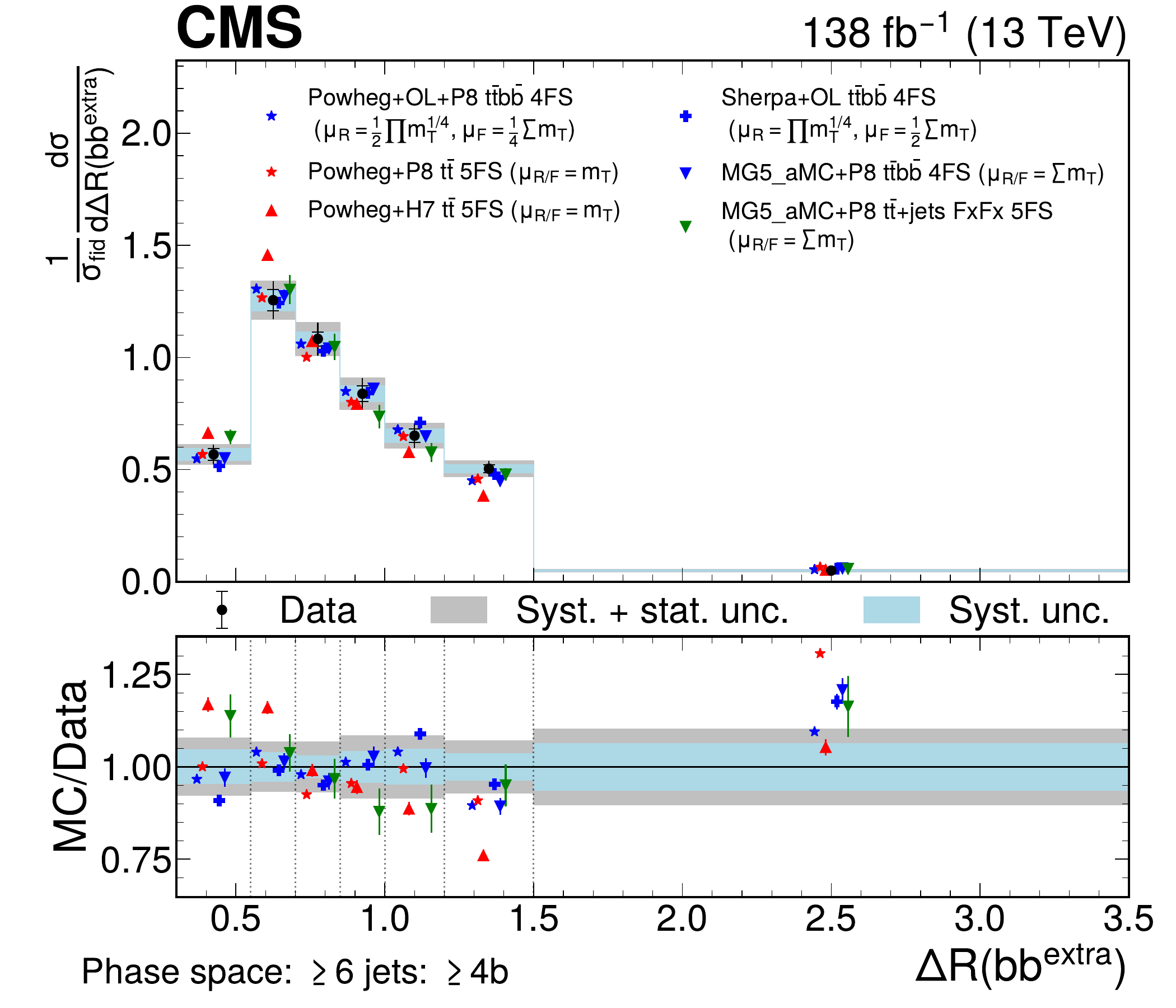} \\
\includegraphics[width=0.5\textwidth]{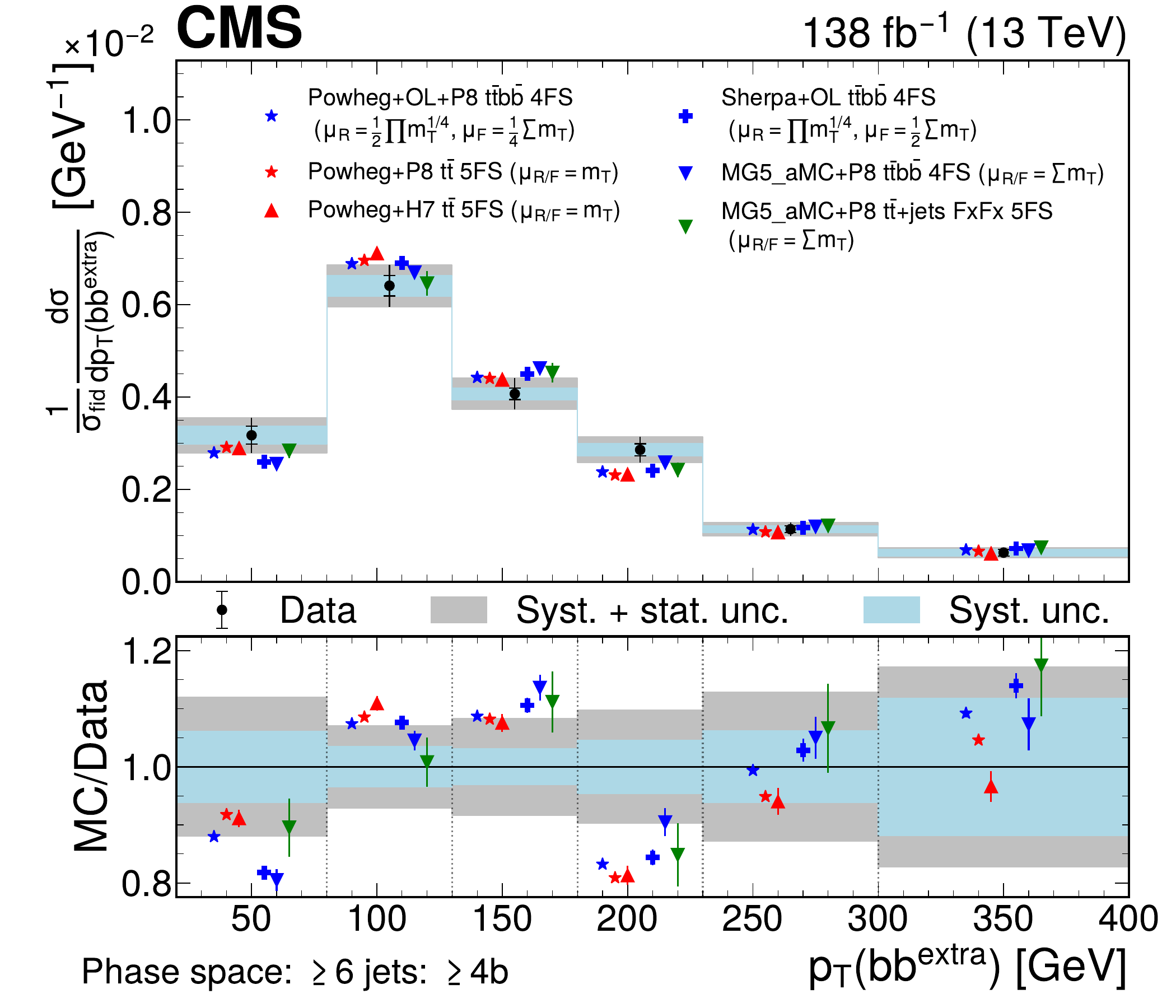}%
\includegraphics[width=0.5\textwidth]{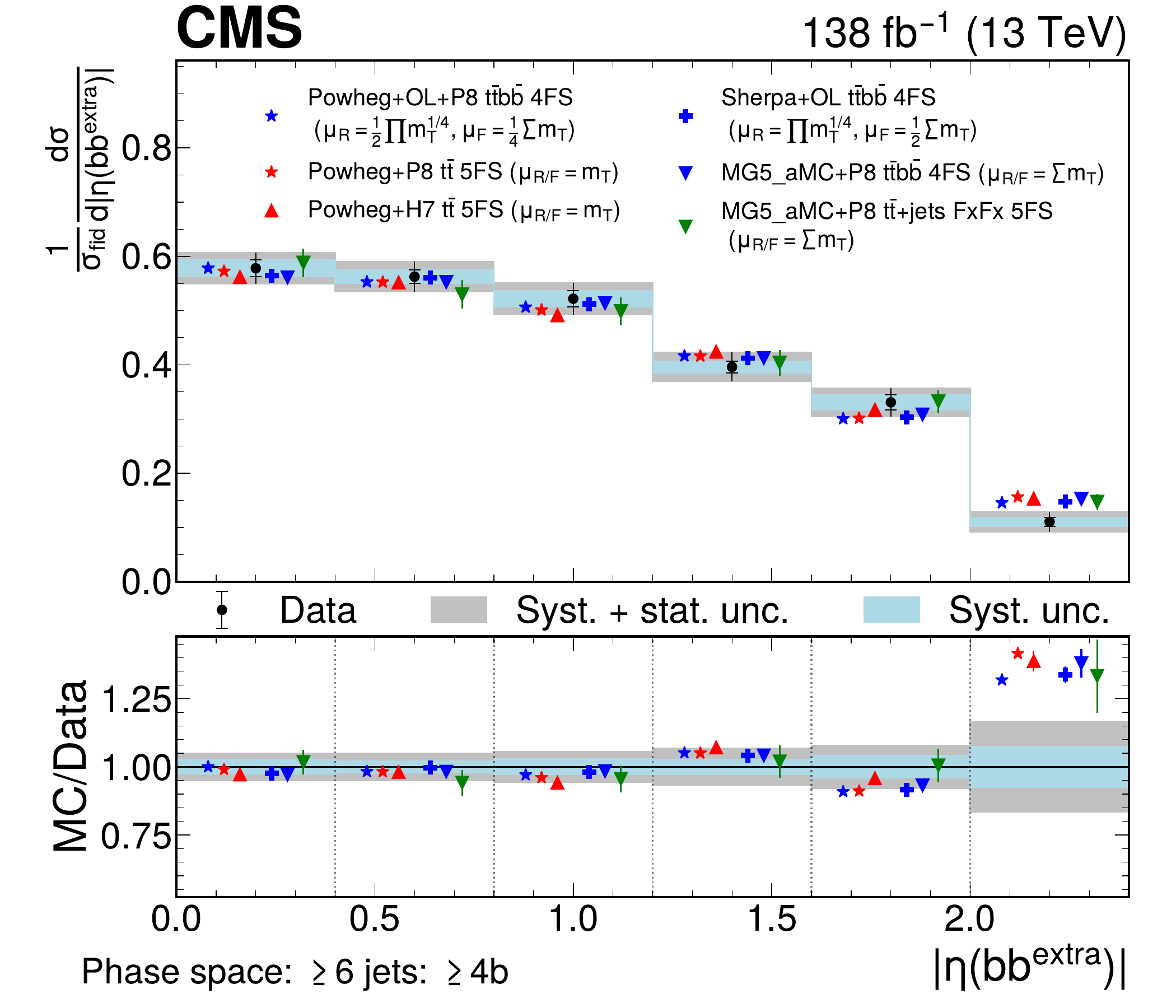}
\caption{%
    Predicted and observed normalized differential cross sections in the \sjfbLONG fiducial phase space, for the average \deltaR of all possible \bb pairs (upper left), the largest invariant mass of any \bb pair (upper right), the invariant mass (middle left), \deltaR (middle right), \pt (lower left), and \abseta (lower right) of the extra \PQb-jet pair.
    The data are represented by points, with inner (outer) vertical bars indicating the systematic (total) uncertainties, also represented as blue (grey) bands.
    Cross section predictions obtained at the particle level from different simulation approaches are shown, including their statistical uncertainties, as coloured symbols.
    For \mbb and \pt, the last bins contain the overflow.
}
\label{fig:results:6j4b:2}
\end{figure}

\begin{figure}[!p]
\centering
\includegraphics[width=0.5\textwidth]{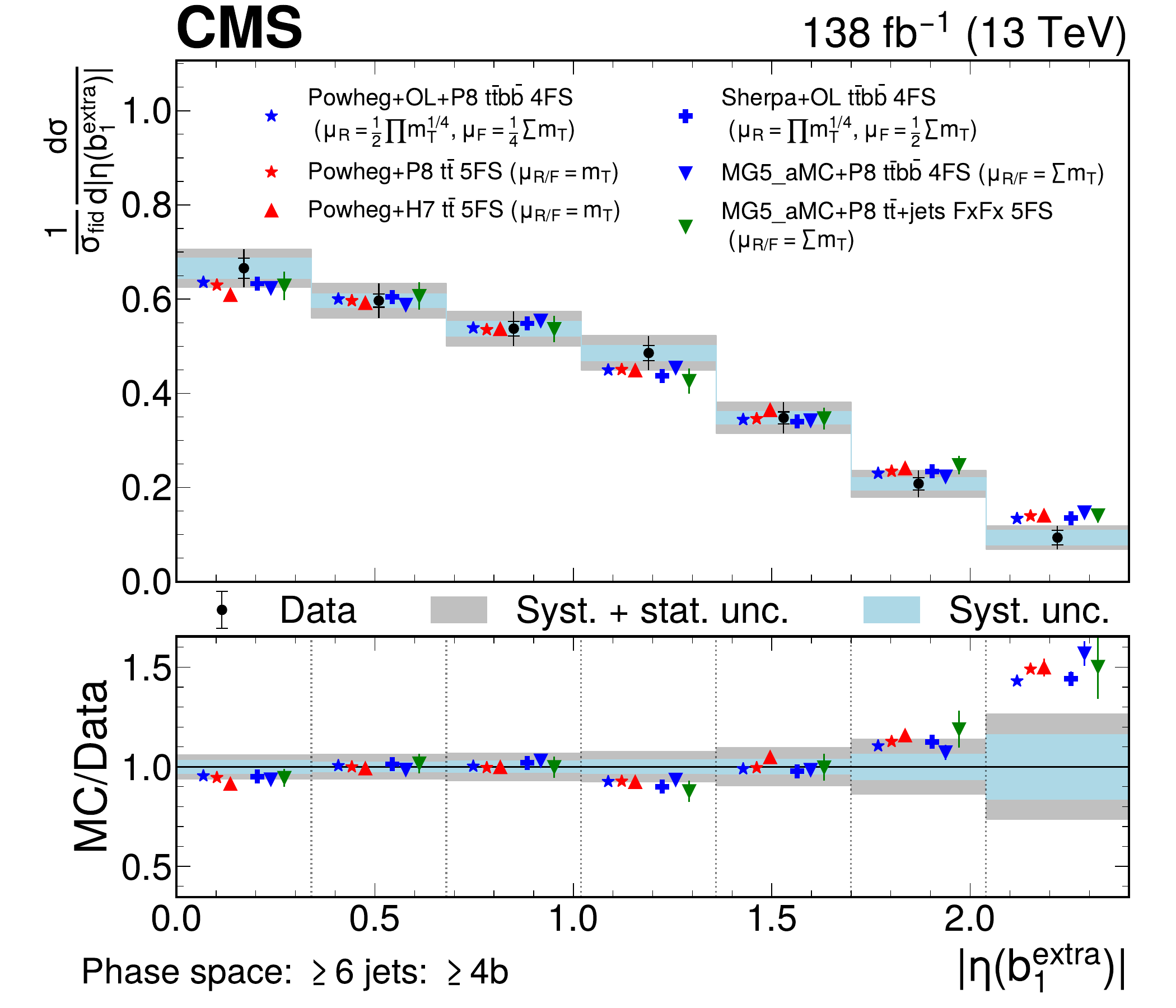}%
\includegraphics[width=0.5\textwidth]{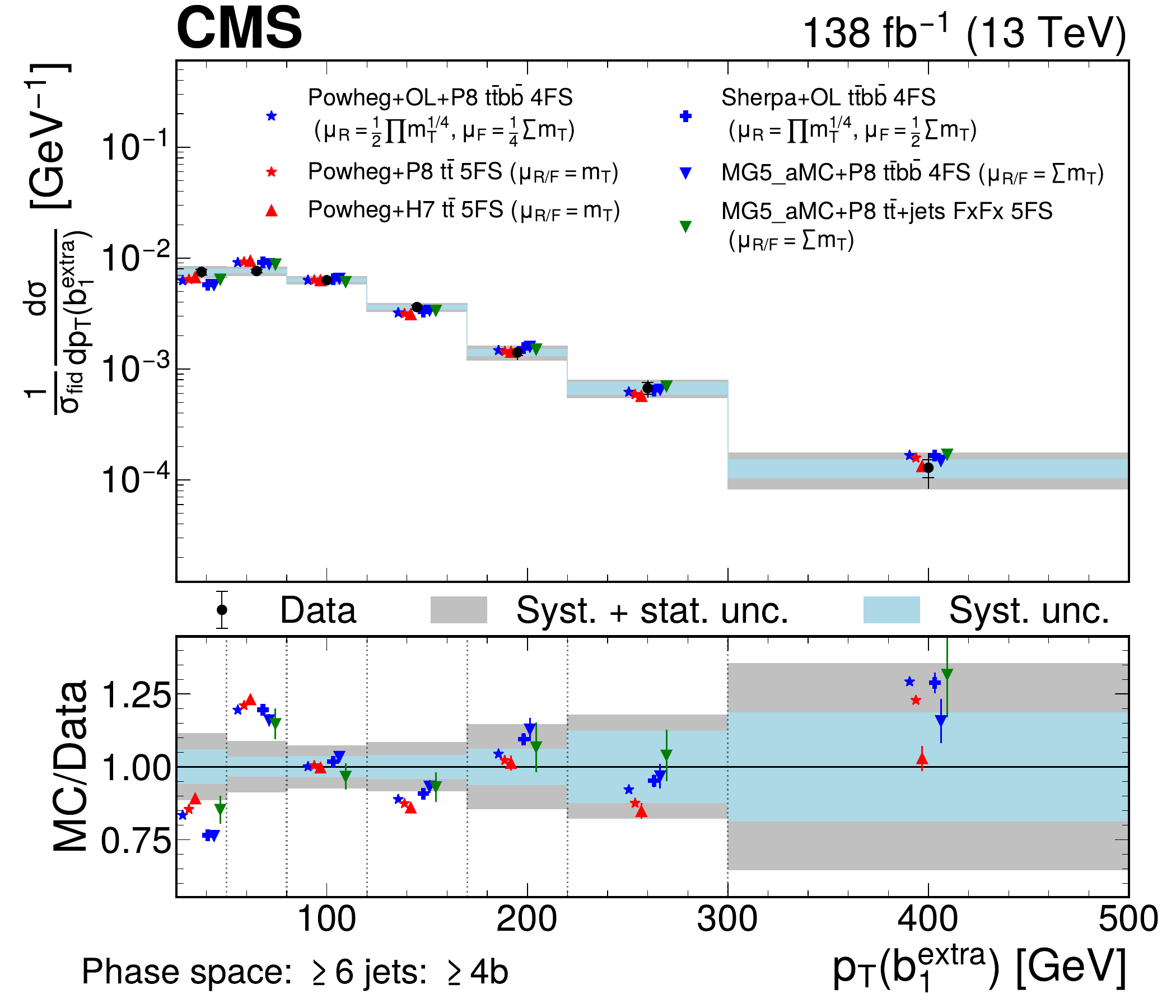} \\
\includegraphics[width=0.5\textwidth]{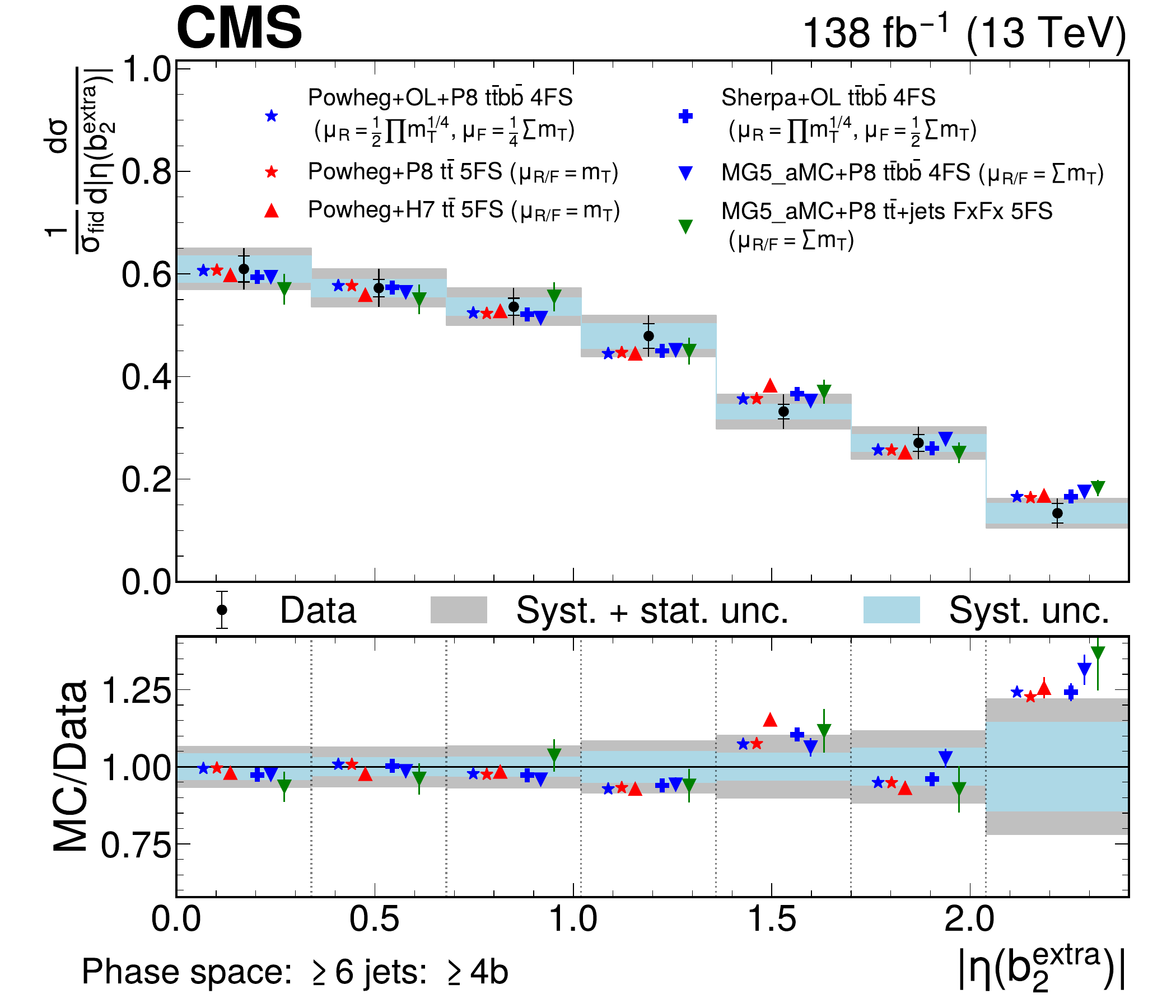}%
\includegraphics[width=0.5\textwidth]{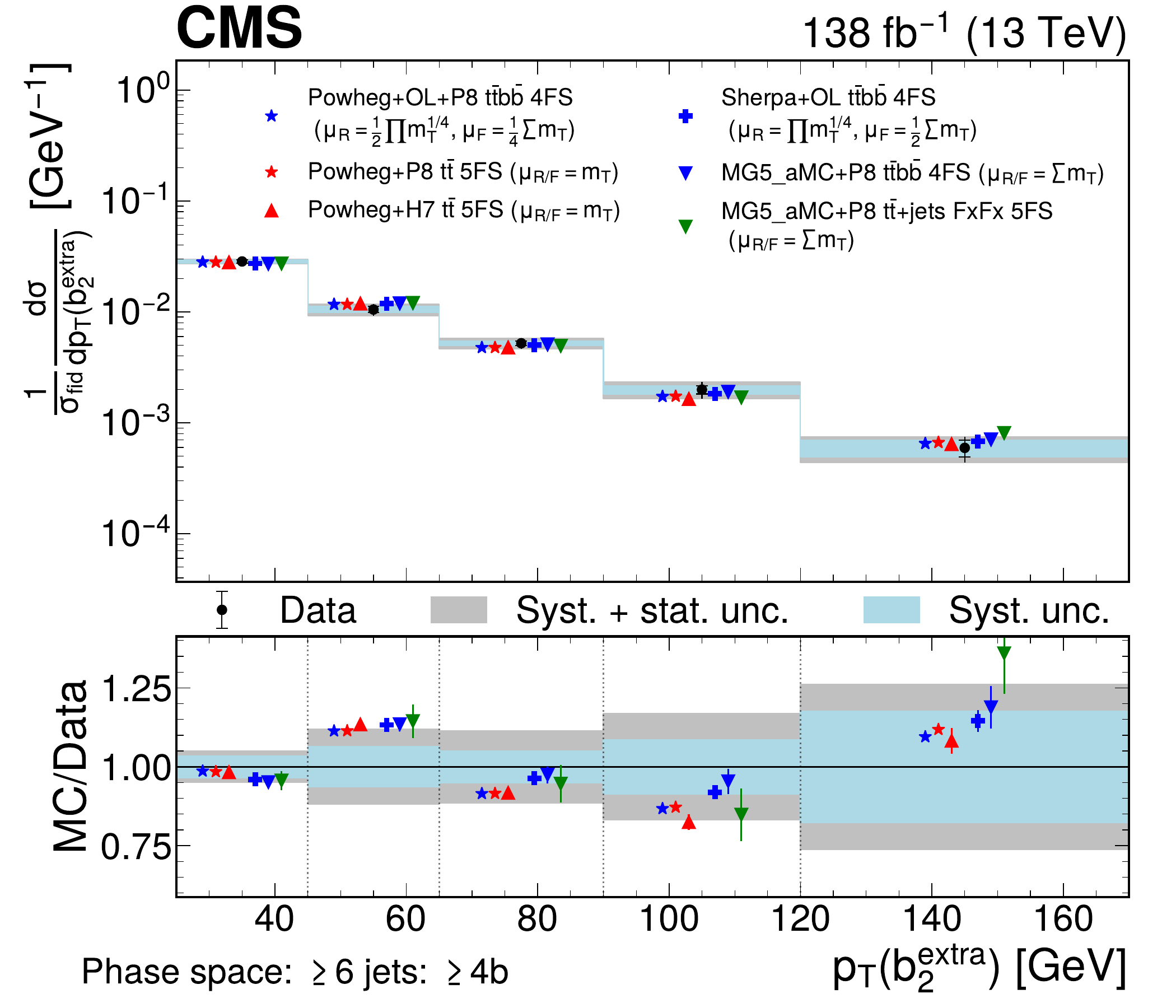} \\
\includegraphics[width=0.5\textwidth]{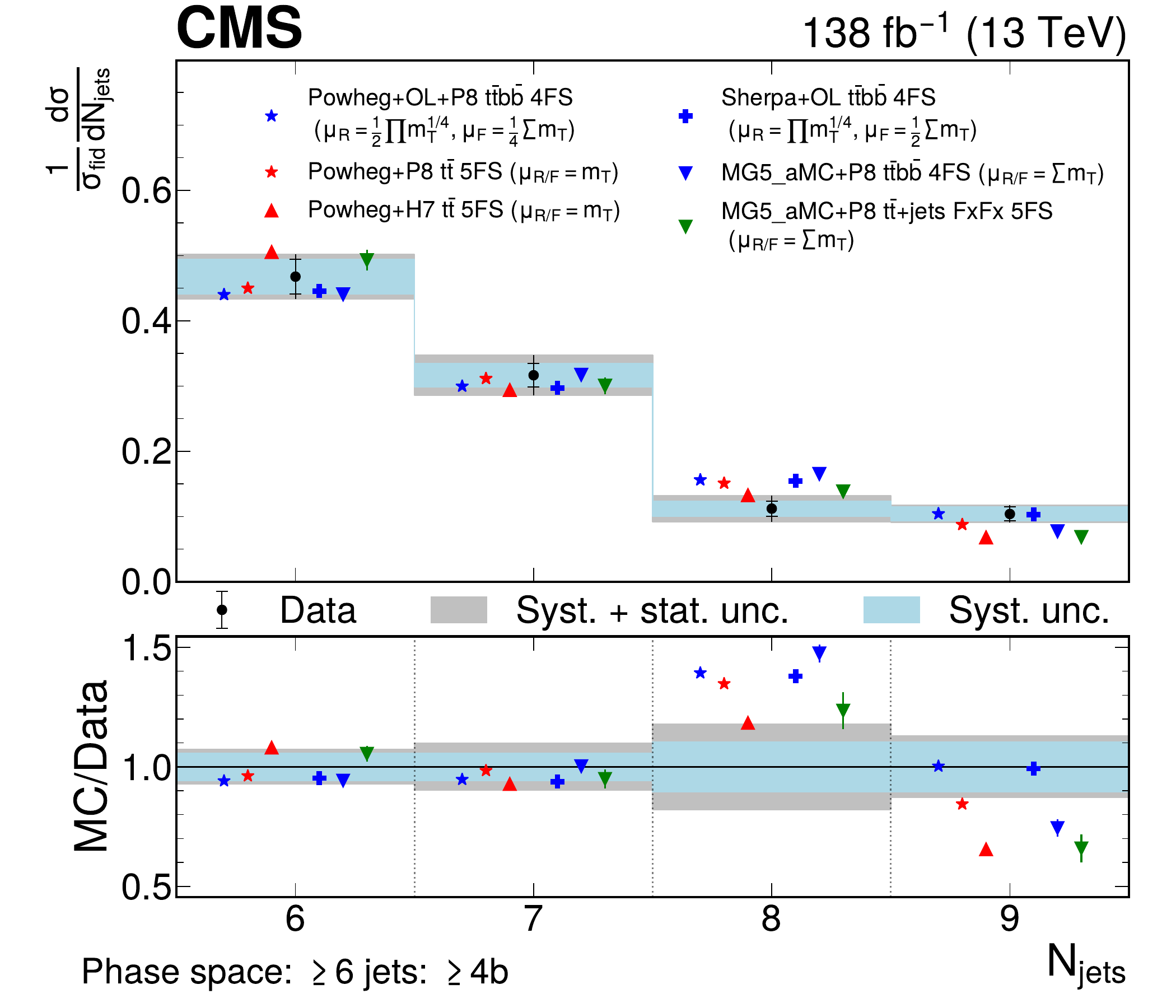}%
\makebox[0.5\textwidth]{}
\caption{%
    Predicted and observed normalized differential cross sections in the \sjfbLONG fiducial phase space, for the \abseta (upper left) and \pt (upper right) of the first extra \PQb jet, the \abseta (middle left) and \pt (middle right) of the second extra \PQb jet, and the inclusive jet multiplicity (lower left).
    The data are represented by points, with inner (outer) vertical bars indicating the systematic (total) uncertainties, also represented as blue (grey) bands.
    Cross section predictions obtained at the particle level from different simulation approaches are shown, including their statistical uncertainties, as coloured symbols.
    For \Nj and \pt, the last bins contain the overflow.
}
\label{fig:results:6j4b:3}
\end{figure}

\begin{figure}[!p]
\centering
\includegraphics[width=0.5\textwidth]{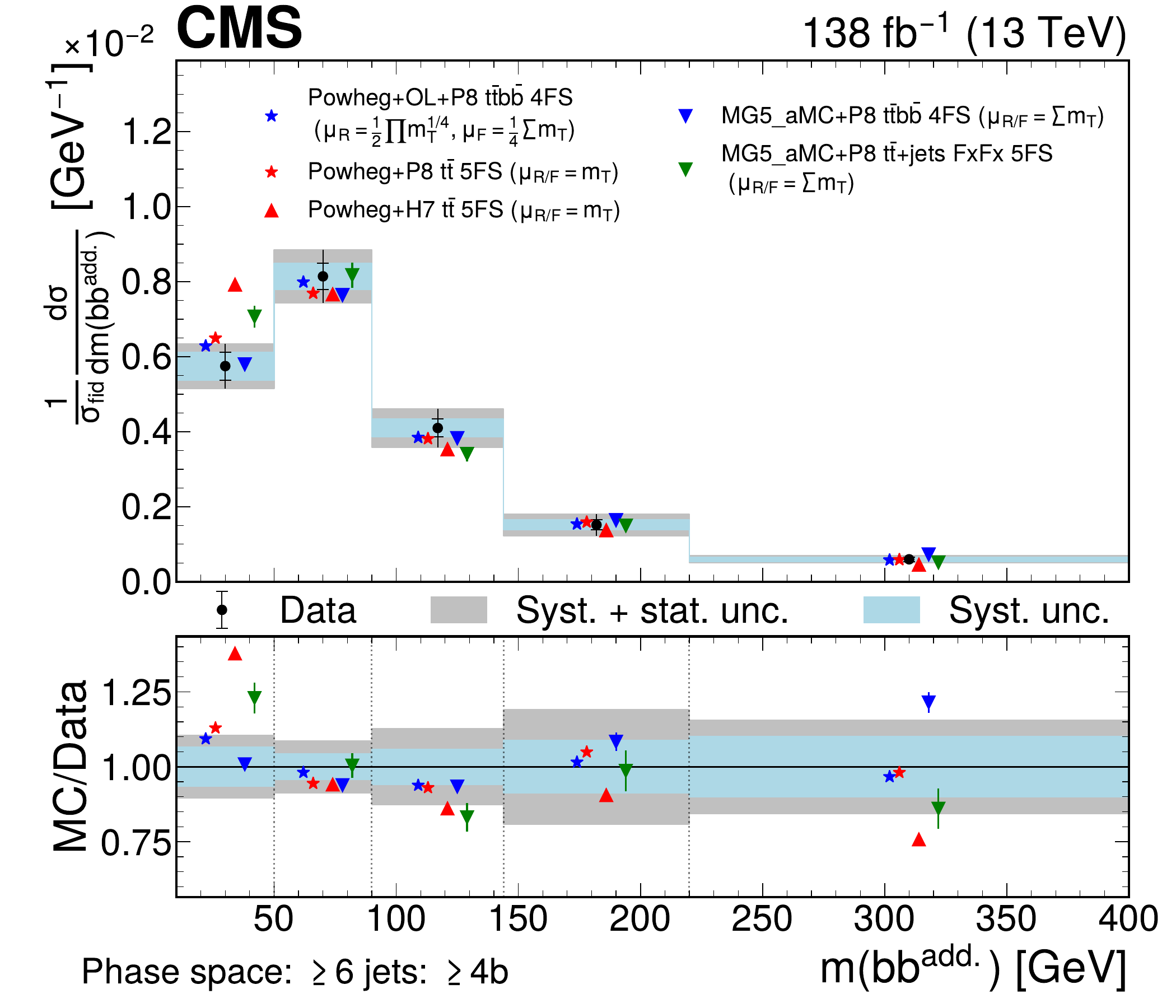}%
\includegraphics[width=0.5\textwidth]{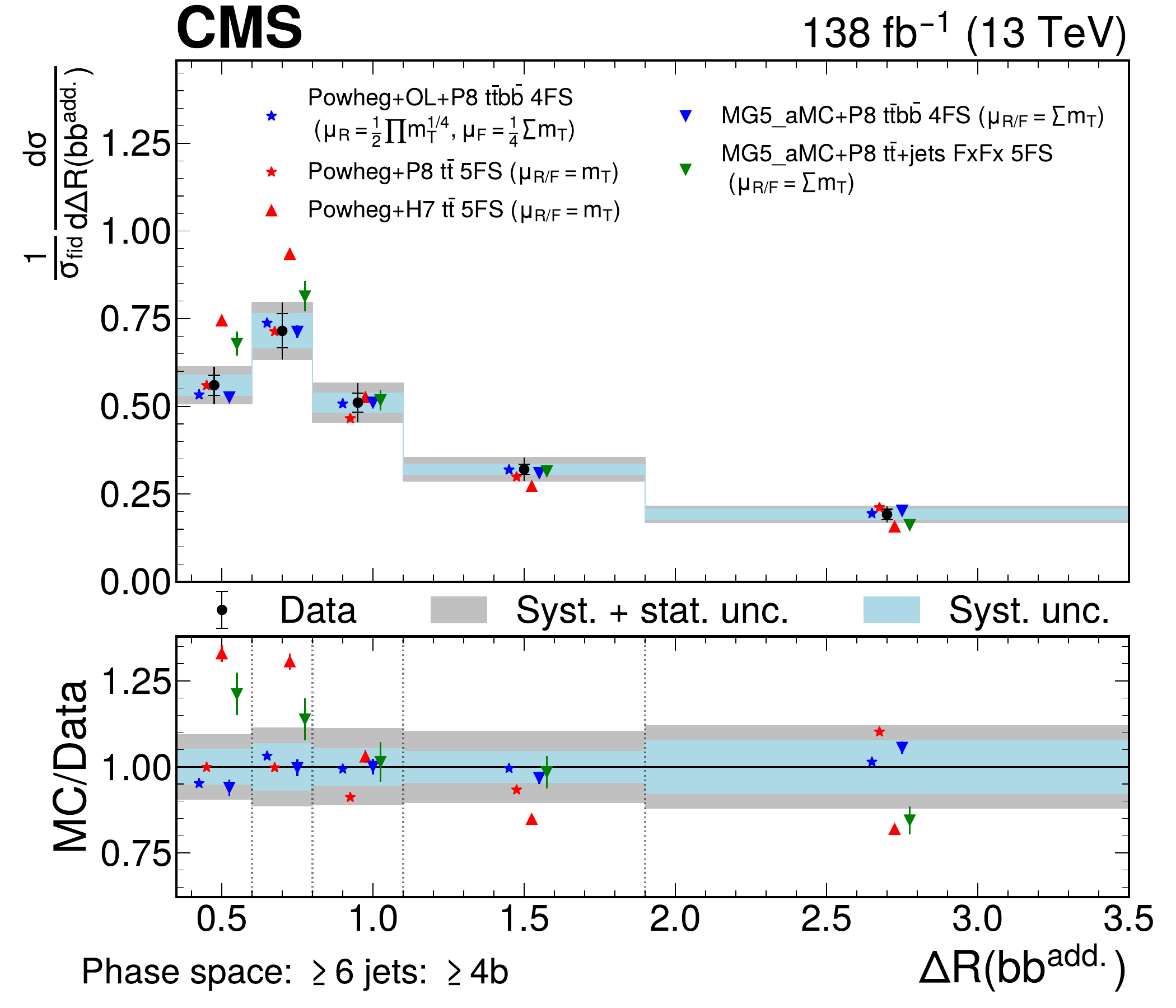} \\
\includegraphics[width=0.5\textwidth]{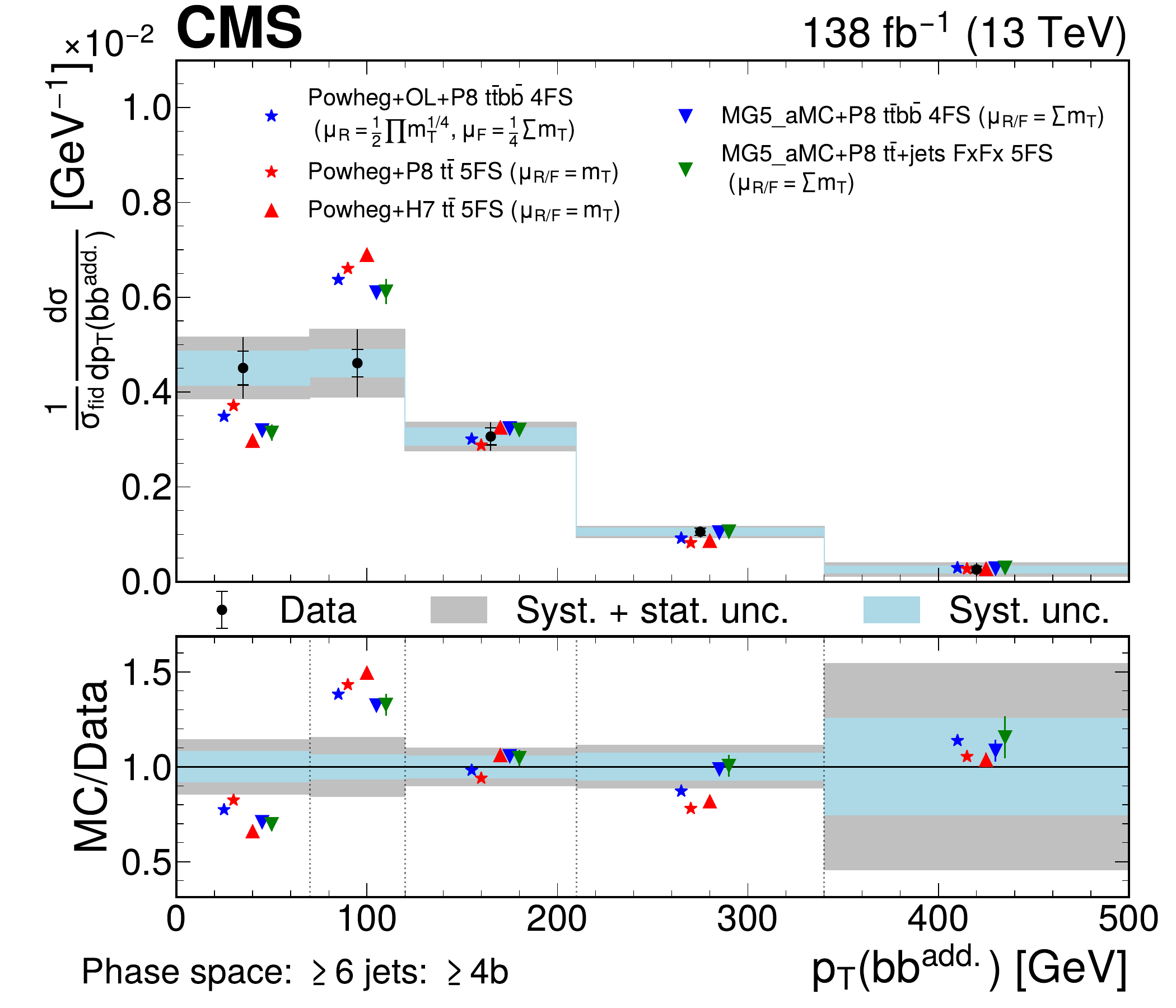}%
\includegraphics[width=0.5\textwidth]{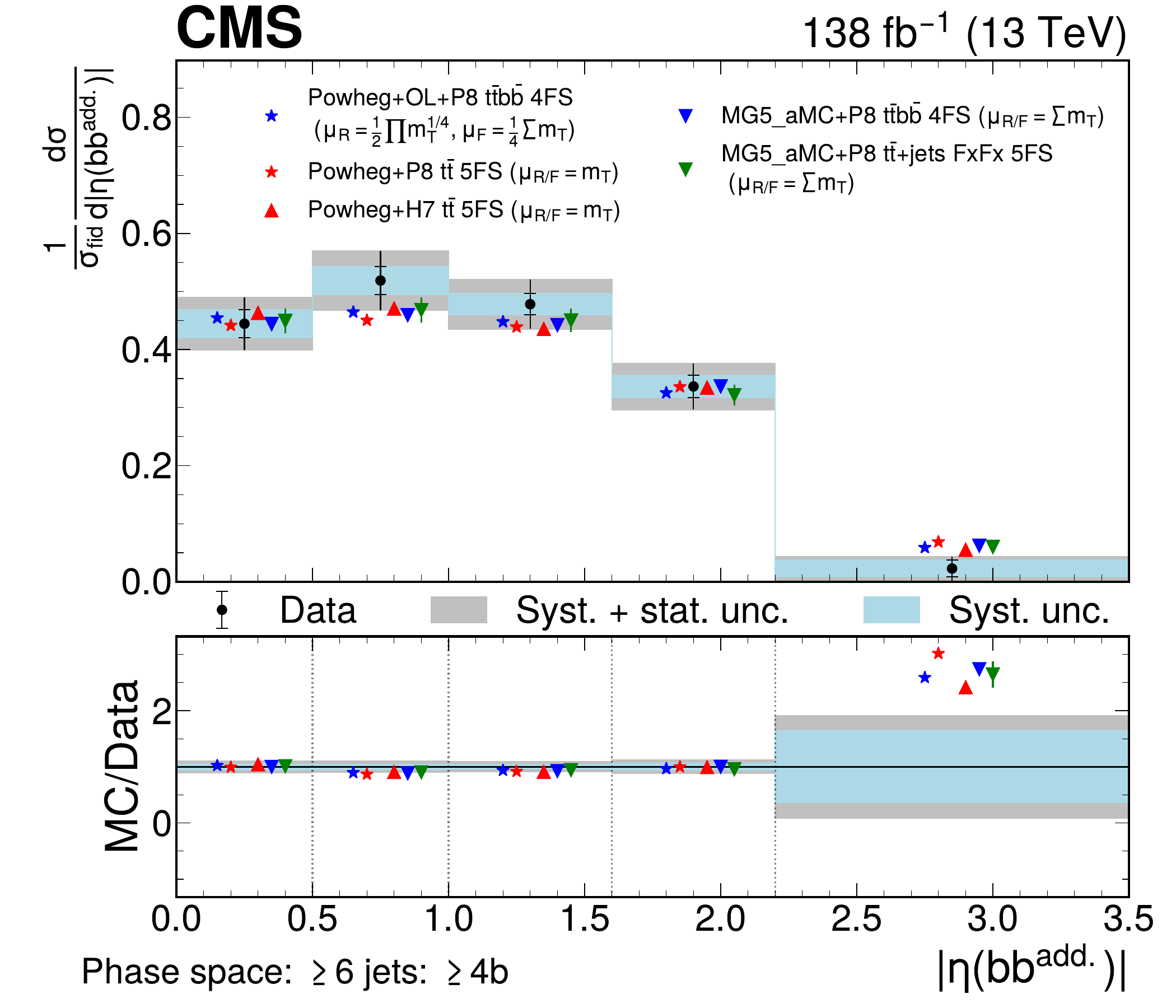}
\caption{%
    Predicted and observed normalized differential cross sections in the \sjfbLONG fiducial phase space, for the invariant mass (upper left), \deltaR (upper right), \pt (lower left), and \abseta (lower right) of the additional \PQb-jet pair not originating from decaying top quarks.
    The data are represented by points, with inner (outer) vertical bars indicating the systematic (total) uncertainties, also represented as blue (grey) bands.
    Cross section predictions obtained at the particle level from different simulation approaches are shown, including their statistical uncertainties, as coloured symbols.
    For \mbb and \pt, the last bins contain the overflow.
}
\label{fig:results:dnn:1}
\end{figure}

\begin{figure}[!p]
\centering
\includegraphics[width=0.5\textwidth]{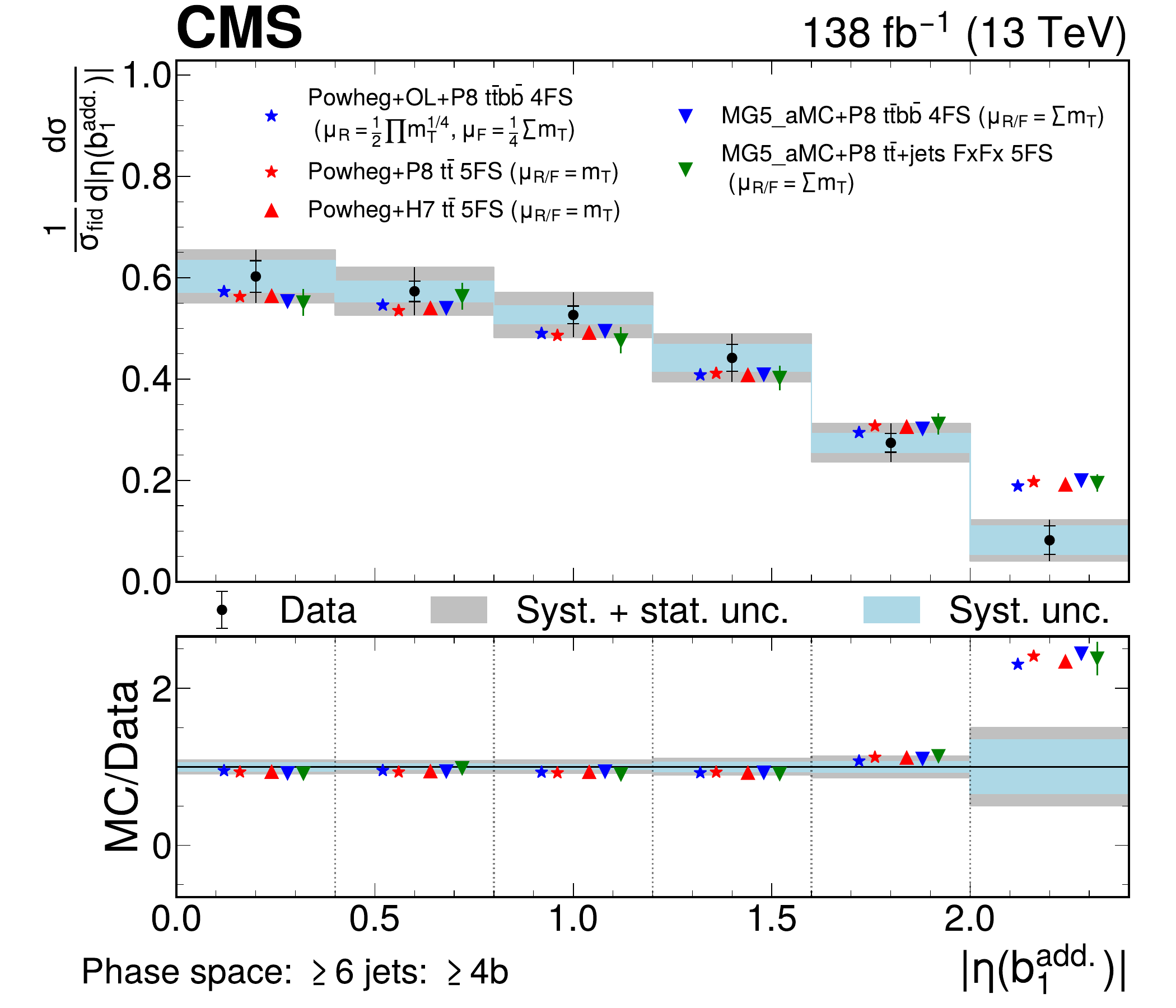}%
\includegraphics[width=0.5\textwidth]{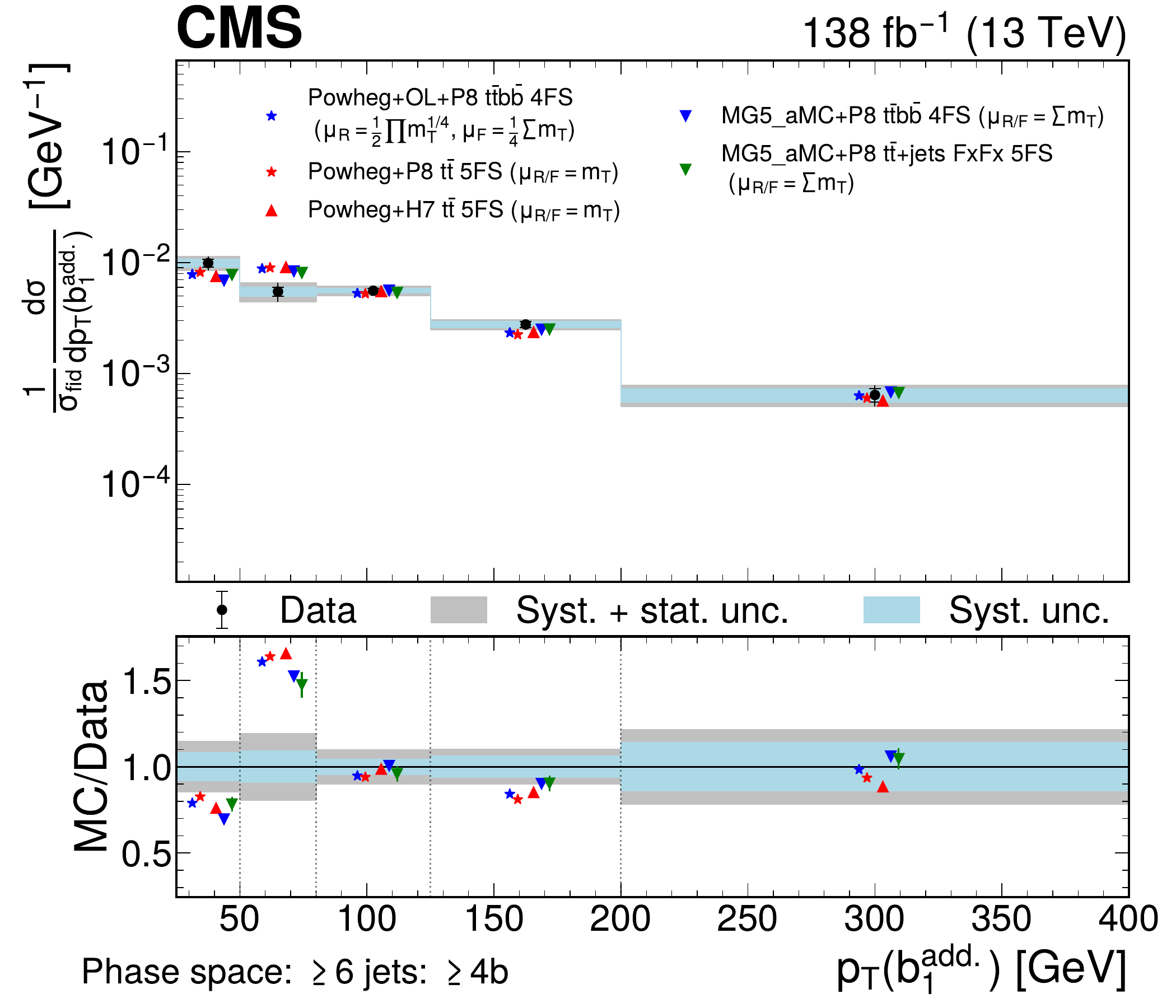} \\
\includegraphics[width=0.5\textwidth]{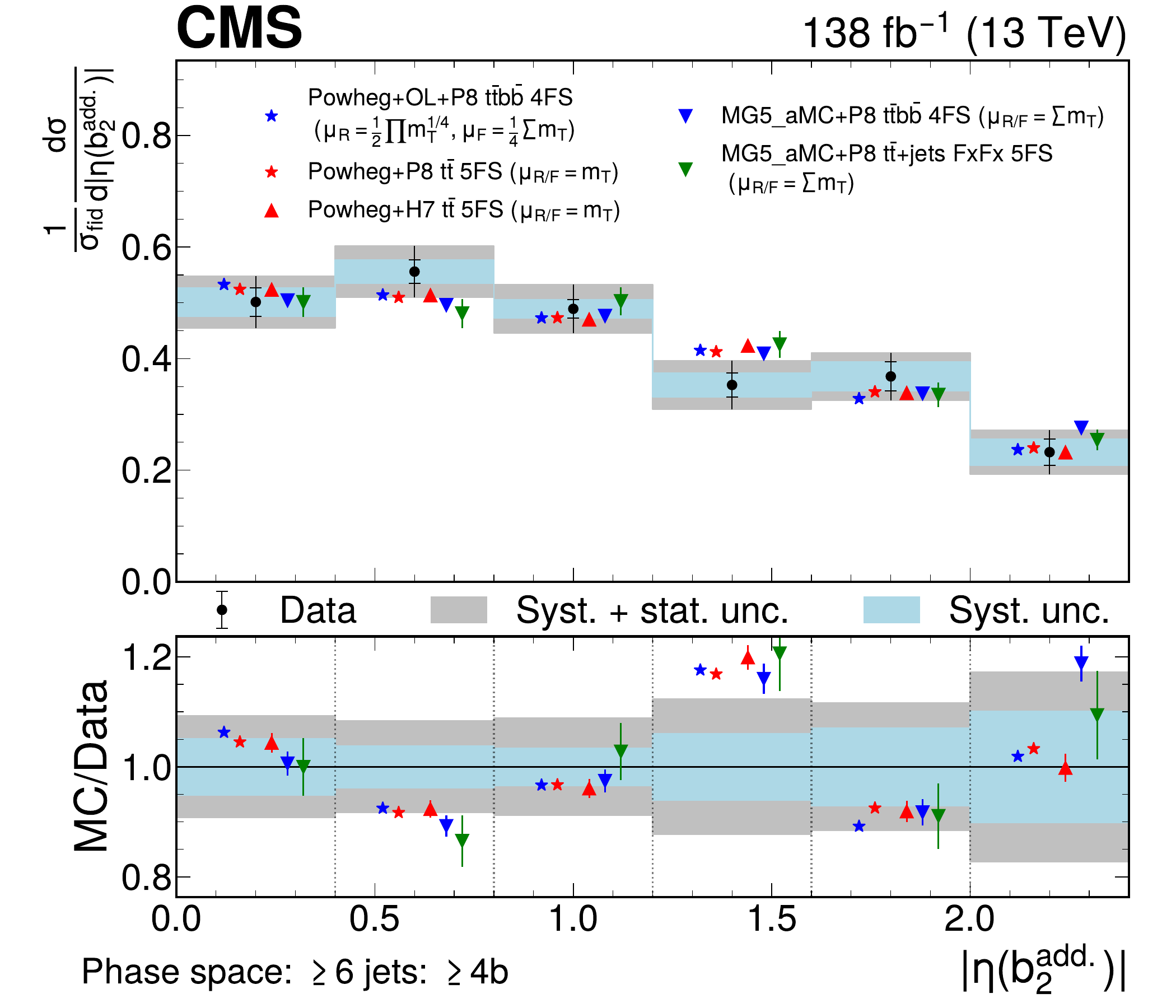}%
\includegraphics[width=0.5\textwidth]{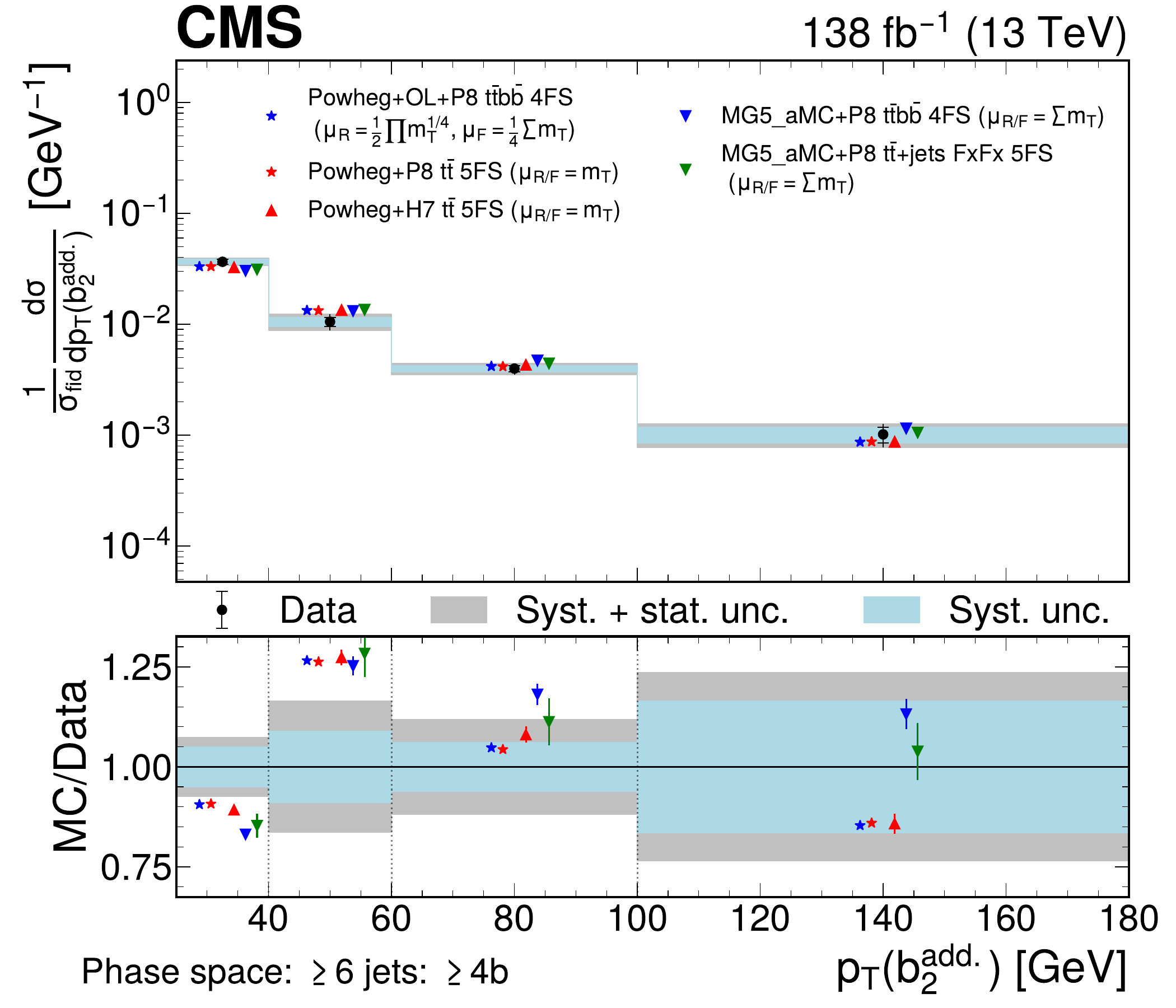}
\caption{%
    Predicted and observed normalized differential cross sections in the \sjfbLONG fiducial phase space, for the \abseta (left) and \pt (right) of the first (upper row) and second (lower row) additional \PQb of the \PQb-jet pair not originating from decaying top quarks.
    The data are represented by points, with inner (outer) vertical bars indicating the systematic (total) uncertainties, also represented as blue (grey) bands.
    Cross section predictions obtained at the particle level from different simulation approaches are shown, including their statistical uncertainties, as coloured symbols.
    For \pt, the last bins contain the overflow.
}
\label{fig:results:dnn:2}
\end{figure}

\begin{figure}[!p]
\centering
\includegraphics[width=0.49\textwidth]{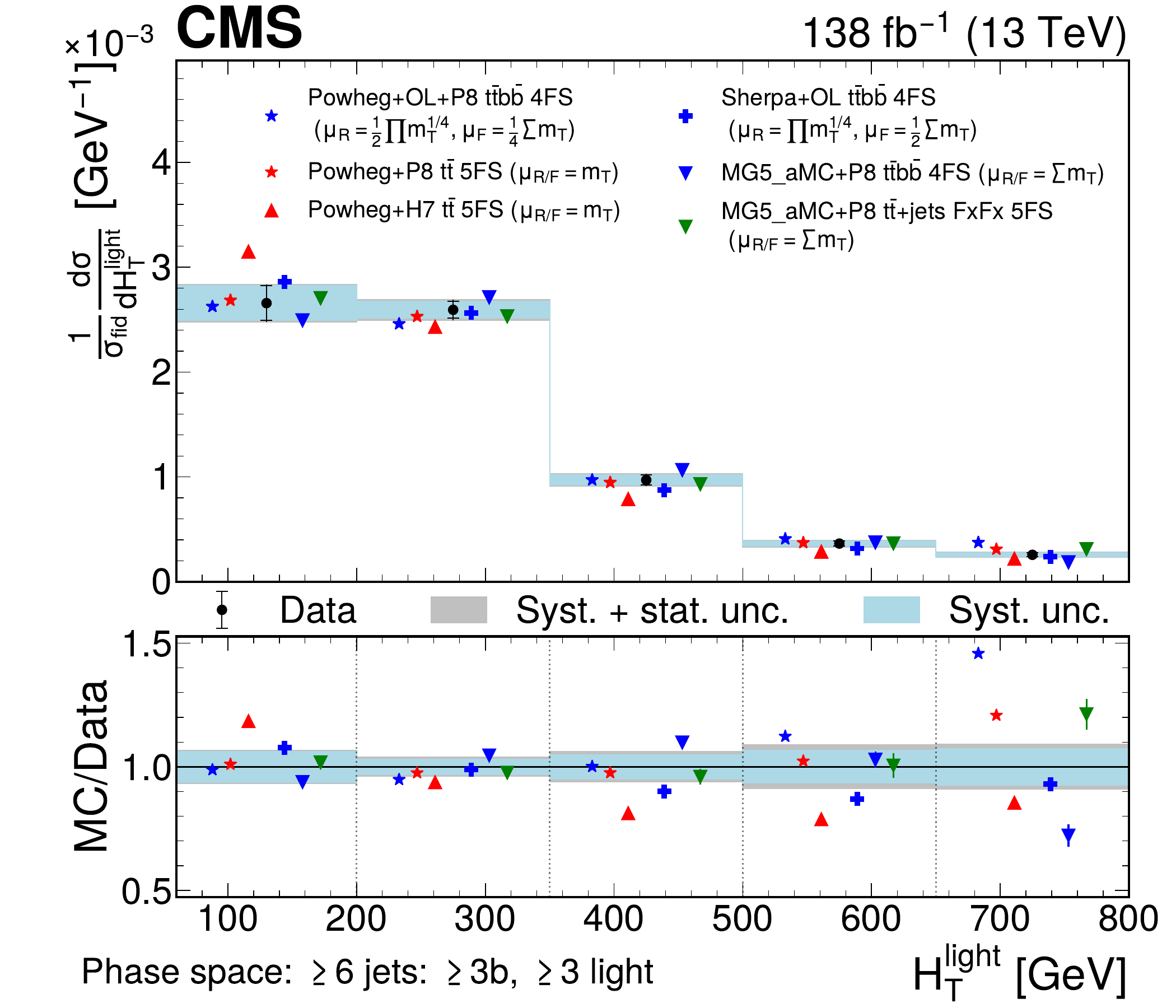}%
\includegraphics[width=0.49\textwidth]{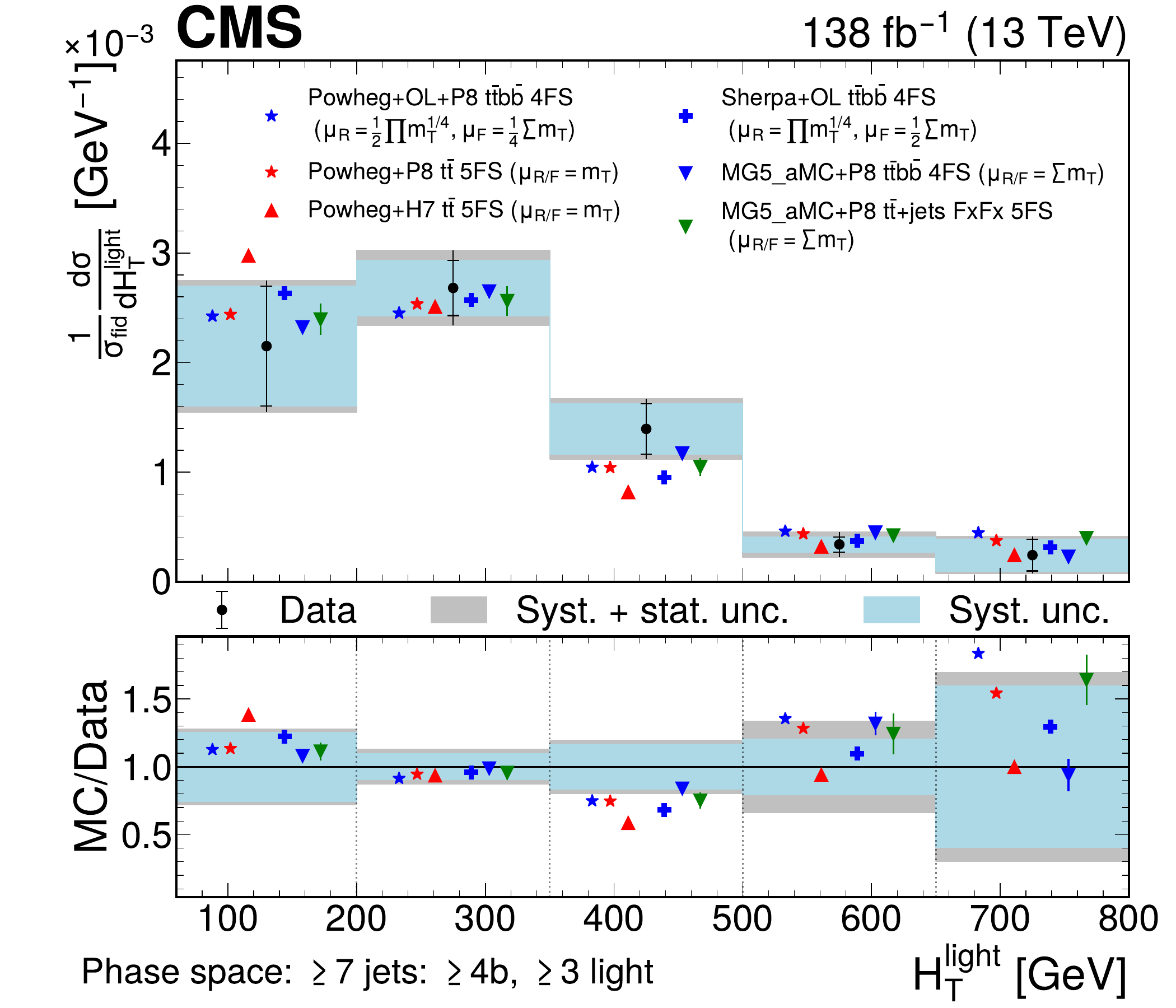} \\
\includegraphics[width=0.49\textwidth]{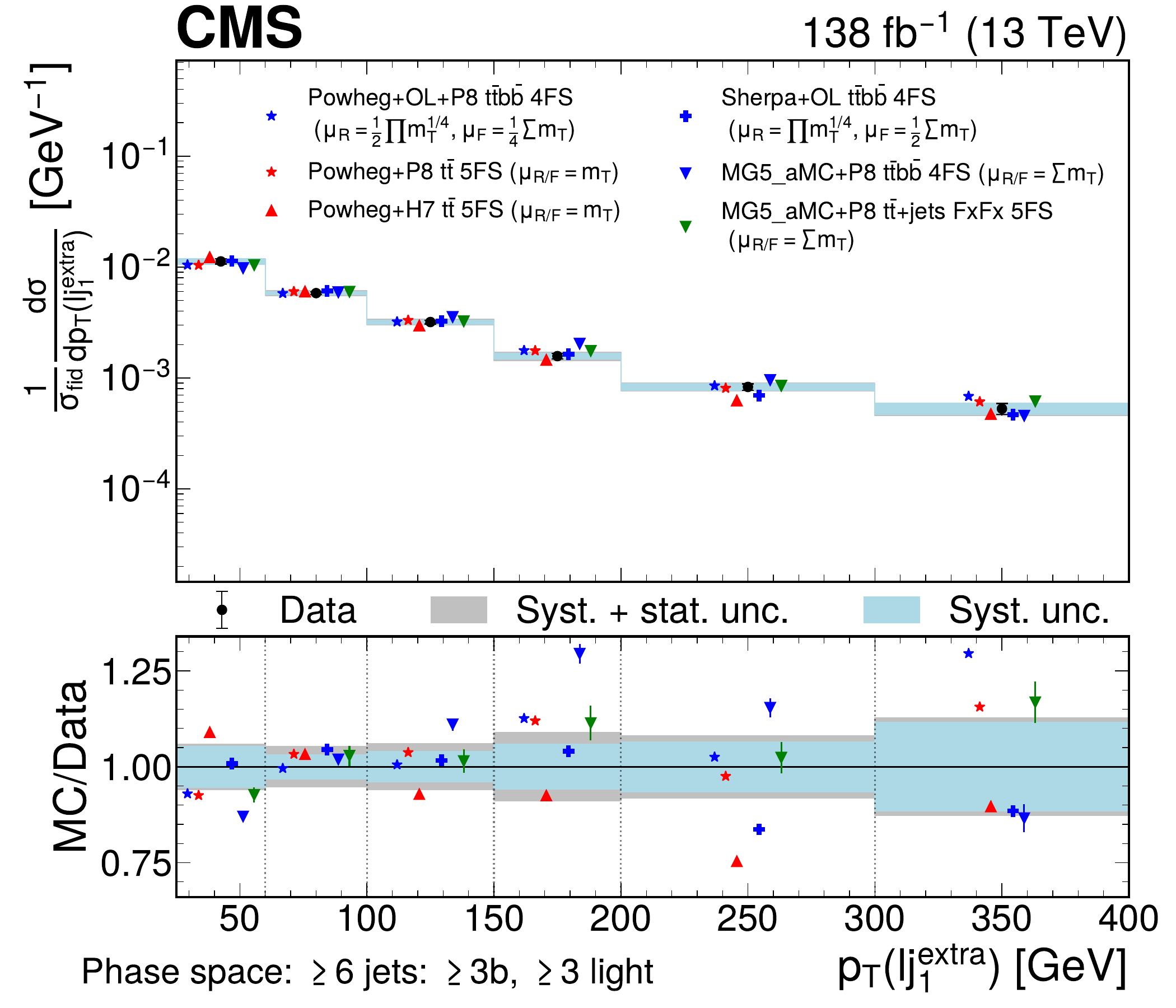}%
\includegraphics[width=0.49\textwidth]{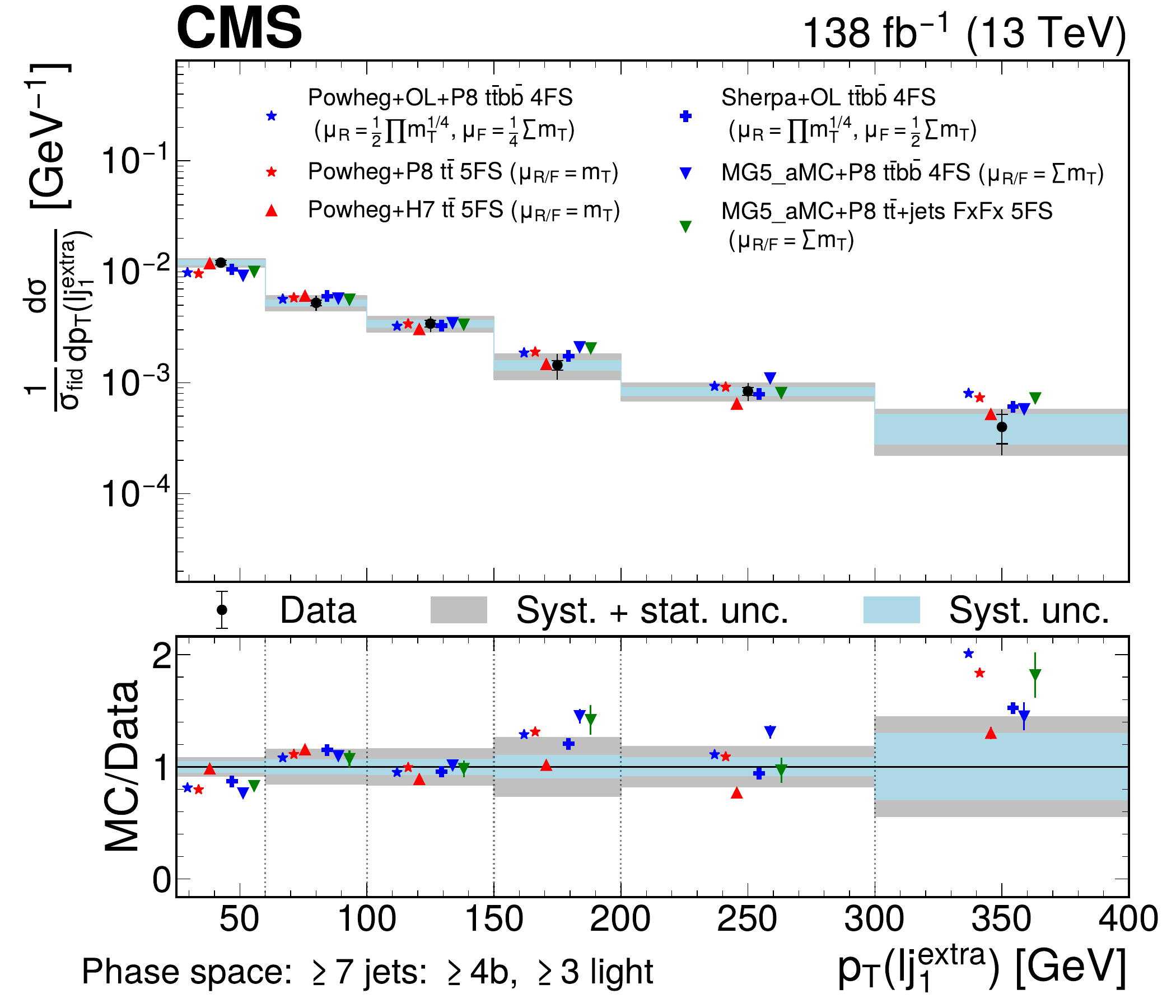} \\
\includegraphics[width=0.49\textwidth]{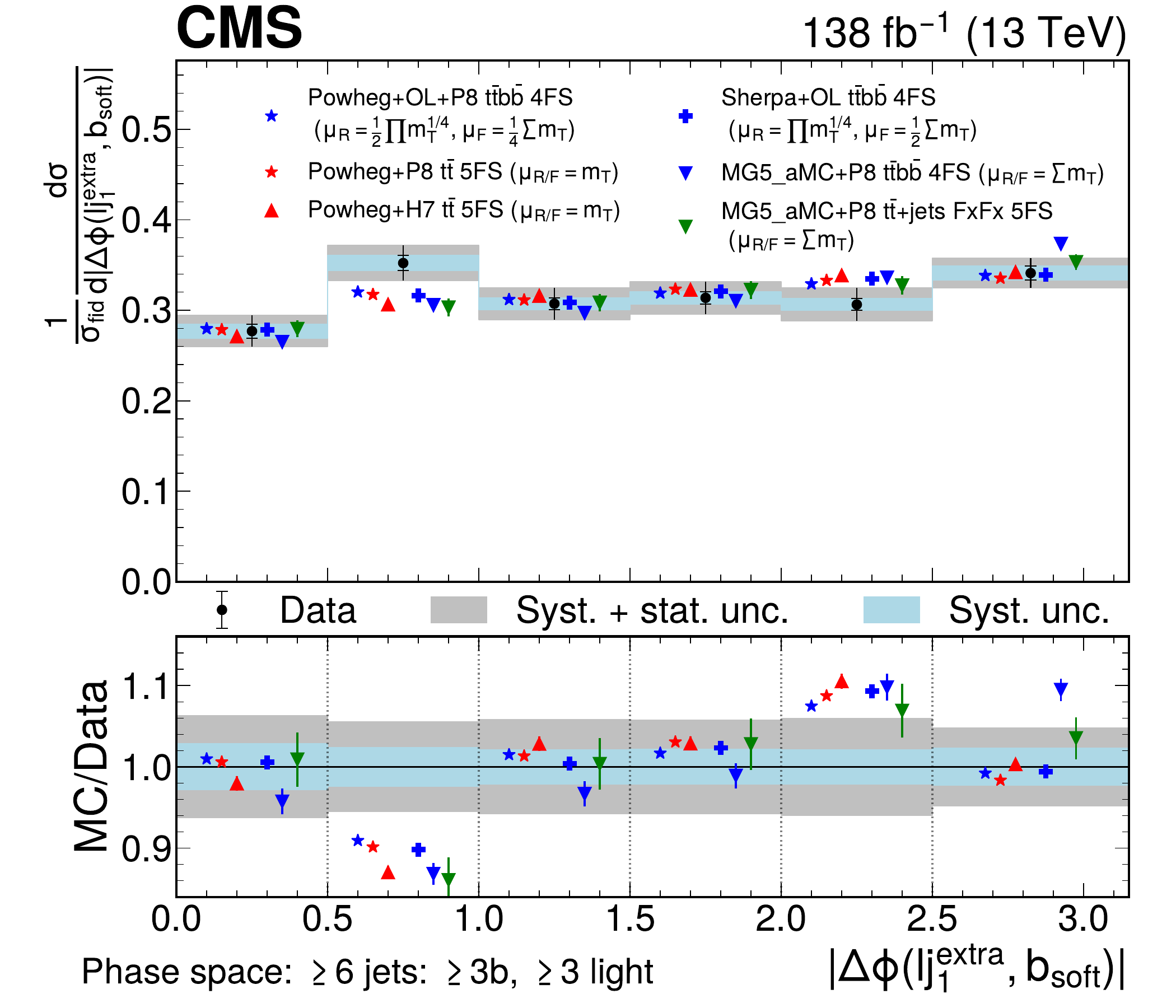}%
\includegraphics[width=0.49\textwidth]{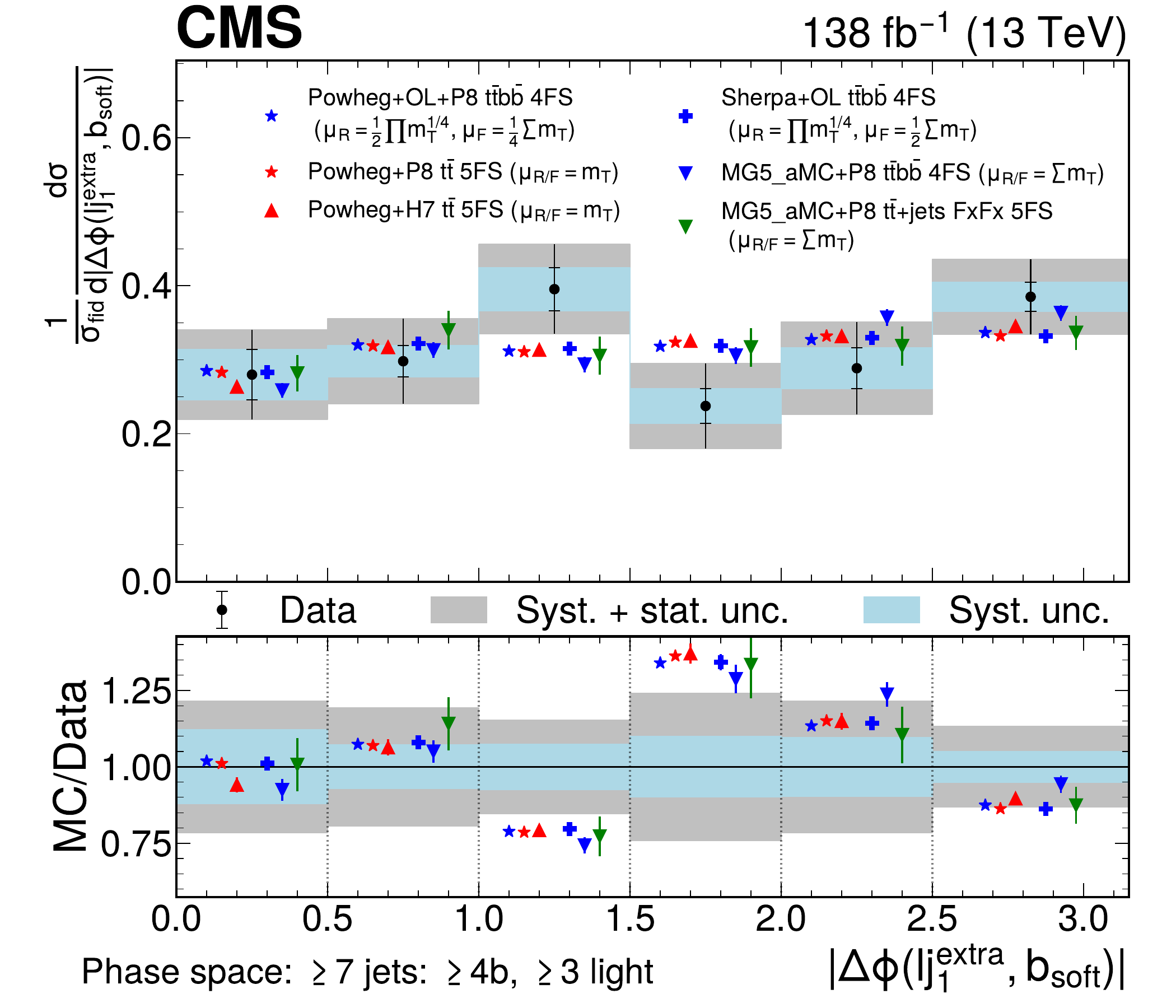}
\caption{%
    Predicted and observed normalized differential cross sections in the \sjtbtlLONG (left) and \sejfbtlLONG (right) fiducial phase space regions, for the \HT of light jets (upper row), the \pt of the extra light jet (middle row), and the \dPhi between the extra light jet and the softest \PQb jet (lower row).
    The data are represented by points, with inner (outer) vertical bars indicating the systematic (total) uncertainties, also represented as blue (grey) bands.
    Cross section predictions obtained at the particle level from different simulation approaches are shown, including their statistical uncertainties, as coloured symbols with different shapes.
    For \HT and \pt, the last bins contain the overflow.
}
\label{fig:results:lj}
\end{figure}

The resulting normalized differential cross sections are shown in Fig.~\ref{fig:results:5j3b} for the observables of the \fjtb phase space, in Figs.~\ref{fig:results:6j4b:1}--\ref{fig:results:6j4b:3} for the observables of the \sjfb phase space, for the observables targeting the \PQb jets that do not originate from decaying top quarks in Figs.~\ref{fig:results:dnn:1} and~\ref{fig:results:dnn:2}, and finally in Fig.~\ref{fig:results:lj} for the observables of the \sjtbtl and \sejfbtl phase space regions.
The measurements are compared to six cross section predictions of the \ttbb process, obtained at the particle level, produced with the different combinations of event generators and PSs introduced in Section~\ref{sec:mc}.
The various predictions are shown as symbols distinguished by colour and shape.
It should be noted that predictions from \ttbbSherpa cannot be compared to the observables related to the additional \PQb jets, since that generator does not provide the necessary information (parton-level top quarks) to assign \PQb jets to the decaying top quarks. The calculations using \ttbb in the matrix element do not include contributions coming from the production of \ttbar in the primary parton scattering with additional \PQb quarks coming only from MPI.

The compatibility of the predictions with the unfolded data in each of the phase space regions is quantified using \chitwo tests, which are converted to $z$ scores~\cite{Cousins:2007bmb}, which quantify the $p$-value in terms of the equivalent number of standard deviations by which each theoretical prediction differs from the mean of a normal distribution centered at the normalized cross-section measurements.
These are shown in Fig.~\ref{fig:results:zscores} and as tables in Appendix~\ref{app:zscores}.
The \chitwo test statistics are computed using the predicted and measured normalized cross sections, as well as the estimated covariance between the bins of the unfolded distributions, but do not include the inclusive cross sections.
No uncertainties in the predictions are considered in these tests.
Varying levels of agreement between the predicted and measured distributions are observed, whereby no event generator setup describes all observables well since they each result in $z\geq 2$ for several observables.

\begin{figure}[!p]
\centering
\includegraphics[width=\textwidth]{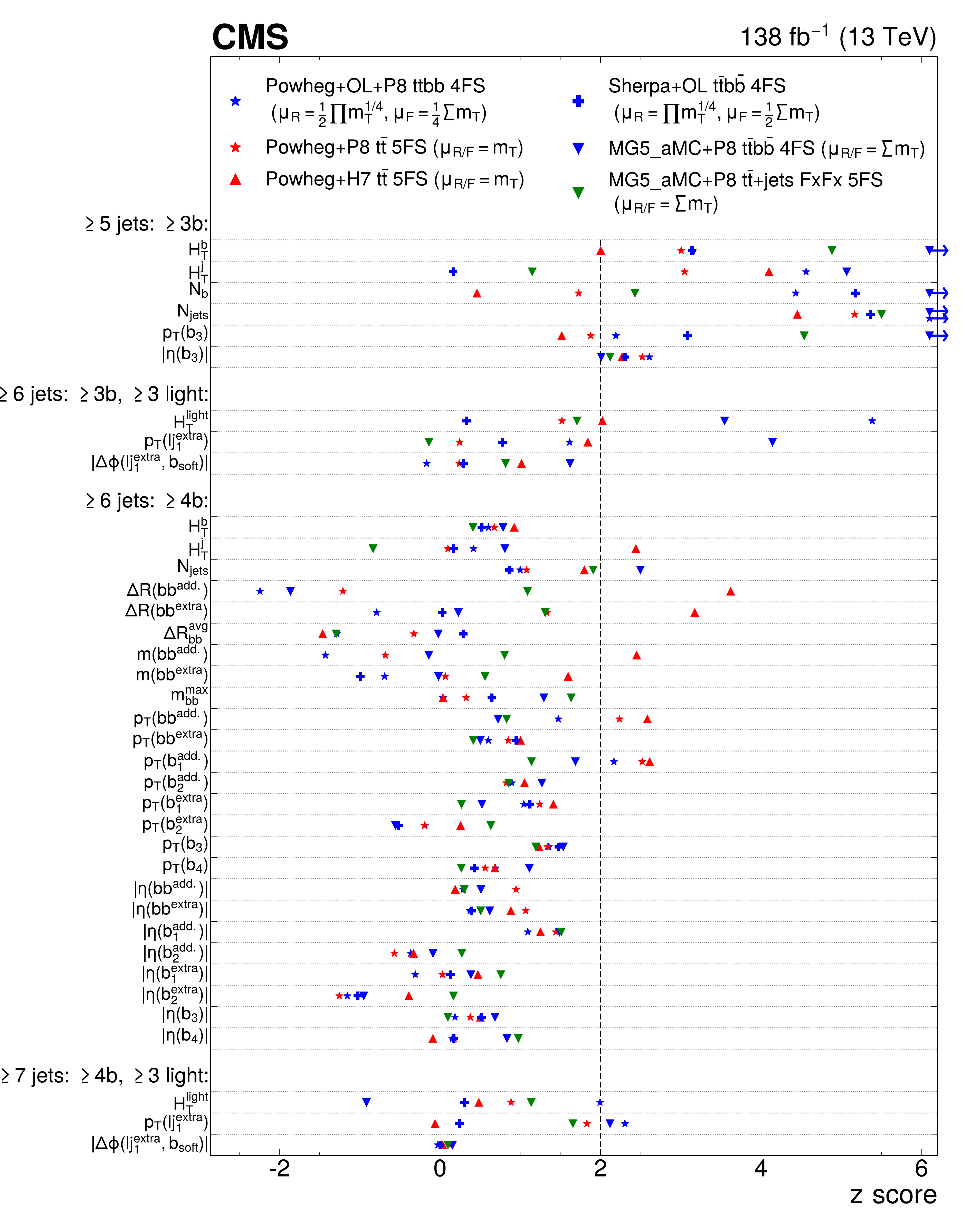}
\caption{%
    Observed $z$ score for each of the theoretical predictions, given the unfolded normalized differential cross sections and their covariances.
    A lower value indicates a better agreement between prediction and measurement.
    The dashed line at $z=2$ indicates a $p$-value of 5\%.
    Predictions for which the $z$ score exceeds the visible range of the figure are marked with arrows ($\to$).
}
\label{fig:results:zscores}
\end{figure}

In the \fjtb phase space (Fig.~\ref{fig:results:5j3b}), with a few exceptions, none of the distributions are described well by any of the considered generator setups, quantified by the $z$ scores without any consideration of the uncertainties on the predictions.
Most generator setups predict a higher number of inclusive jets than what is observed, with the exception of \ttbarPH and \ttjetsAMC showing better agreement.
The \PQb jet multiplicity distribution, which can be interpreted as a measure of the ratio between the cross sections of \ttbb and \ttb, is only well described by the inclusive \POWHEG{} \ttbar simulations matched to \PYTHIA or \HERWIG.
The NLO \ttbb simulations all predict a higher ratio of events with at least four \PQb jets over exactly three \PQb jets.
The \HTj distribution shows a significant trend towards higher values than what is observed in data for \ttbbAMC, and towards lower values for \ttbarPH, while the other generator setups better describe the distribution.
The \HTb distribution shows a trend towards lower values in data compared to most predictions, with the exception of \ttbarPH and \ttbbSherpa.
The predictions of the different simulations for \absetabN{3} are all very similar to each other and have $z$ scores around 2.
All generator setups predict somewhat larger values of the \pt of the third \PQb jet.

Observables which are not fully described by the NLO \ttbb ME, such as observables related to the radiation of light jets, are expected to have strong dependencies on the \muR and \muF scale choices when described with a NLO \ttbb ME simulation setup~\cite{Jezo:2018yaf}.
Hence, in Appendix~\ref{app:qcdscales}, the measured normalized differential cross sections of the \Nj and \HTj observables are compared with alternative \muR and \muF scale choices of the \ttbbPP simulation approach.
These comparisons show that increased \muR and \muF scales relative to the nominal scale choices (see Table~\ref{tab:generatorsettings}) tend to better describe these observables.

{\tolerance=800
The agreement between data and predictions is generally better in the \sjfb phase space (Figs.~\ref{fig:results:6j4b:1}--\ref{fig:results:6j4b:3}), at least in part due to the larger uncertainties in the measurements.
With a few exceptions, the predictions between the various generator setups are also closer to each other than in the \fjtb phase space.
\HTb and \HTj are generally modelled well, while for the \ttbarPH simulation a trend toward lower values is observed, similar to the \fjtb phase space.
However, most generator setups underpredict the threshold region at low values of \HT.
The predictions for \ptbN{3}, \absetabN{3}, \ptbN{4}, and \absetabN{4} are generally compatible with the data, with only a slight trend visible in data towards more central (lower \abseta) and softer (lower \pt) \PQb jets.
Similarly, predictions for the \pt and \abseta for the extra \PQb jets all describe the data well within the measurement uncertainties.
The total number of jets is reasonably described by all predictions and does not show any clear trend.
\par}

The measured \dRbbavg shows a small trend towards lower values for all simulation approaches, but is compatible with the data within the uncertainties of the measurement, as is the maximum invariant mass of any \bbbar pair.

The observables related to the \bbextra pair are modelled well by most of the generator setups, except for \ttbarPH, which predicts a softer \mbbextra, and a smaller \dRbbextra.
The variables related to the \bbadd pair (Figs.~\ref{fig:results:dnn:1}--\ref{fig:results:dnn:2}), \ie the \PQb jets not part of the top quark decay chain, show similar behaviour, whereby \dRbbadd, \mbbadd and \ptbbadd are not described by \ttbarPH.
Predictions for the \pt and \abseta of the individual additional \PQb jets all tend to describe the data well.

In the \sjtbtl and \sejfbtl phase space regions (Fig.~\ref{fig:results:lj}), the \HTl and \dPhiljb observables are modelled well by most generator setups, with only \ttbbPP predicting a higher \HTl.
With the exception of \ttbarPH, \ptljXN{1} is observed to be softer than what is predicted by most generator setups.
In Appendix~\ref{app:qcdscales}, the measured normalized differential cross sections of the \ptljXN{1} and \HTl observables in the \sjtbtl phase space are compared with alternative \muR and \muF scale choices of the \ttbbPP simulation approach.
These comparisons show that increased \muR and \muF scales relative to the nominal scale choices (see Table~\ref{tab:generatorsettings}) tend to better describe these observables.

\section{Summary}
\label{sec:summary}

Measurements of inclusive and normalized differential cross sections of the associated production of top quark-antiquark and bottom quark-antiquark pairs, \ttbb, for events containing an electron or a muon, have been presented.
These measurements use proton-proton collision data recorded by the CMS detector at $\sqrt{s} = 13\TeV$ and correspond to an integrated luminosity of 138\fbinv.

{\tolerance=800
The inclusive cross sections are measured in four fiducial phase space regions requiring different jet, \PQb jet, and light jet multiplicities.
With total uncertainties of 6--17\%, depending on the phase space, these are the most precise measurements of the \ttbb cross section to date.
The uncertainties are dominated by systematic sources, with the leading uncertainties originating from the calibration of the \PQb tagging and of the jet energy scale, and from the choice of renormalization scale in the signal \ttbb and background \ttbar processes.
In most cases, the measured inclusive cross sections exceed the predictions with the chosen generator settings. The only exception is when using a particular choice of dynamic renormalization scale, $\smash[b]{\muR=\frac12 \prodttbb \mTi^{1/4}}$, where \mTisqdef are the transverse masses of top and bottom quarks.
\par}

Differential cross section measurements are performed as a function of several observables in the aforementioned phase space regions.
These observables mainly target \PQb jets as well as additional light jets produced in association with the top quark pairs.
In the phase space containing events with at least six jets, of which at least four are \PQb tagged, the additional \PQb-jet radiation is probed with two different approaches.
The first approach uses observables defined purely at the particle level, without any reference to the top quark decay chains, by selecting the two \PQb jets with the smallest angular separation.
The second approach uses explicitly the \PQb jets at the generator level that do not originate from top quark decays and identifies those jets at the detector level with a neural network discriminant.
The differential measurements have relative uncertainties in the range of 2--50\%, depending on the phase space and the observable.

The results are compared to the predictions of several event generator setups, and it is found that none of them simultaneously describe all measured distributions in the various phase space regions.
In the more inclusive phase space with five jets and three \PQb jets, the agreement between data and predictions is generally poor, while in the phase space with six jets and four \PQb jets, corresponding to the case in which the two additional \PQb jets in \ttbb production are resolved, most predictions are compatible with the data within the larger experimental uncertainties.
These measurements will help to further tune and refine the theoretical predictions and better assess the validity of the theoretical uncertainties estimated from the various \ttbb event generators.

\begin{acknowledgments}
We congratulate our colleagues in the CERN accelerator departments for the excellent performance of the LHC and thank the technical and administrative staffs at CERN and at other CMS institutes for their contributions to the success of the CMS effort. In addition, we gratefully acknowledge the computing centres and personnel of the Worldwide LHC Computing Grid and other centres for delivering so effectively the computing infrastructure essential to our analyses. Finally, we acknowledge the enduring support for the construction and operation of the LHC, the CMS detector, and the supporting computing infrastructure provided by the following funding agencies: SC (Armenia), BMBWF and FWF (Austria); FNRS and FWO (Belgium); CNPq, CAPES, FAPERJ, FAPERGS, and FAPESP (Brazil); MES and BNSF (Bulgaria); CERN; CAS, MoST, and NSFC (China); MINCIENCIAS (Colombia); MSES and CSF (Croatia); RIF (Cyprus); SENESCYT (Ecuador); MoER, ERC PUT and ERDF (Estonia); Academy of Finland, MEC, and HIP (Finland); CEA and CNRS/IN2P3 (France); SRNSF (Georgia); BMBF, DFG, and HGF (Germany); GSRI (Greece); NKFIH (Hungary); DAE and DST (India); IPM (Iran); SFI (Ireland); INFN (Italy); MSIP and NRF (Republic of Korea); MES (Latvia); LAS (Lithuania); MOE and UM (Malaysia); BUAP, CINVESTAV, CONACYT, LNS, SEP, and UASLP-FAI (Mexico); MOS (Montenegro); MBIE (New Zealand); PAEC (Pakistan); MES and NSC (Poland); FCT (Portugal); MESTD (Serbia); MCIN/AEI and PCTI (Spain); MOSTR (Sri Lanka); Swiss Funding Agencies (Switzerland); MST (Taipei); MHESI and NSTDA (Thailand); TUBITAK and TENMAK (Turkey); NASU (Ukraine); STFC (United Kingdom); DOE and NSF (USA).

\hyphenation{Rachada-pisek} Individuals have received support from the Marie-Curie programme and the European Research Council and Horizon 2020 Grant, contract Nos.\ 675440, 724704, 752730, 758316, 765710, 824093, and COST Action CA16108 (European Union); the Leventis Foundation; the Alfred P.\ Sloan Foundation; the Alexander von Humboldt Foundation; the Science Committee, project no. 22rl-037 (Armenia); the Belgian Federal Science Policy Office; the Fonds pour la Formation \`a la Recherche dans l'Industrie et dans l'Agriculture (FRIA-Belgium); the Agentschap voor Innovatie door Wetenschap en Technologie (IWT-Belgium); the F.R.S.-FNRS and FWO (Belgium) under the ``Excellence of Science -- EOS" -- be.h project n.\ 30820817; the Beijing Municipal Science \& Technology Commission, No. Z191100007219010 and Fundamental Research Funds for the Central Universities (China); the Ministry of Education, Youth and Sports (MEYS) of the Czech Republic; the Shota Rustaveli National Science Foundation, grant FR-22-985 (Georgia); the Deutsche Forschungsgemeinschaft (DFG), under Germany's Excellence Strategy -- EXC 2121 ``Quantum Universe" -- 390833306, and under project number 400140256 - GRK2497; the Hellenic Foundation for Research and Innovation (HFRI), Project Number 2288 (Greece); the Hungarian Academy of Sciences, the New National Excellence Program - \'UNKP, the NKFIH research grants K 124845, K 124850, K 128713, K 128786, K 129058, K 131991, K 133046, K 138136, K 143460, K 143477, 2020-2.2.1-ED-2021-00181, and TKP2021-NKTA-64 (Hungary); the Council of Science and Industrial Research, India; the Latvian Council of Science; the Ministry of Education and Science, project no. 2022/WK/14, and the National Science Center, contracts Opus 2021/41/B/ST2/01369 and 2021/43/B/ST2/01552 (Poland); the Funda\c{c}\~ao para a Ci\^encia e a Tecnologia, grant CEECIND/01334/2018 (Portugal); the National Priorities Research Program by Qatar National Research Fund; MCIN/AEI/10.13039/501100011033, ERDF ``a way of making Europe", and the Programa Estatal de Fomento de la Investigaci{\'o}n Cient{\'i}fica y T{\'e}cnica de Excelencia Mar\'{\i}a de Maeztu, grant MDM-2017-0765 and Programa Severo Ochoa del Principado de Asturias (Spain); the Chulalongkorn Academic into Its 2nd Century Project Advancement Project, and the National Science, Research and Innovation Fund via the Program Management Unit for Human Resources \& Institutional Development, Research and Innovation, grant B05F650021 (Thailand); the Kavli Foundation; the Nvidia Corporation; the SuperMicro Corporation; the Welch Foundation, contract C-1845; and the Weston Havens Foundation (USA).
\end{acknowledgments}

\bibliography{auto_generated}

\clearpage
\appendix
\numberwithin{figure}{section}
\numberwithin{table}{section}

\section{Leading nuisance parameter impacts}
\label{app:fidXS_impacts}

\begin{figure}[!hp]
\centering
\includegraphics[width=\textwidth]{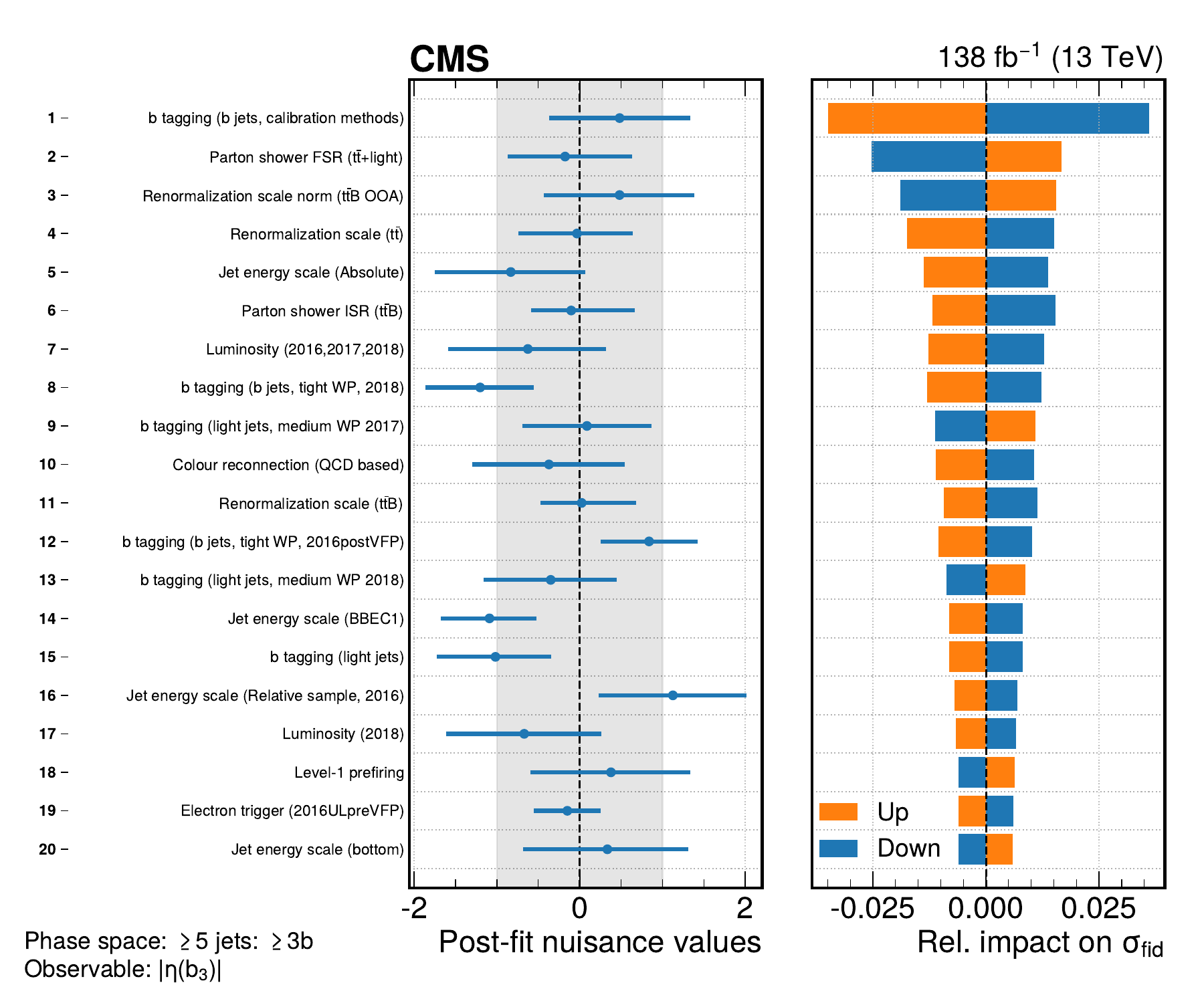}
\caption{%
    Post-fit nuisance parameter values and relative impacts on the fiducial cross section, for the fit of \abseta of the \PQb jet with third-highest \pt in the \fjtbLONG phase space. The nuisance parameters are defined such that the prefit value is zero with unity uncertainty.
}
\end{figure}

\begin{figure}[!hp]
\centering
\includegraphics[width=\textwidth]{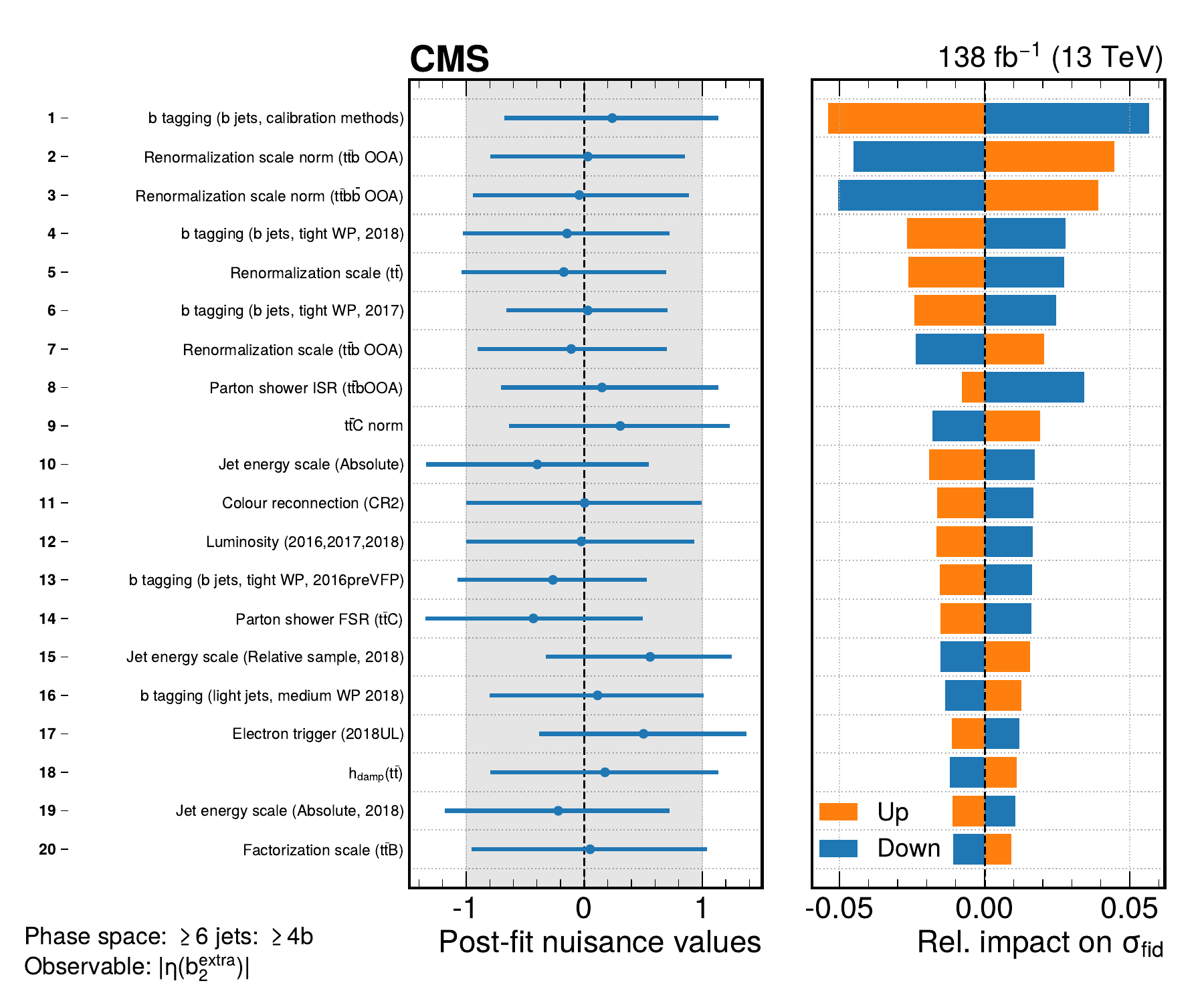}
\caption{%
    Post-fit nuisance parameter values and relative impacts on the fiducial cross section, for the fit of \abseta of the subleading extra \PQb jet in the \sjfbLONG phase space. The nuisance parameters are defined such that the prefit value is zero with unity uncertainty.
}
\end{figure}

\begin{figure}[!hp]
\centering
\includegraphics[width=\textwidth]{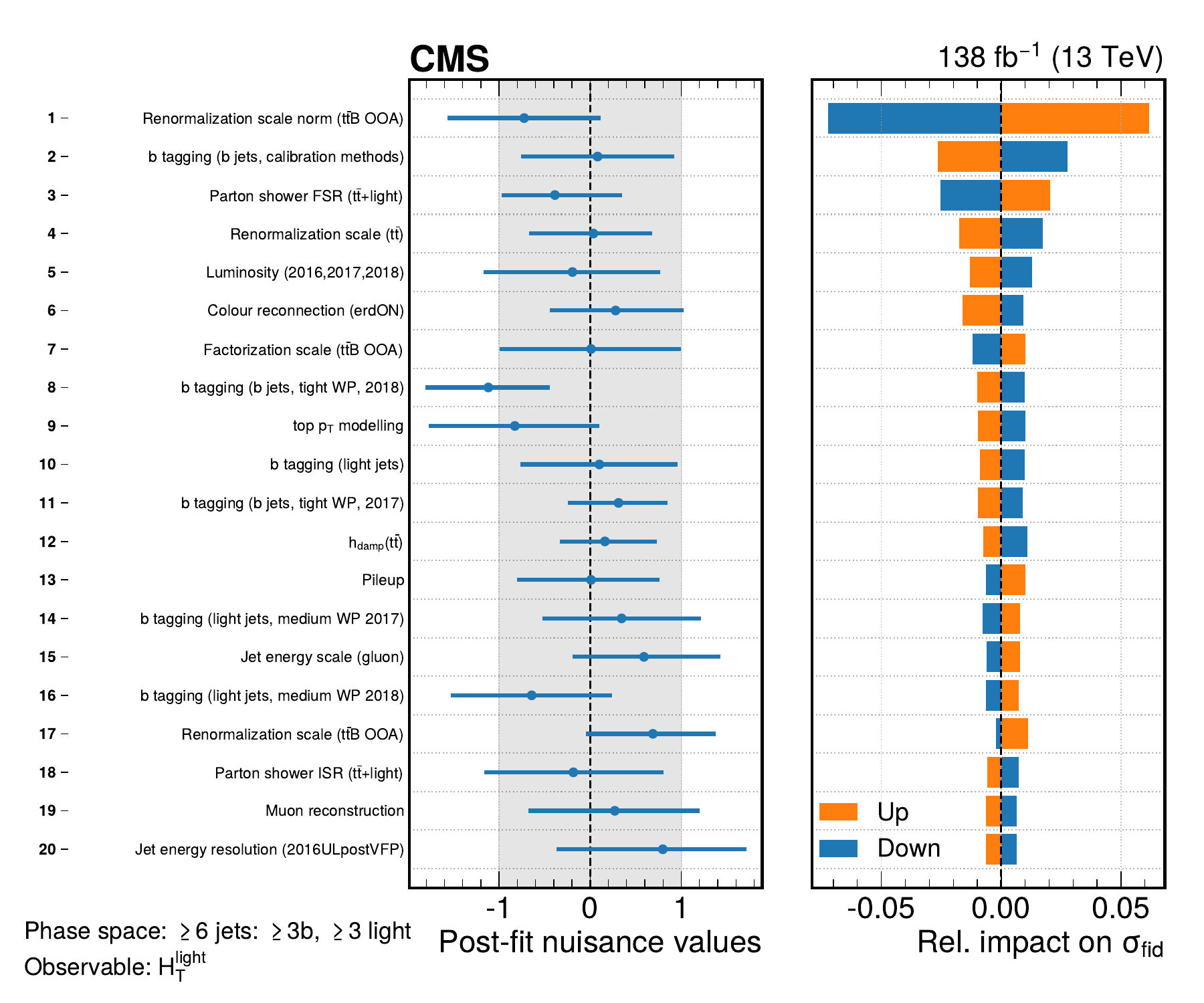}
\caption{%
    Post-fit nuisance parameter values and relative impacts on the fiducial cross section, for the fit of \HT of light jets in the \sjtbtlLONG phase space. The nuisance parameters are defined such that the prefit value is zero with unity uncertainty.
}
\end{figure}

\begin{figure}[!hp]
\centering
\includegraphics[width=\textwidth]{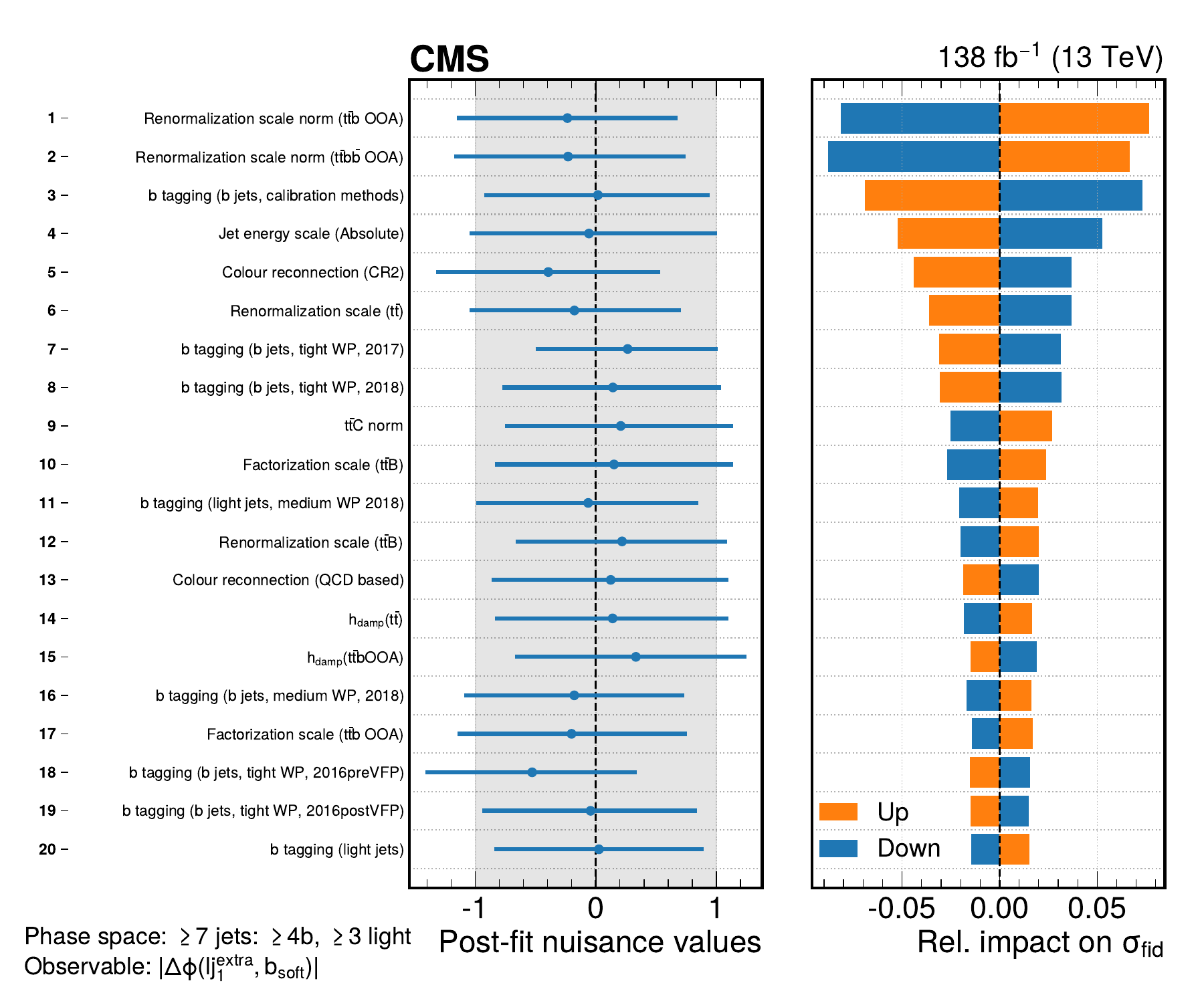}
\caption{%
    Post-fit nuisance parameter values and relative impacts on the fiducial cross section, for the fit of \dPhi between leading light jet and softest \PQb jet in the \sejfbtlLONG phase space. The nuisance parameters are defined such that the prefit value is zero with unity uncertainty.
}
\end{figure}

\clearpage
\section{Groups of impacts}
\label{app:diffXS_impacts}

\begin{figure}[!hp]
\centering
\includegraphics[width=\textwidth]{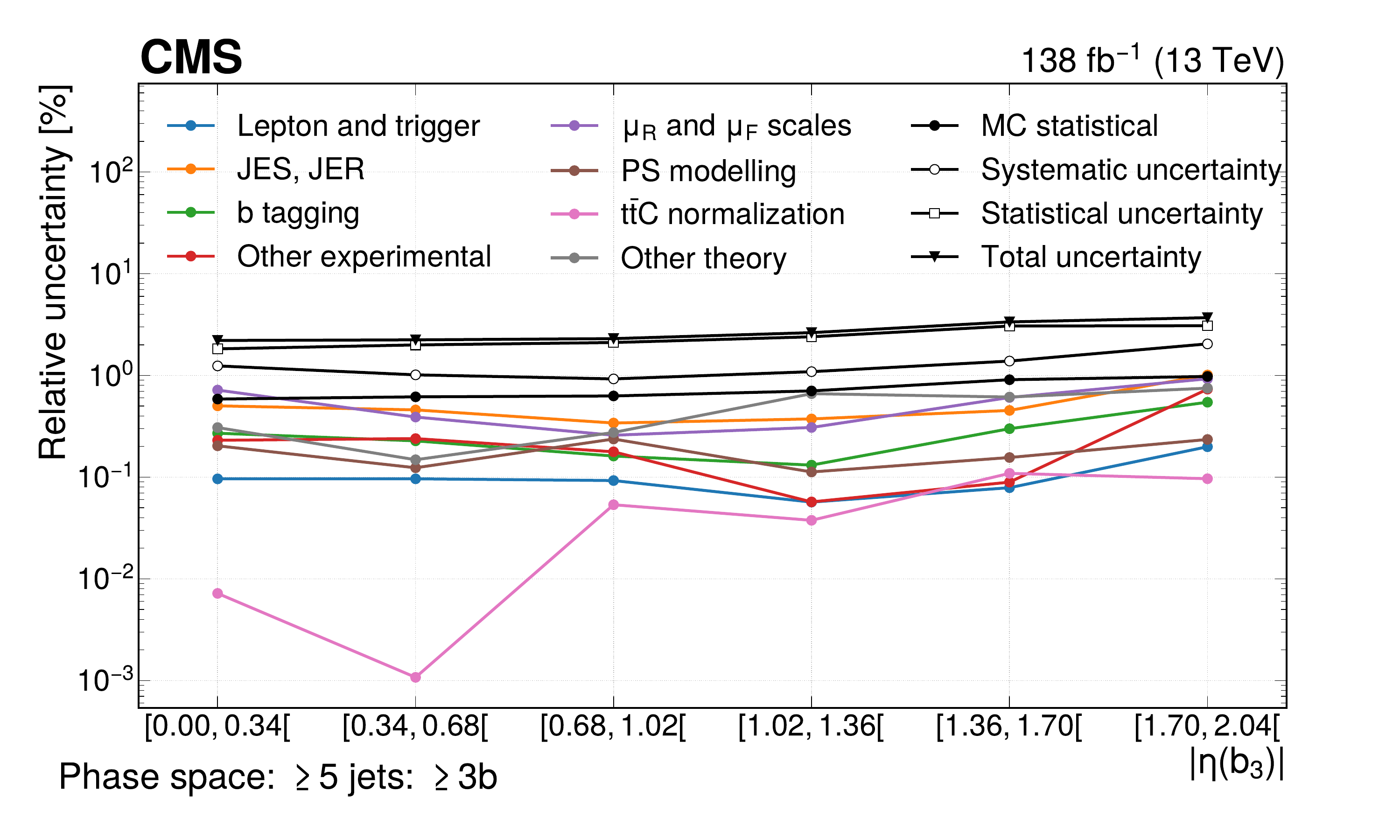}
\caption{%
    Effect of the considered sources of uncertainties on the measurement of the normalized differential cross section of the \abseta of the \PQb jet with third-highest \pt in the \fjtbLONG phase space, obtained by combining the impacts of associated nuisance parameters. The last bin of the distribution is not shown, since it has no associated parameter of interest but is constrained by the other bins. The category ``other theory'' includes \PQb quark fragmentation, top quark \pt modelling, PDF, \hdamp, colour reconnection, and underlying event uncertainties. The category ``other experimental'' includes pileup and the integrated luminosity uncertainties.
}
\end{figure}

\begin{figure}[!hp]
\centering
\includegraphics[width=\textwidth]{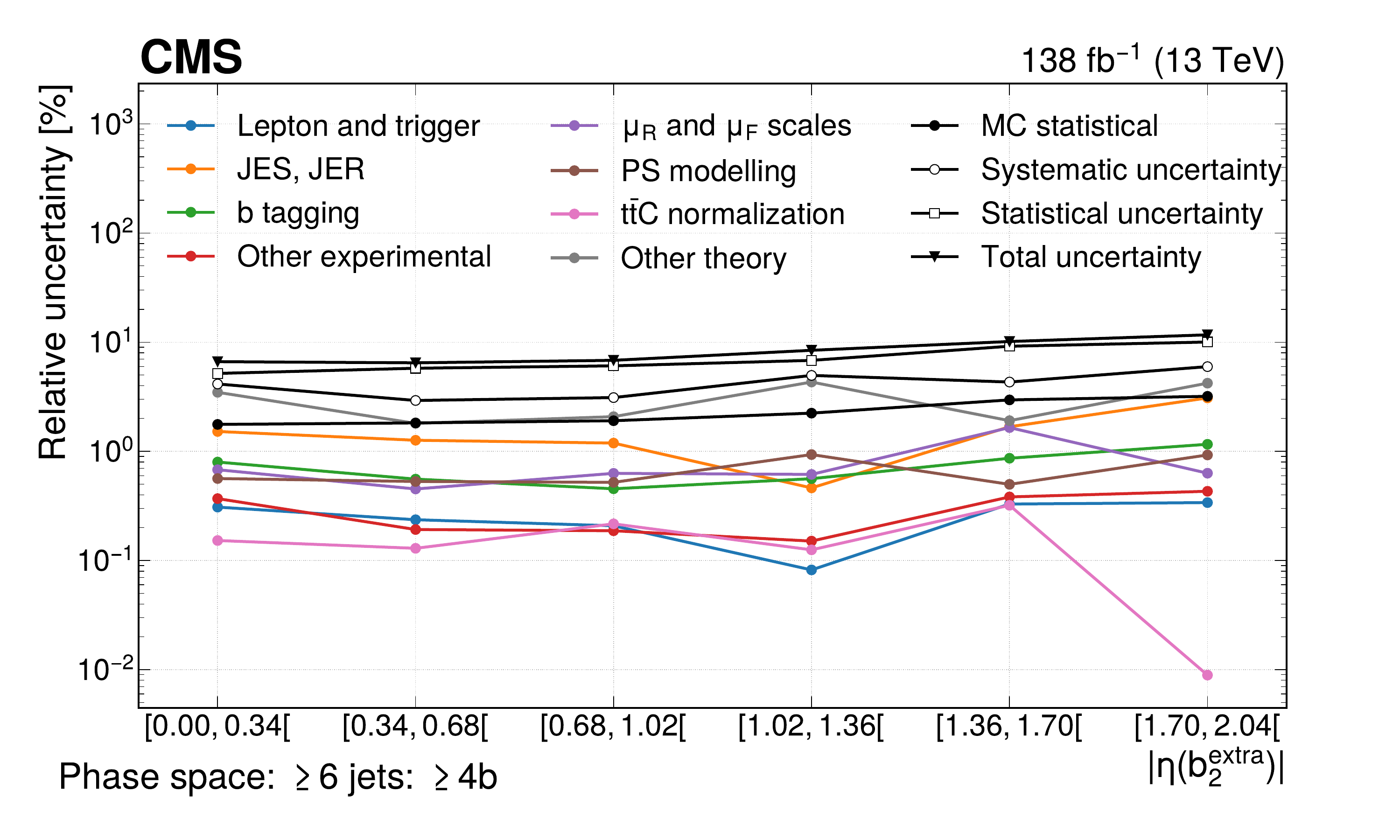}
\caption{%
    Effect of the considered sources of uncertainties on the measurement of the normalized differential cross section of the \abseta of the subleading extra \PQb jet in the \sjfbLONG phase space, obtained by combining the impacts of associated nuisance parameters. The last bin of the distribution is not shown, since it has no associated parameter of interest but is constrained by the other bins. The category ``other theory'' includes \PQb quark fragmentation, top quark \pt modelling, PDF, \hdamp, colour reconnection, and underlying event uncertainties. The category ``other experimental'' includes pileup and the integrated luminosity uncertainties.
}
\end{figure}

\begin{figure}[!hp]
\centering
\includegraphics[width=\textwidth]{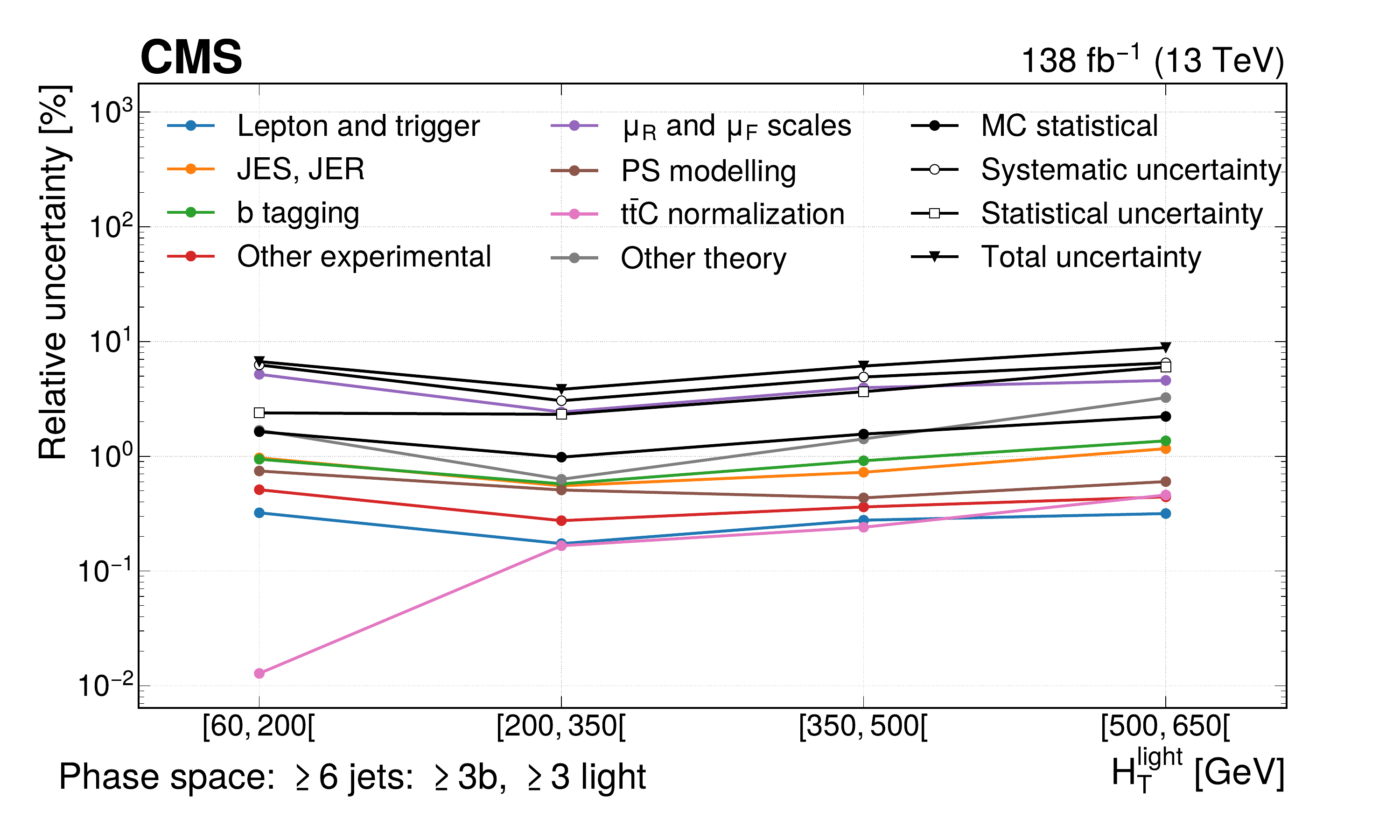}
\caption{%
    Effect of the considered sources of uncertainties on the measurement of the normalized differential cross section of the \HT of light jets in the \sjtbtlLONG phase space, obtained by combining the impacts of associated nuisance parameters. The last bin of the distribution is not shown, since it has no associated parameter of interest but is constrained by the other bins. The category ``other theory'' includes \PQb quark fragmentation, top quark \pt modelling, PDF, \hdamp, colour reconnection, and underlying event uncertainties. The category ``other experimental'' includes pileup and the integrated luminosity uncertainties.
}
\end{figure}

\begin{figure}[!hp]
\centering
\includegraphics[width=\textwidth]{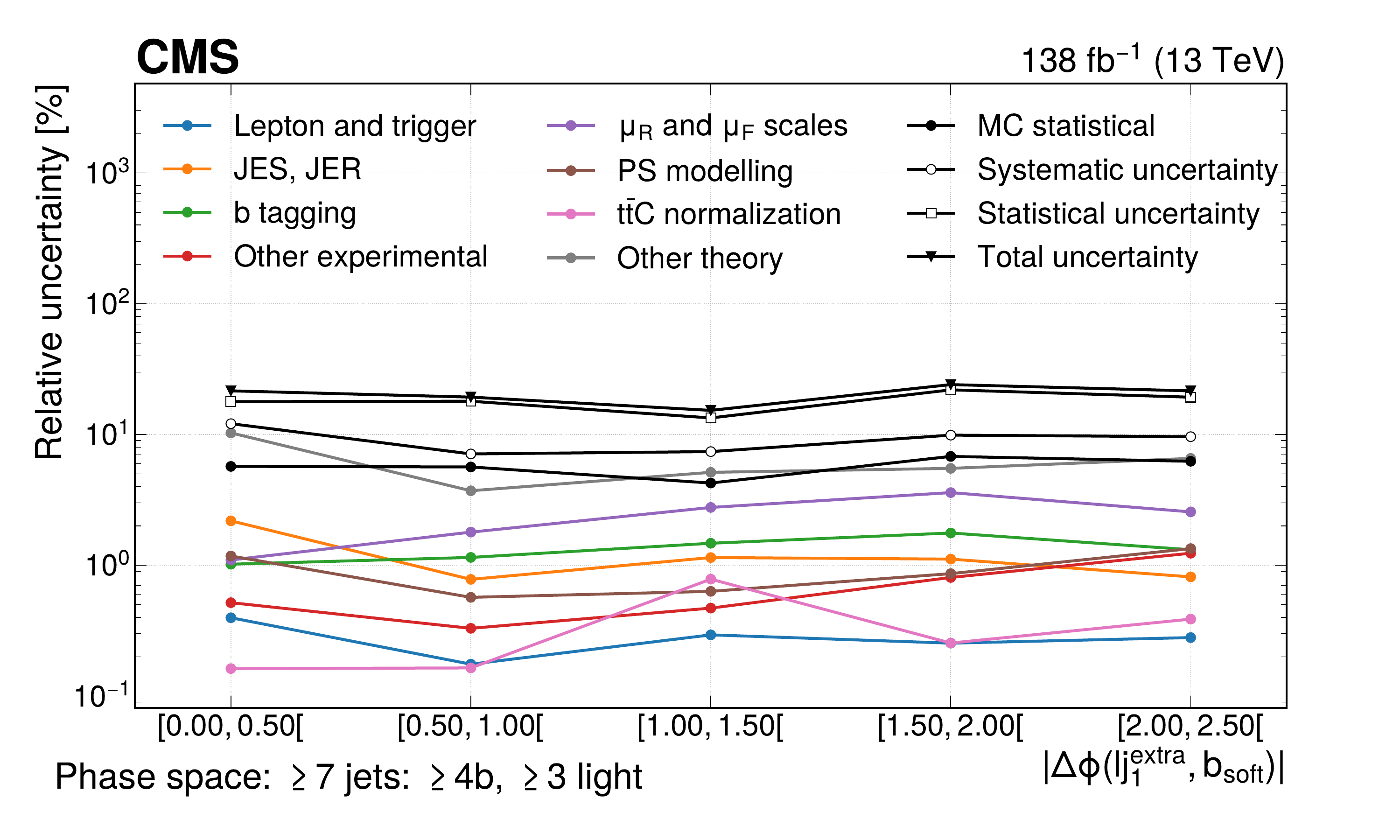}
\caption{%
    Effect of the considered sources of uncertainties on the measurement of the normalized differential cross section of the \dPhi between leading light jet and softest \PQb jet in the \sejfbtlLONG phase space, obtained by combining the impacts of associated nuisance parameters. The last bin of the distribution is not shown, since it has no associated parameter of interest but is constrained by the other bins. The category ``other theory'' includes \PQb quark fragmentation, top quark \pt modelling, PDF, \hdamp, colour reconnection, and underlying event uncertainties. The category ``other experimental'' includes pileup and the integrated luminosity uncertainties.
}
\end{figure}

\clearpage
\section{Correlation of POIs}
\label{app:correlations}

\begin{figure}[!hp]
\centering
\includegraphics[width=\textwidth]{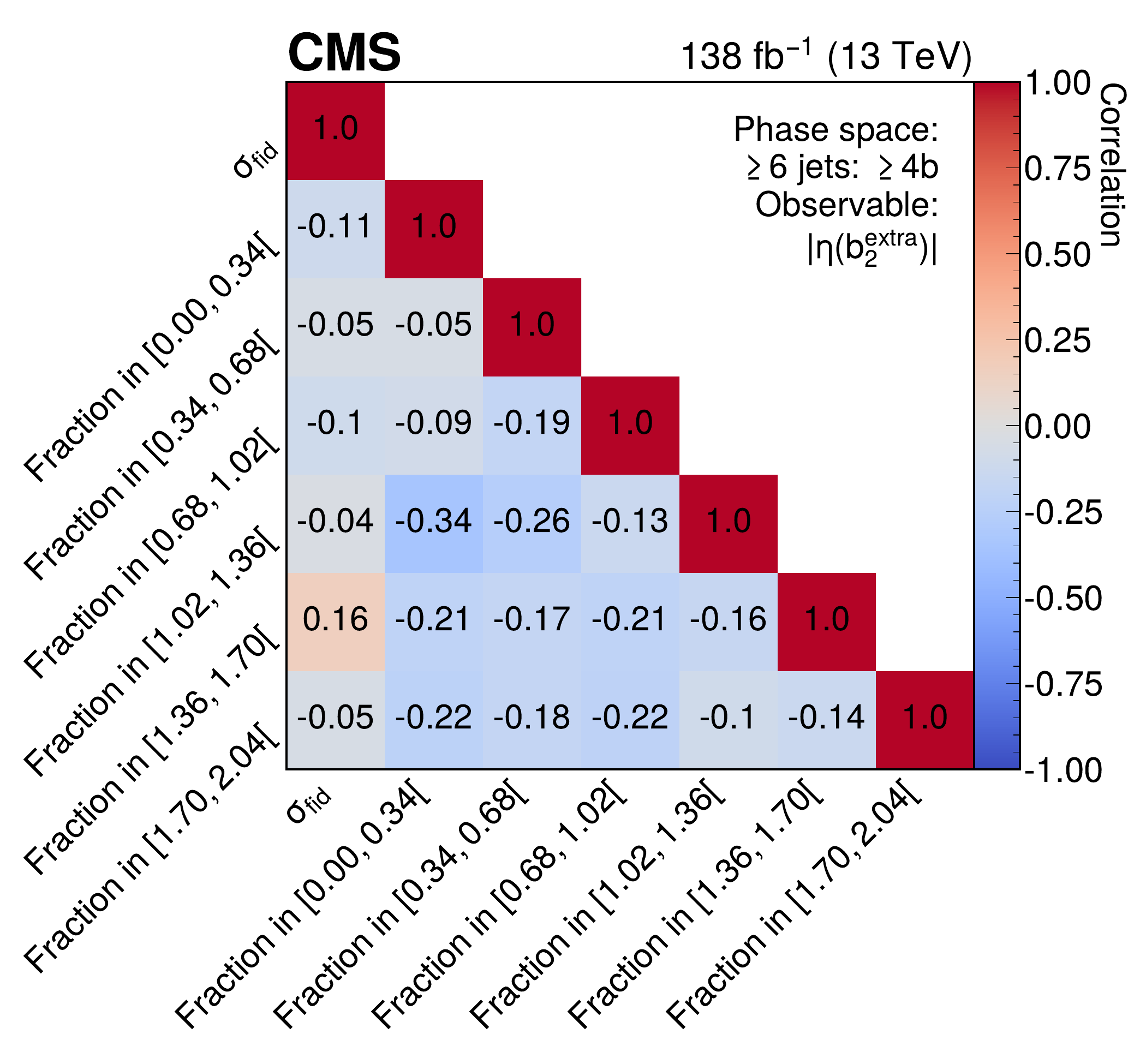}
\caption{%
    Correlations between the parameters of interest \vecmu in the fit of the \abseta of the subleading extra \PQb jet in the \sjfbLONG phase space.
}
\end{figure}

\begin{figure}[!hp]
\centering
\includegraphics[width=\textwidth]{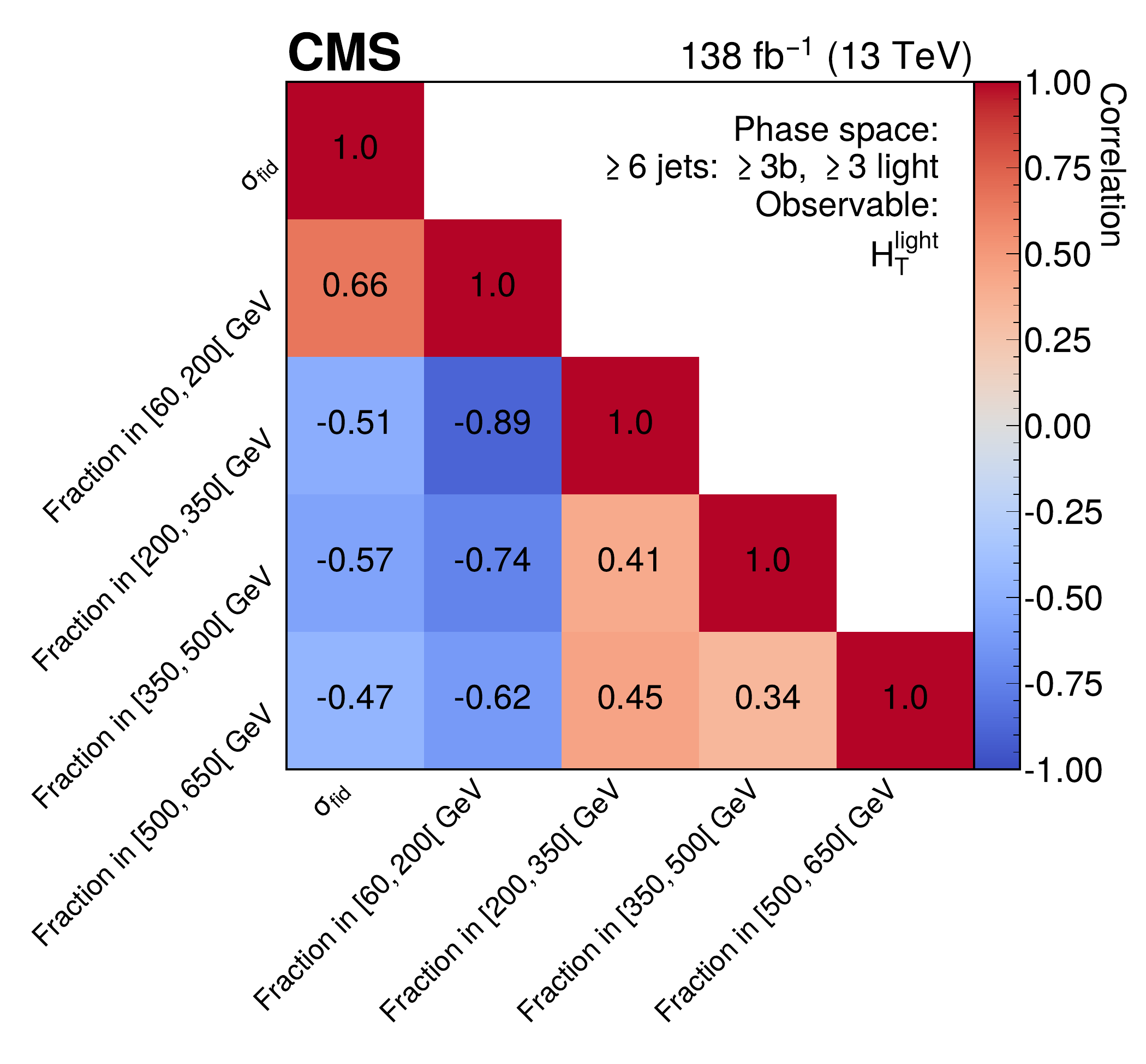}
\caption{%
    Correlations between the parameters of interest \vecmu in the fit of the \HT of light jets in the \sjtbtlLONG phase space.
}
\end{figure}

\begin{figure}[!hp]
\centering
\includegraphics[width=\textwidth]{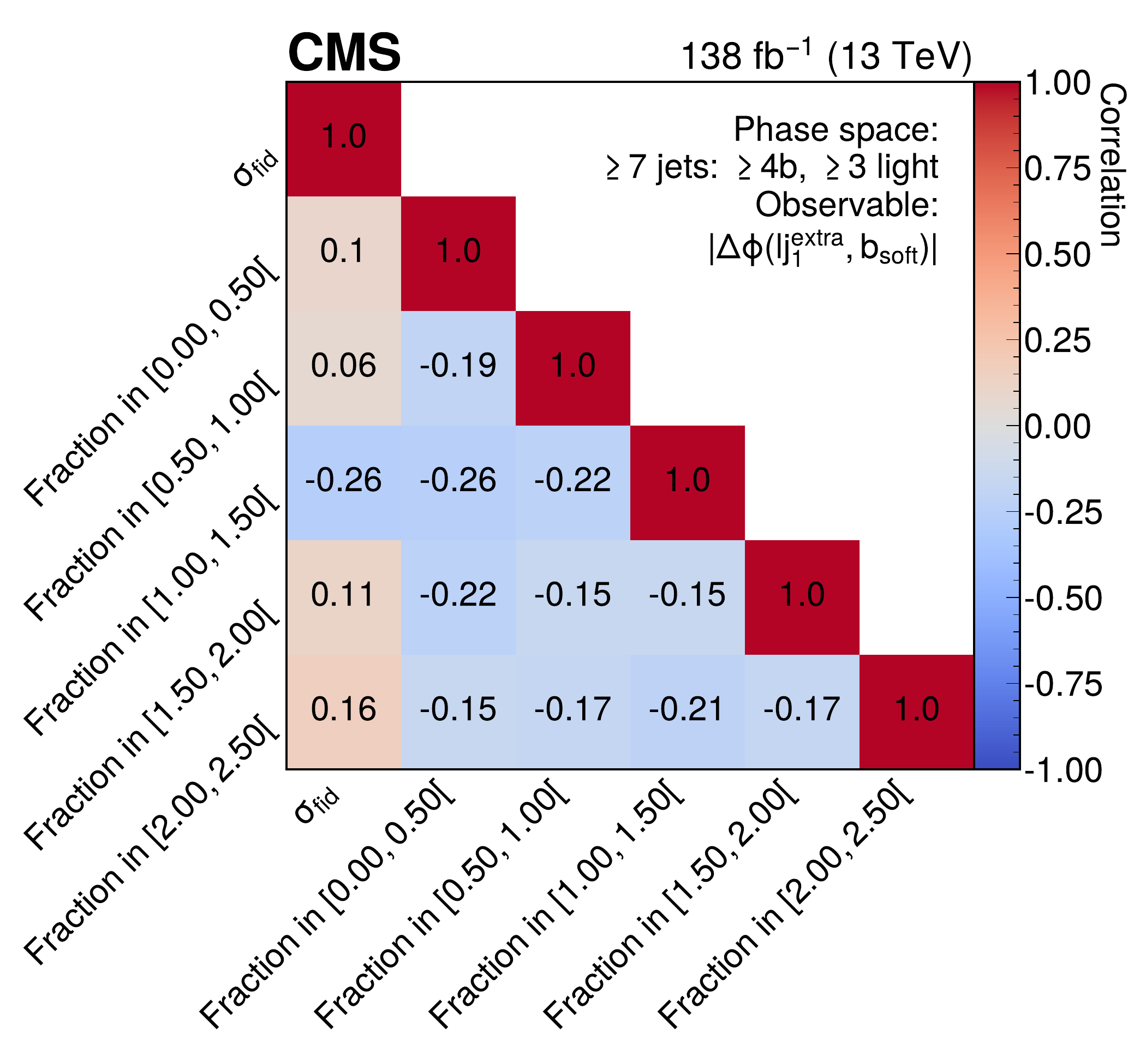}
\caption{%
    Correlations between the parameters of interest \vecmu in the fit of the \dPhi between leading light jet and softest \PQb jet in the \sejfbtlLONG phase space.
}
\end{figure}

\clearpage
\section{Variation of renormalization and factorization scales}
\label{app:qcdscales}

\begin{figure}[!hp]
\centering
\includegraphics[width=\textwidth]{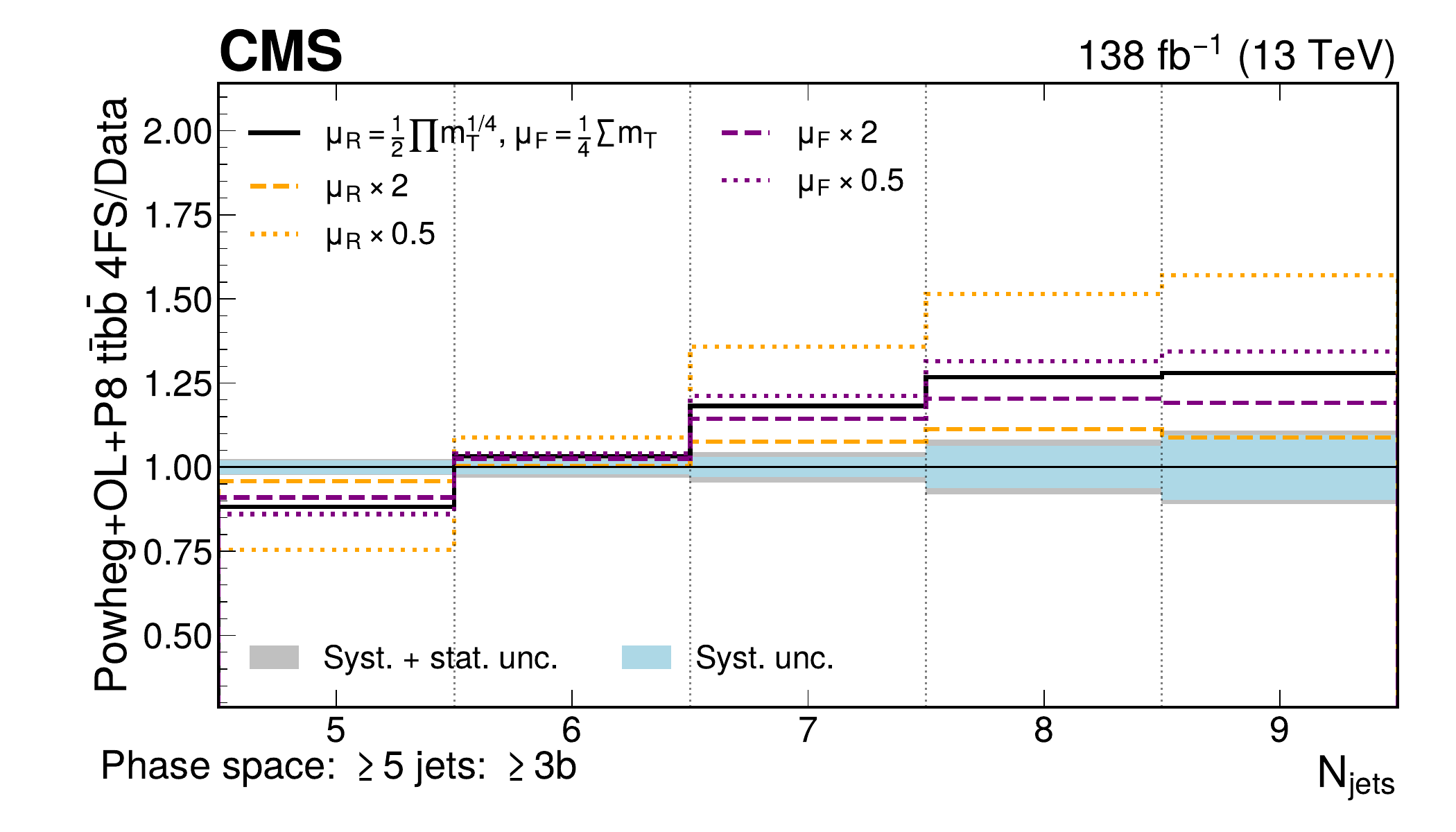} \\
\includegraphics[width=\textwidth]{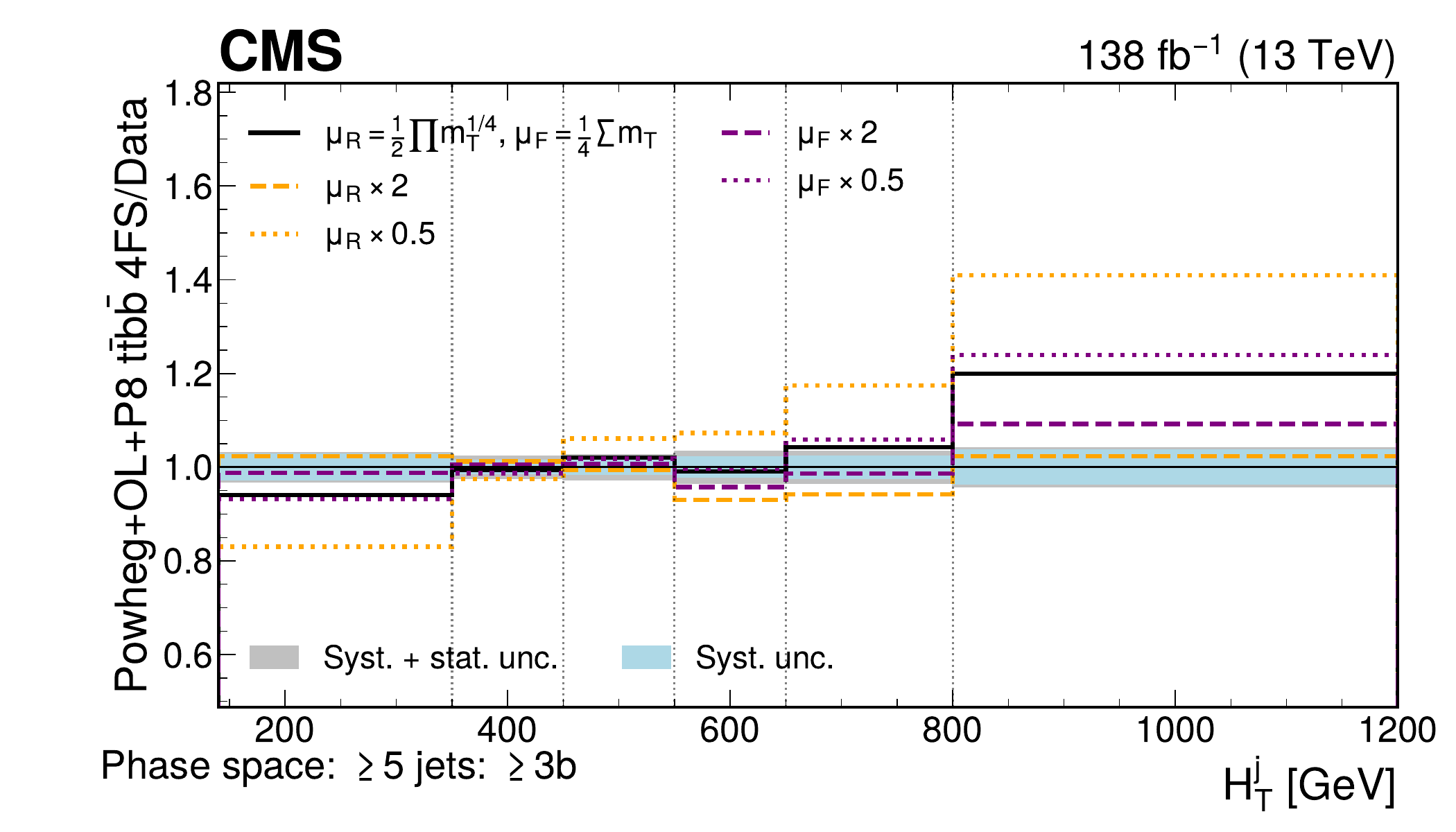}
\caption{%
    Ratio of normalized differential cross section predictions of the \ttbbPP modelling approach with different \muR and \muF scale settings relative to the measured normalized differential cross sections for the number of jets (upper) and \HT of jets (lower) in the \fjtbLONG phase space. The systematic (total) uncertainties of the measurement are represented as grey (blue) bands. Variations of the \muR (\muF) scale relative to the nominal scale setting are shown in orange (purple). The last bin in the distributions contains the overflow.
}
\end{figure}

\begin{figure}[!hp]
\centering
\includegraphics[width=\textwidth]{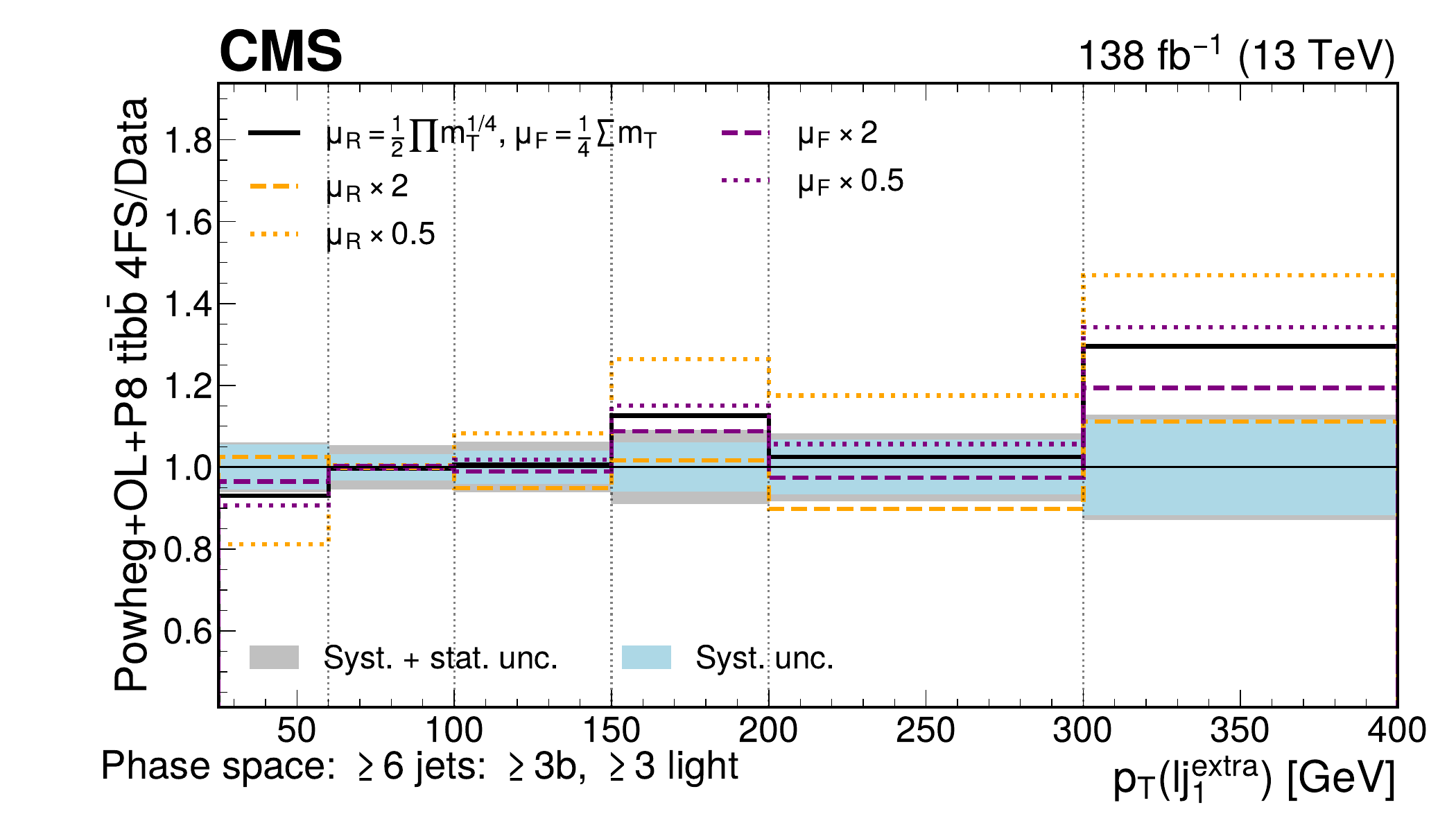} \\
\includegraphics[width=\textwidth]{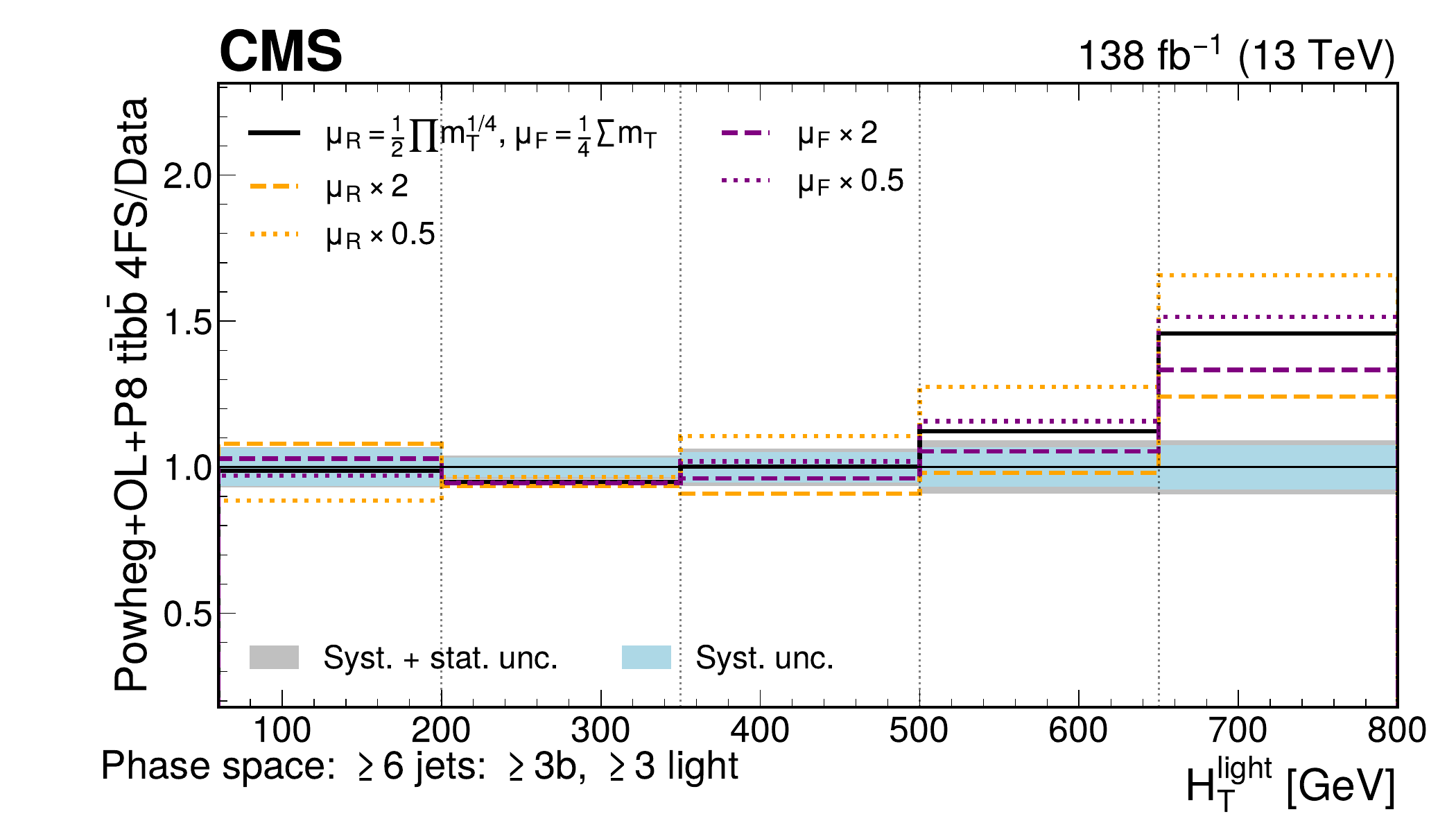}
\caption{%
    Ratio of normalized differential cross section predictions of the \ttbbPP modelling approach with different \muR and \muF scale settings relative to the measured normalized differential cross sections for the extra light jet (upper) and \HT of light jets (lower) in the \sjtbtlLONG phase space. The systematic (total) uncertainties of the measurement are represented as grey (blue) bands. The systematic (total) uncertainties of the measurement are represented as grey (blue) bands. Variations of the \muR (\muF) scale relative to the nominal scale setting are shown in orange (purple). The last bin in the distributions contains the overflow.
}
\end{figure}

\clearpage
\section{Normalized differential cross section compatibility}
\label{app:zscores}

\begin{table}[!hp]
\centering
\topcaption{%
    Observed $z$ score for each of the theoretical predictions in the \fjtb phase space, given the unfolded data and covariance matrix. For the determination of the $z$ score, only the measurement uncertainties are considered.
}
\renewcommand{\arraystretch}{1.2}
\begin{tabular}{lrrrrrr}
    & \multicolumn{6}{c}{Observed $z$ score} \\
    \fjtb phase space & \HTb & \HTj & \Nb & \Nj & \ptbN{3} & \absetabN{3} \\
    \hline
    \ttjetsAMC & 4.88 & 1.15 & 2.43 & 5.50 & 4.54 & 2.12 \\
    \ttbbAMC & $>$6 & 5.07 & $>$6 & $>$6 & $>$6 & 2.01 \\
    \ttbarPH & 2.00 & 4.10 & 0.46 & 4.45 & 1.53 & 2.27 \\
    \ttbbPP & 3.14 & 4.56 & 4.43 & $>$6 & 2.19 & 2.61 \\
    \ttbarPP & 3.01 & 3.05 & 1.73 & 5.17 & 1.88 & 2.52 \\
    \ttbbSherpa & 3.14 & 0.16 & 5.18 & 5.37 & 3.08 & 2.31 \\
\end{tabular}
\end{table}

\begin{table}[!hp]
\centering
\topcaption{%
    Observed $z$ score for each of the theoretical predictions in the \sjfb phase space, given the unfolded data and covariance matrix. For the determination of the $z$ score, only the measurement uncertainties are considered.
}
\renewcommand{\arraystretch}{1.2}
\cmsTable{\begin{tabular}{lrrrrrrrrr}
    & \multicolumn{9}{c}{Observed $z$ score} \\
    \sjfb phase space &  \HTb &  \HTj &  \Nj  &  \dRbbavg &  \mbbmax &  \ptbN{3} &  \ptbN{4} &  \absetabN{3} &  \absetabN{4} \\
    \hline
    \ttjetsAMC & 0.41 & $-0.83$ & 1.91 & $-1.30$ & 1.63 & 1.19 & 0.27 & 0.10 & 0.98 \\
    \ttbbAMC & 0.79 & 0.81 & 2.50 & $-0.02$ & 1.30 & 1.54 & 1.11 & 0.69 & 0.83 \\
    \ttbarPH & 0.92 & 2.44 & 1.80 & $-1.46$ & 0.04 & 1.24 & 0.68 & 0.50 & $-0.09$ \\
    \ttbbPP & 0.60 & 0.42 & 1.00 & $-1.29$ & 0.04 & 1.36 & 0.69 & 0.19 & 0.15 \\
    \ttbarPP & 0.68 & 0.10 & 1.08 & $-0.33$ & 0.34 & 1.34 & 0.56 & 0.38 & 0.18 \\
    \ttbbSherpa & 0.52 & 0.17 & 0.86 & 0.29 & 0.65 & 1.48 & 0.43 & 0.52 & 0.17 \\
\end{tabular}}
\end{table}

\begin{table}[!hp]
\centering
\topcaption{%
    Observed $z$ score for each of the theoretical predictions in the \sjfb phase space of the observables related to the \bbextra pair, given the unfolded data and covariance matrix. For the determination of the $z$ score, only the measurement uncertainties are considered.
}
\renewcommand{\arraystretch}{1.2}
\cmsTable{\begin{tabular}{lrrrrrrrr}
    & \multicolumn{8}{c}{Observed $z$ score} \\
    \sjfb phase space &  \dRbbextra & \mbbextra &  \ptbbextra &  \ptbXN{1} &  \ptbXN{2} &  \absetabXN{1} & \absetabXN{2} &  \absetabbextra \\
    \hline
    \ttjetsAMC & 1.31 & 0.56 & 0.42 & 0.27 & 0.63 & 0.76 & 0.17 & 0.51 \\
    \ttbbAMC & 0.23 & $-0.02$ & 0.50 & 0.52 & $-0.56$ & 0.38 & $-0.95$ & 0.62 \\
    \ttbarPH & 3.17 & 1.60 & 1.00 & 1.41 & 0.26 & 0.47 & $-0.39$ & 0.88 \\
    \ttbbPP & $-0.79$ & $-0.69$ & 0.60 & 1.05 & $-0.19$ & $-0.31$ & $-1.15$ & 0.38 \\
    \ttbarPP & 1.33 & 0.07 & 0.85 & 1.24 & $-0.19$ & 0.03 & $-1.25$ & 1.07 \\
    \ttbbSherpa & 0.03 & $-0.99$ & 0.95 & 1.12 & $-0.52$ & 0.13 & $-1.02$ & 0.40 \\
\end{tabular}}
\end{table}

\begin{table}[!hp]
\centering
\topcaption{%
    Observed $z$ score for each of the theoretical predictions in the \sjfb phase space of the observables related to the \bbadd pair, given the unfolded data and covariance matrix. For the determination of the $z$ score, only the measurement uncertainties are considered.
}
\renewcommand{\arraystretch}{1.2}
\cmsTable{\begin{tabular}{lrrrrrrrr}
    & \multicolumn{8}{c}{Observed $z$ score} \\
    \sjfb phase space &  \dRbbadd &  \mbbadd & \ptbbadd &  \ptbAN{1} & \ptbAN{2} &  \absetabbadd &  \absetabAN{1} &  \absetabAN{2} \\
    \hline
    \ttjetsAMC & 1.09 & 0.81 & 0.83 & 1.14 & 0.86 & 0.30 & 1.51 & 0.27 \\
    \ttbbAMC & $-1.86$ & $-0.14$ & 0.72 & 1.69 & 1.27 & 0.51 & 1.49 & $-0.09$ \\
    \ttbarPH & 3.62 & 2.45 & 2.58 & 2.61 & 1.05 & 0.19 & 1.25 & $-0.33$ \\
    \ttbbPP & $-2.24$ & $-1.43$ & 1.48 & 2.17 & 0.90 & 0.29 & 1.09 & $-0.36$ \\
    \ttbarPP & $-1.21$ & $-0.68$ & 2.24 & 2.52 & 0.83 & 0.95 & 1.44 & $-0.57$ \\
\end{tabular}}
\end{table}

\begin{table}[!hp]
\centering
\topcaption{%
    Observed $z$ score for each of the theoretical predictions in the \sjtbtl phase space, given the unfolded data and covariance matrix. For the determination of the $z$ score, only the measurement uncertainties are considered.
}
\renewcommand{\arraystretch}{1.2}
\begin{tabular}{lrrr}
    & \multicolumn{3}{c}{Observed $z$ score} \\
    \sjtbtl phase space & \HTl & \ptljXN{1} & \dPhiljb \\
    \hline
    \ttjetsAMC & 1.71 & $-0.14$ & 0.81 \\
    \ttbbAMC & 3.55 & 4.14 & 1.62 \\
    \ttbarPH & 2.02 & 1.84 & 1.03 \\
    \ttbbPP & 5.39 & 1.62 & $-0.17$ \\
    \ttbarPP & 1.52 & 0.24 & 0.24 \\
    \ttbbSherpa & 0.33 & 0.78 & 0.30 \\
\end{tabular}
\end{table}

\begin{table}[!hp]
\centering
\topcaption{%
    Observed $z$ score for each of the theoretical predictions in the \sejfbtl phase space, given the unfolded data and covariance matrix. For the determination of the $z$ score, only the measurement uncertainties are considered.
}
\renewcommand{\arraystretch}{1.2}
\begin{tabular}{lrrr}
    & \multicolumn{3}{c}{Observed $z$ score} \\
    \sejfbtl phase space & \HTl & \ptljXN{1} & \dPhiljb \\
    \hline
    \ttjetsAMC & 1.14 & 1.66 & 0.10 \\
    \ttbbAMC & $-0.92$ & 2.12 & 0.16 \\
    \ttbarPH & 0.49 & $-0.06$ & 0.03 \\
    \ttbbPP & 2.00 & 2.30 & $-0.03$ \\
    \ttbarPP & 0.89 & 1.83 & 0.13 \\
    \ttbbSherpa & 0.31 & 0.24 & 0.00 \\
\end{tabular}
\end{table}
\cleardoublepage \section{The CMS Collaboration \label{app:collab}}\begin{sloppypar}\hyphenpenalty=5000\widowpenalty=500\clubpenalty=5000
\cmsinstitute{Yerevan Physics Institute, Yerevan, Armenia}
{\tolerance=6000
A.~Hayrapetyan, A.~Tumasyan\cmsAuthorMark{1}\cmsorcid{0009-0000-0684-6742}
\par}
\cmsinstitute{Institut f\"{u}r Hochenergiephysik, Vienna, Austria}
{\tolerance=6000
W.~Adam\cmsorcid{0000-0001-9099-4341}, J.W.~Andrejkovic, T.~Bergauer\cmsorcid{0000-0002-5786-0293}, S.~Chatterjee\cmsorcid{0000-0003-2660-0349}, K.~Damanakis\cmsorcid{0000-0001-5389-2872}, M.~Dragicevic\cmsorcid{0000-0003-1967-6783}, A.~Escalante~Del~Valle\cmsorcid{0000-0002-9702-6359}, P.S.~Hussain\cmsorcid{0000-0002-4825-5278}, M.~Jeitler\cmsAuthorMark{2}\cmsorcid{0000-0002-5141-9560}, N.~Krammer\cmsorcid{0000-0002-0548-0985}, D.~Liko\cmsorcid{0000-0002-3380-473X}, I.~Mikulec\cmsorcid{0000-0003-0385-2746}, J.~Schieck\cmsAuthorMark{2}\cmsorcid{0000-0002-1058-8093}, R.~Sch\"{o}fbeck\cmsorcid{0000-0002-2332-8784}, D.~Schwarz\cmsorcid{0000-0002-3821-7331}, M.~Sonawane\cmsorcid{0000-0003-0510-7010}, S.~Templ\cmsorcid{0000-0003-3137-5692}, W.~Waltenberger\cmsorcid{0000-0002-6215-7228}, C.-E.~Wulz\cmsAuthorMark{2}\cmsorcid{0000-0001-9226-5812}
% ...
\par}
\vskip\cmsinstskip
\dag:~Deceased\\
$^{1}$Also at Yerevan State University, Yerevan, Armenia\\
$^{2}$Also at TU Wien, Vienna, Austria\\
% ...
\end{sloppypar}
\end{document}